\documentclass{article}

\usepackage{amssymb,amsfonts,amsmath}
\usepackage{enumerate,float}
\usepackage{color}
\usepackage{tikz}
\usepackage{subfigure}
\usetikzlibrary{arrows,snakes,backgrounds}
\usepackage{graphicx,filecontents,stackengine}

\def\be{\begin{eqnarray}}
\def\ee{\end{eqnarray}}
\def\nn{\nonumber}

\date{}

\def\tr{{\rm tr}\,}
\def\Tr{{\rm Tr}\,}

\def\PPhi{\phi}
\def\PPPhi{\Phi}
\def\horr{{{\smallsmile}\atop{\smallfrown}}}

\definecolor{red}{rgb}{1,0,0}
\definecolor{orange}{rgb}{1,0.5,0}
\definecolor{violet}{rgb}{0.7,0,1}

\usepackage{setspace}
\usepackage[utf8]{inputenc}
\usepackage[english]{babel}
\usepackage{amsfonts,amsmath,amssymb,mathtools}
\usepackage[left=2cm,right=2cm,top=2cm,bottom=2cm,bindingoffset=0cm]{geometry}
\usepackage{graphicx}
\usepackage{authblk}
\usepackage{xcolor}
\usepackage{hyperref}
\usepackage{dsfont}
\usepackage{authblk}
\usepackage{slashed}
\usepackage{braket}
\usepackage{comment}
\usepackage{pb-diagram}
\usepackage{wasysym}
\usepackage{amsthm}
\usepackage{cancel}
\usepackage{tikz}\usetikzlibrary{tikzmark}
\usetikzlibrary{fadings}
\usetikzlibrary{positioning}% To get more advances positioning options
\usetikzlibrary{backgrounds} % LATEX and plain TEX
\usetikzlibrary{decorations.pathreplacing}
\usetikzlibrary{arrows}% To get more arrow heads
\definecolor{airforceblue}{rgb}{0.36, 0.54, 0.66}	
\definecolor{beige}{rgb}{0.96, 0.96, 0.86}
\definecolor{bittersweet}{rgb}{1.0, 0.44, 0.37}
\definecolor{melon}{rgb}{0.99, 0.74, 0.71}
\definecolor{mustard}{rgb}{1.0, 0.86, 0.35}
\definecolor{lava}{rgb}{0.81, 0.06, 0.13}
\definecolor{magnolia}{rgb}{0.97, 0.96, 1.0}
\definecolor{lavendermist}{rgb}{0.9, 0.9, 0.98}
\definecolor{lavendergray}{rgb}{0.77, 0.76, 0.82}
\definecolor{palepink}{rgb}{0.98, 0.85, 0.87}
\definecolor{palesilver}{rgb}{0.79, 0.75, 0.73}
\definecolor{cadetgrey}{rgb}{0.57, 0.64, 0.69}
\definecolor{anti-flashwhite}{rgb}{0.95, 0.95, 0.96}
\colorlet{Light0anti-flashwhite}{anti-flashwhite!70!white}
\colorlet{Lightanti-flashwhite}{anti-flashwhite!50!white}
\colorlet{Light2anti-flashwhite}{anti-flashwhite!30!white}
\definecolor{linkcolor}{rgb}{0,0,1}
\definecolor{urlcolor}{rgb}{0,0,1}

\usepackage{tikz-cd}
\usetikzlibrary{decorations.markings}
\usetikzlibrary{positioning}
\usepackage{braids}
\usepackage{array}
\usepackage{ytableau}
\usepackage{longtable}
\usepackage{empheq}
\usepackage{multicol}
\usepackage{mathrsfs}
\usepackage{diagbox}
\usepackage[vcentermath]{youngtab}
\usepackage[natbib=true, backend = bibtex, style=numeric-comp, sorting=none,maxbibnames=99]{biblatex}
\hypersetup{pdfstartview=FitH,  linkcolor=linkcolor,urlcolor=urlcolor,citecolor=urlcolor, colorlinks=true}

\makeatletter
\let\save@mathaccent\mathaccent
\newcommand*\if@single[3]{%
  \setbox0\hbox{${\mathaccent"0362{#1}}^H$}%
  \setbox2\hbox{${\mathaccent"0362{\kern0pt#1}}^H$}%
  \ifdim\ht0=\ht2 #3\else #2\fi
  }
\newcommand*\rel@kern[1]{\kern#1\dimexpr\macc@kerna}
\newcommand*\widebar[1]{\@ifnextchar^{{\wide@bar{#1}{0}}}{\wide@bar{#1}{1}}}
\newcommand*\wide@bar[2]{\if@single{#1}{\wide@bar@{#1}{#2}{1}}{\wide@bar@{#1}{#2}{2}}}
\newcommand*\wide@bar@[3]{%
  \begingroup
  \def\mathaccent##1##2{%
    \let\mathaccent\save@mathaccent
    \if#32 \let\macc@nucleus\first@char \fi
    \setbox\z@\hbox{$\macc@style{\macc@nucleus}_{}$}%
    \setbox\tw@\hbox{$\macc@style{\macc@nucleus}{}_{}$}%
    \dimen@\wd\tw@
    \advance\dimen@-\wd\z@
    \divide\dimen@ 3
    \@tempdima\wd\tw@
    \advance\@tempdima-\scriptspace
    \divide\@tempdima 10
    \advance\dimen@-\@tempdima
    \ifdim\dimen@>\z@ \dimen@0pt\fi
    \rel@kern{0.6}\kern-\dimen@
    \if#31
      \overline{\rel@kern{-0.6}\kern\dimen@\macc@nucleus\rel@kern{0.4}\kern\dimen@}%
      \advance\dimen@0.4\dimexpr\macc@kerna
      \let\final@kern#2%
      \ifdim\dimen@<\z@ \let\final@kern1\fi
      \if\final@kern1 \kern-\dimen@\fi
    \else
      \overline{\rel@kern{-0.6}\kern\dimen@#1}%
    \fi
  }%
  \macc@depth\@ne
  \let\math@bgroup\@empty \let\math@egroup\macc@set@skewchar
  \mathsurround\z@ \frozen@everymath{\mathgroup\macc@group\relax}%
  \macc@set@skewchar\relax
  \let\mathaccentV\macc@nested@a
  \if#31
    \macc@nested@a\relax111{#1}%
  \else
    \def\gobble@till@marker##1\endmarker{}%
    \futurelet\first@char\gobble@till@marker#1\endmarker
    \ifcat\noexpand\first@char A\else
      \def\first@char{}%
    \fi
    \macc@nested@a\relax111{\first@char}%
  \fi
  \endgroup
}
\makeatother

\begin{document}

\title{\bf Planar decomposition of bipartite HOMFLY polynomials in symmetric representations
}

\author[2,3]{{\bf A. Anokhina}\thanks{\href{mailto:anokhina@itep.ru}{anokhina@itep.ru}}}
\author[1,2,3]{{\bf E. Lanina}\thanks{\href{mailto:lanina.en@phystech.edu}{lanina.en@phystech.edu}}}
\author[1,2,3]{{\bf A. Morozov}\thanks{\href{mailto:morozov@itep.ru}{ morozov@itep.ru}}}

\vspace{5cm}

\affil[1]{Moscow Institute of Physics and Technology, 141700, Dolgoprudny, Russia}
%\affil[2]{Institute for Theoretical and Experimental Physics, 117218, Moscow, Russia}
\affil[2]{Institute for Information Transmission Problems, 127051, Moscow, Russia}
\affil[3]{NRC "Kurchatov Institute", 123182, Moscow, Russia}
\affil[4]{Institute for Theoretical and Experimental Physics, 117218, Moscow, Russia}
\renewcommand\Affilfont{\itshape\small}

\maketitle

\vspace{-7.5cm}

\begin{center}
	\hfill MIPT/TH-19/24\\
	\hfill ITEP/TH-25/24\\
	\hfill IITP/TH-20/24
\end{center}

\vspace{4.5cm}

\begin{abstract}

{
We generalize the recently discovered  planar decomposition (Kauffman bracket) for the HOMFLY polynomials
of bipartite knot/link diagrams to (anti)symmetrically colored HOMFLY polynomials.
Cabling destroys planarity, but it is restored after projection to (anti)symmetric representations.
This allows to go beyond arborescent calculus, which so far produced the majority of results
for colored polynomials.
%??????In particular, we show that HOMFLY polynomials for Kanenobu knot and its link generalizations
%remain dependent on the sum of two evolution parameters for all symmetric representations --
%despite these are not mutants.??????
Technicalities include combinations of projectors,
and these can be handled rigorously, without any guess-work --
what can be also useful for other considerations, where reliable quantization was so far unavailable. We explicitly provide simple examples of calculation of the HOMFLY polynomials in symmetric representations with the use of our planar technique. These examples reveal what we call the bipartite evolution and the bipartite decomposition of squares of $\mathcal{R}$-matrices eigenvalues in the antiparallel channel.
}
%beyond/
\end{abstract}

\tableofcontents

\section{Introduction}

In this paper, we show that the planar resolution technique is applicable to the symmetrically colored HOMFLY polynomial at generic $N$ for a special class of {\it bipartite} links made from the lock tangles (see Fig.\,\ref{fig:pladeco}), but not just for $N=2$ and for the fundamental representation as stated by the Kauffman bracket in Fig.\,\ref{fig:Kauff}.

\subsection{Non-perturbative calculations in the Chern--Simons theory}\label{sec:RT-CS}

We study the gauge invariant observables -- the Wilson loops 
\begin{equation}
     \label{WilsonLoopExpValue}
         H_{R}^{\mathcal{K}}(q, A) = \left\langle \text{tr}_{R} \ P \exp \left( \oint_{\mathcal{K}} {\cal A} \right) \right\rangle_{\text{CS}}
     \end{equation}
in the 3-dimensional Chern--Simons theory~\cite{CS,Witten,MoSmi} with $SU(N)$ gauge group and the following action:
\begin{equation}
         S_{\text{CS}}[{\cal A}] = \frac{\kappa}{4 \pi} \int_{S^3} \text{tr} \left( {\cal A} \wedge d{\cal A} +  \frac{2}{3} {\cal A} \wedge {\cal A} \wedge {\cal A} \right).
\end{equation}
Here we consider the gauge fields taken in an arbitrary finite-dimensional irreducible $\mathfrak{sl}_N$ representation ${\cal A}={\cal A}_\mu^a T_a^R dx^\mu$ with $T_a$ being generators of $\mathfrak{sl}_N$ Lie algebra and the integration contour tied in an arbitrary knot or link. In the last case, there are just several Wilson operators corresponding to different link components. 

Due to the theory being topological, the observables~\eqref{WilsonLoopExpValue} depend on an isotopy class of a knot but not on its special embedding in the 3d Euclidean space. %Such objects are usually called knot invariants. 
Moreover, it turned out~\cite{Witten} that these Wilson loops are polynomial in the special variables $q = \exp\left(\frac{2\pi i}{\kappa +N}\right)$ and $A=q^N$ and coincide with the quantum link invariants\footnote{The colored HOMFLY polynomials are called quantum link invariants as being connected with the quantum universal enveloping algebra $U_q(\mathfrak{sl}_N)$.} -- the so-called colored (by a representation $R$) {\it HOMFLY polynomials}. This observation was a breakthrough as it has provided tools for non-perturbative calculation of observables of the 3d Chern--Simons theory. 

\begin{figure}[h!]
\begin{picture}(300,110)(-50,-80)

    \put(0,0){\line(1,-1){50}}
    \put(0,-50){\line(1,1){22}}
    \put(27,-22){\line(1,1){22}}
    \put(16,-16){\vector(-1,1){10}}
    \put(40,-40){\vector(-1,1){10}}
    \put(33,-16){\vector(1,1){10}}
    \put(10,-40){\vector(1,1){10}}
    \put(60,-30){\mbox{$\longrightarrow\quad {\cal R}^{ab}_{cd}\;:\; U\otimes V\; \rightarrow \; V\otimes U$}}
    \put(47,7){\mbox{$b$}}
\put(0,7){\mbox{$a$}}
\put(47,-60){\mbox{$d$}}
\put(0,-60){\mbox{$c$}}

\put(250,0){\line(0,1){15}}
\put(270,0){\line(0,1){15}}
\qbezier(260,-15)(270,-15)(270,0)
\qbezier(260,-15)(250,-15)(250,0)
\put(250,0){\vector(0,1){10}}
\put(270,10){\vector(0,-1){8}}
\put(268,20){\mbox{$\bar{b}$}}
\put(248,20){\mbox{$a$}}
\put(285,0){\mbox{$\longrightarrow\quad {\cal M}^{a\bar{b}}\;:\; U\otimes U^*\; \rightarrow \; \mathbb{C}$}}

\put(250,-60){\line(0,1){15}}
\put(270,-60){\line(0,1){15}}
\qbezier(260,-30)(270,-30)(270,-45)
\qbezier(260,-30)(250,-30)(250,-45)
\put(250,-45){\vector(0,-1){10}}
\put(270,-55){\vector(0,1){10}}
\put(268,-72){\mbox{$b$}}
\put(248,-72){\mbox{$\bar{a}$}}
\put(285,-50){\mbox{$\longrightarrow\quad \bar{\cal M}_{\bar{a}b}\;:\; \mathbb{C}\; \rightarrow \; U^*\otimes U$}}
\end{picture}
    \caption{\footnotesize The colored HOMFLY polynomial~\eqref{WilsonLoopExpValue} can be non-pertubatively calculated by contraction of matrices ${\cal R}^{ab}_{cd}$ and turning point operators ${\cal M}^{a\bar{b}}$ and their inverse while going along a link. Actually, matrices ${\cal R}^{ab}_{cd}$ are quantum $\cal R$-matrices of $U_q(\mathfrak{sl}_N)$ quantum group and $U$, $V$ are $U_q(\mathfrak{sl}_N)$-modules.}
    \label{fig:R-M}
\end{figure}
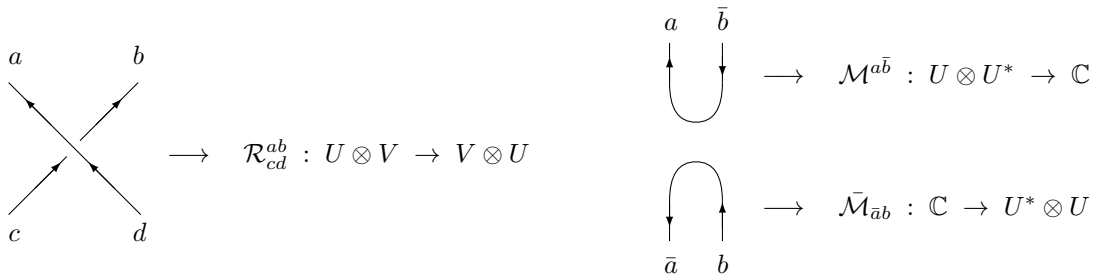

One of such non-perturbative techniques can be derived in the Hamiltonian approach\footnote{There is still lack of straightforward derivation in the Lagrangian approach, see works~\cite{MoSmi,smirnov2012notes} for some attempts.}~\cite{guadagnini1990chern,guadagnini1990chern-2}. There is one distinguished coordinate (arrow of evolution), and it turns out\footnote{The actual calculation is effectively done in 2d space, thus leading to an interesting duality between the 3d Chern--Simons theory and the conformal Wess--Zumino--Novikov--Witten theory, which provides more methods~\cite{moore1989classical,Gu_2015,gerasimov1990wess} of non-perturbative calculation of the Chern--Simons correlators.} that only critical and crossing points contribute the final answer for~\eqref{WilsonLoopExpValue}, see Fig.\,\ref{fig:R-M}. The tensor ${\cal R}^{ab}_{cd}$ can be organised in a matrix called {\it $\cal R$-matrix} and the matrix ${\cal M}^{a \bar{b}}$ is called the {\it turning point} or {\it grading} operator. These objects themselves depend on the representation $R$ \vspace{0.08cm} and dependence on a knot is revealed through a way of contraction of tensors ${\cal R}^{ab}_{cd}$, ${\cal M}^{a\bar{b}}$ and $\bar{\cal M}_{\bar{a} b}$ being elements of the inverse matrix $\hat{\cal M}^{-1}$. We provide only the expressions for the grading matrices (being $N\times N$ diagonal matrix) needed in the further calculations in Sections~\ref{sec:quantum-proj},~\ref{sec:proj-prop-[2]},~\ref{sec:proj-calc-[3]}\,:
\begin{equation}\label{M-op}
\begin{aligned}
    \hat {\cal M}&:= {\rm diag} \Big(q^{2i-N-1},\ \  i = 1,\ldots, N\Big). %, \\
    %\hat{\bar{\cal M}} &= \hat {\cal M}^{-1} = {\rm diag} \Big(q^{N+1-2i},\ \  i = 1,\ldots, N\Big)
\end{aligned}
\end{equation}

The described non-perturbative method of calculation of the Wilson loops in the 3d Chern--Simons theory forms the celebrated Reshetikhin--Turaev approach~\cite{Reshetikhin,reshetikhin1990ribbon,turaev1990yang,reshetikhin1991invariants} to calculation of quantum knot invariants. However, this approach becomes extremely difficult with the growth of complication of the knot (as one needs to store huge-rank tensors) and of the representation (as $\cal R$-matrix is an intricate representation dependent matrix of size equal to square of dimension of the representation). Moreover, the Reshetikhin--Turaev approach can be applied only to each knot/link separately what makes the internal structure of the colored HOMFLY polynomials hidden. 

Thus, different methods for computation of the Wilson loop observables should be developed. Among them are the well developed arborescent  calculus~\cite{mironov2015colored}, applicable to a wide, still limited set of knots,
and the more universal, but still hopeful tangle calculus~\cite{kaul1998chern, MMMtangles, morozov2018knot, anokhina2021khovanov,
anokhina2024towards} operating with concatenation of simple tangles including the lock element (in Fig.\,\ref{fig:pladeco}) utilised in this paper. The lock tangle deserves separate attention as it allows to extend the Kauffman bracket calculation of~\eqref{WilsonLoopExpValue} for the fundamental representation from $N=2$ to an arbitrary value of $N$~\cite{ALM} but for the special class of {\it bipartite} links~\cite{BipKnots,Lewark} (glued entirely from locks), and moreover, to symmetrically colored HOMFLY polynomials being the subject of this paper.

Despite locks are not sufficient to describe all links, and thus, the HOMFLY planar decomposition is also applicable
to a limited set of links, this set seems quite big -- e.g. currently there were found definitely just 12 exceptions with up to 10 crossings in the Rolfsen table, what is now less than there
are non-arborescent knots in it.
More important is the limitation to particular representations.
What we do in this paper, we lift this restriction for all the symmetric (and thus, antisymmetric) ones.
Non-(anti)symmetric representations remain a true challenge in this context.

\subsection{Kauffman bracket and the colored Jones polynomial}\label{sec:colored-Kauff}

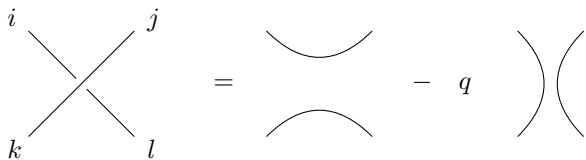
\begin{figure}[h!]
\begin{picture}(100,80)(-200,-40)

\put(-60,-20){\line(1,1){40}}
\put(-20,-20){\line(-1,1){18}}
\put(-60,20){\line(1,-1){18}}

\put(10,-2){\mbox{$=$}}

\qbezier(30,20)(50,0)(70,20)
\qbezier(30,-20)(50,0)(70,-20)

\put(85,-2){\mbox{$- \ \ \ q $}}

\qbezier(125,20)(145,0)(125,-20)
\qbezier(150,20)(130,0)(150,-20)

\put(-68,22){\mbox{$i$}}  \put(-15,22){\mbox{$j$}}  \put(-68,-28){\mbox{$k$}}  \put(-15,-28){\mbox{$l$}}

\end{picture}
\caption{\footnotesize The  celebrated ``Kauffman bracket'' -- the planar decomposition
of the ${\cal R}$-matrix vertex for the fundamental representation of $U_q(\mathfrak{sl}_2)$.
In this case ($N=2$) the conjugate of the fundamental representation is isomorphic to it,
thus, tangles in the picture has no orientation.
}
\label{fig:Kauff}
\end{figure}

The colored HOMFLY polynomial~\eqref{WilsonLoopExpValue} in the case of $SU(2)$ gauge group is called the colored {\it Jones polynomial}. In the case of the fundamental representation, the very efficient way of non-perturbative calculation is provided by the Kauffman bracket\footnote{We write the Kauffman bracket in the framing with unity standing by the resolution $=\,$. This framing does not preserve topological invariance but is convenient for computations. The topological invariance is restored by just a proper multiplier that should present in the final answer for the Jones polynomial.}, see Fig.\,\ref{fig:Kauff}. %leading to a planar decomposition of a knot. 
The final answer for the Jones polynomial has the form of a cycle decomposition with each cycle contributing as the $[2]:=q+q^{-1}$ multiplier. The crucial property here is that the resolutions of a crossing are both planar.

To proceed to higher representations, one can consider a {\it cabled} knot. The $n$-cabled knot is a knot resulting from an initial knot by the substitution of a single strand to a cable of $n$ strands. If one strand carry the fundamental representation\footnote{Finite-dimensional irreducible representations of highest weights of $\mathfrak{sl}_N$ and $U_q(\mathfrak{sl}_N)$ are enumerated by Young diagrams.} $\Box$, then the whole cable carries the representation $\Box^{\otimes n}$. Due to the diagonality of $\cal R$-matrix in the basis of irreducible representation, one can use projectors to separate one concrete irreducible representation from other ones and get the colored quantum knot invariant. For details see~\cite{AnoAnd}. 

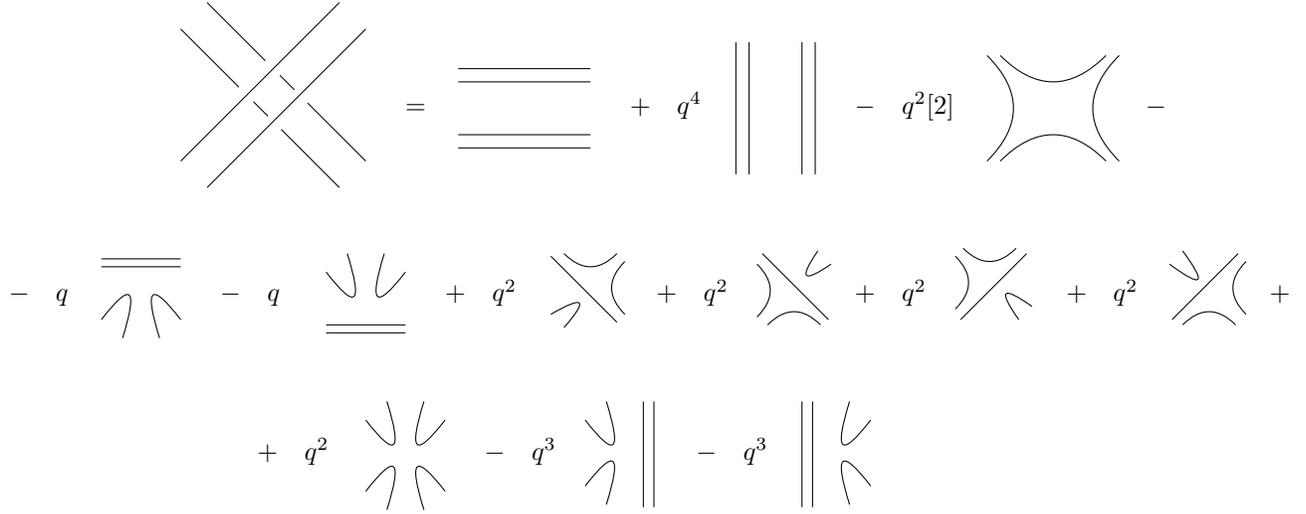
\begin{figure}[H]
\begin{picture}(100,200)(-260,-160)

\put(-125,0){
\put(-60,-20){\line(1,1){60}}
\put(-50,-30){\line(1,1){60}}

\put(0,-30){\line(-1,1){22}}
\put(-27,-3){\line(-1,1){5.5}}
\put(-38,8){\line(-1,1){22}}

\put(10,-20){\line(-1,1){22}}
\put(-17,7){\line(-1,1){5.5}}
\put(-28,18){\line(-1,1){22}}
}

\put(-100,-2){\mbox{$=$}}

\put(-80,10){\line(1,0){50}}
\put(-80,15){\line(1,0){50}}

\put(-80,-10){\line(1,0){50}}
\put(-80,-15){\line(1,0){50}}

\put(-15,-2){\mbox{$+ \ \ \ q^4 $}}

\put(25,-25){\line(0,1){50}}
\put(30,-25){\line(0,1){50}}

\put(55,-25){\line(0,1){50}}
\put(50,-25){\line(0,1){50}}

\put(70,-2){\mbox{$- \ \ \ q^2[2] $}}

\put(95,0){
\qbezier(30,20)(50,0)(70,20)
\qbezier(30,-20)(50,0)(70,-20)

\put(-75,0){
\put(-25,0){
\qbezier(125,20)(145,0)(125,-20)}
\qbezier(150,20)(130,0)(150,-20)
}}

\put(180,-2){\mbox{$-$}}

\put(-250,-73){\mbox{$- \ \ \ q $}}

\put(-215,-60){\line(1,0){30}}
\put(-215,-57){\line(1,0){30}}
\qbezier(-215,-80)(-198,-58)(-207,-87)
\qbezier(-185,-80)(-202,-58)(-193,-87)

\put(-170,-73){\mbox{$- \ \ \ q $}}

\put(85,-7){
\put(-215,-75){\line(1,0){30}}
\put(-215,-78){\line(1,0){30}}
\qbezier(-215,-55)(-198,-77)(-207,-48)
\qbezier(-185,-55)(-202,-77)(-193,-48)
}

\put(-85,-73){\mbox{$+ \ \ \ q^2 $}}

\put(-45,-56){\line(1,-1){25}}
\qbezier(-40,-55)(-30,-65)(-20,-55)
\put(-2,-3){\qbezier(-15,-55)(-25,-65)(-15,-75)}
\put(170,0){\qbezier(-215,-78)(-196,-67)(-210,-83)}

\put(-5,-73){\mbox{$+ \ \ \ q^2 $}}

\put(5,0){
\put(30,-55){\line(1,-1){25}}
\qbezier(28,-59)(38,-69)(28,-79)
\qbezier(32,-82)(42,-72)(52,-82)
\put(5,-5){\qbezier(46,-49)(35,-65)(51,-54)}
}

\put(70,-73){\mbox{$+ \ \ \ q^2 $}}

\put(135,-55){\line(-1,-1){25}}
\put(80,3){\qbezier(28,-59)(38,-69)(28,-79)}
\put(151,2){\qbezier(-40,-55)(-30,-65)(-20,-55)}
\put(-28,0){\qbezier(160,-80)(148,-63)(165,-75)}

\put(150,-73){\mbox{$+ \ \ \ q^2 $}}

\put(215,-55){\line(-1,-1){25}}
\put(162,0){\qbezier(32,-82)(42,-72)(52,-82)}
\put(233,-3){\qbezier(-15,-55)(-25,-65)(-15,-75)}
\put(-26,-4){\qbezier(215,-55)(234,-69)(220,-50)}

\put(227,-73){\mbox{$+$}}

\put(40,-10){
\put(-196,-123){\mbox{$+ \ \ \ q^2 $}}

%\put(-350,-60){\qbezier(215,-55)(234,-69)(220,-50)}

\put(60,-55){
\qbezier(-215,-80)(-198,-58)(-207,-87)
\qbezier(-185,-80)(-202,-58)(-193,-87)
}
\put(60,-53){
\qbezier(-215,-55)(-198,-77)(-207,-48)
\qbezier(-185,-55)(-202,-77)(-193,-48)
}

\put(-110,-123){\mbox{$- \ \ \ q^3$}}

\put(-46,-141){\line(0,1){40}}
\put(-50,-141){\line(0,1){40}}
\put(143,-53){
\qbezier(-215,-80)(-198,-58)(-207,-87)
\qbezier(-215,-55)(-198,-77)(-207,-48)
}

\put(-30,-123){\mbox{$- \ \ \ q^3$}}
\put(60,0){
\put(-46,-141){\line(0,1){40}}
\put(-50,-141){\line(0,1){40}}
}
\put(221,-53){
\qbezier(-185,-80)(-202,-58)(-193,-87)
\qbezier(-185,-55)(-202,-77)(-193,-48)
}

}

\end{picture}
    \caption{\footnotesize Planar decomposition via the Kauffman bracket of a 2-cabled vertex. The smaller diagrams in the second and the third lines vanish when projecting to the first symmetric representation as all of them carry only the singlet representation on the closed end.}
    \label{fig:2-cabled-vert}
\end{figure}

Let us focus on obtaining the Jones polynomial in the first symmetric representation. We take $2$-cabled crossing carrying the representation $\Box\otimes\Box=\varnothing + [2]$ and apply the Kauffman bracket decomposition as demonstrated in Fig.\,\ref{fig:2-cabled-vert}. We end up with 12 planar diagrams. However, 9 of them (in the second and third rows) carry only the singlet representation in one or more of four of the diagrams channels. Thus, after projection on the representation $[2]$ only three bigger diagrams in the first line do contribute the answer for the Jones polynomial in the first symmetric representation. Therefore, it is natural to expect that the same story takes place for an arbitrary $N$ but for a lock vertex (as it is in the case of the representation $\Box$), and it is actually the case as discussed in Section~\ref{sec:rep-[2]}. 

Moreover, it is clear that the application of the Kauffman bracket to the $r$-cabled crossing and further projection of each of four ends on the representation $[r]$ gives exactly $r+1$ summands. The same situation holds in the HOMFLY case for the antiparallel combination of two crossings, see Sections~\ref{sec:rep-[3]},~\ref{sec:rep-[r]}. 

%\newpage
\subsection{Results and organisation of this paper}

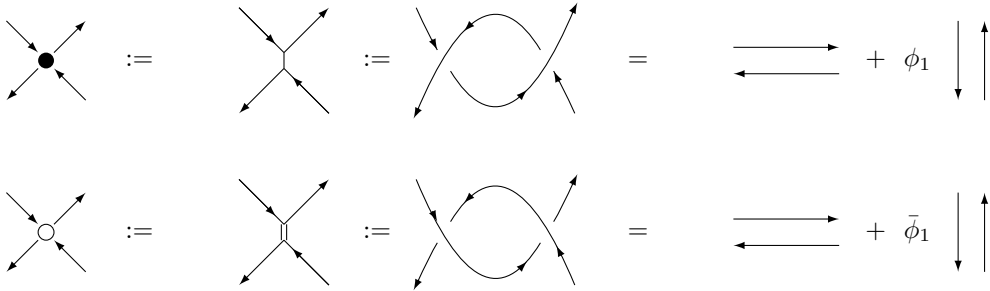
\begin{figure}[h]
\begin{picture}(100,130)(-150,-180)

\put(0,-85){
\put(-105,15){\vector(1,-1){12}} \put(-87,3){\vector(1,1){12}}
\put(-93,-3){\vector(-1,-1){12}} \put(-75,-15){\vector(-1,1){12}}
\put(-90,0){\circle*{6}}

\put(-60,-2){\mbox{$:=$}}
}

\put(0,-150){
\put(-105,15){\vector(1,-1){12}} \put(-87,3){\vector(1,1){12}}
\put(-93,-3){\vector(-1,-1){12}} \put(-75,-15){\vector(-1,1){12}}
\put(-90,0){\circle{6}}

\put(-60,-2){\mbox{$:=$}}
}

\put(0,-85){
%\put(-20,17){\line(1,-1){17}}\put(-20,17){\vector(1,-1){14}}   \put(3,0){\vector(1,1){17}}
%\put(-3,0){\vector(-1,-1){17}}   \put(20,-17){\line(-1,1){17}} \put(20,-17){\vector(-1,1){14}}
%\put(-3,0){\line(1,0){6}}
\put(-17,20){\line(1,-1){17}}\put(-17,20){\vector(1,-1){14}}   \put(0,3){\vector(1,1){17}}
\put(0,-3){\vector(-1,-1){17}}   \put(17,-20){\line(-1,1){17}} \put(17,-20){\vector(-1,1){14}}
\put(0,-3){\line(0,1){6}}

\put(30,-2){\mbox{$:=$}}

\qbezier(50,20)(55,9)(58,4) \qbezier(63,-4)(85,-40)(110,20)
\put(56,8){\vector(1,-2){2}} \put(90,-13){\vector(1,1){2}} \put(109,18){\vector(1,2){2}}
\qbezier(50,-20)(75,40)(97,4)  \qbezier(102,-4)(105,-9)(110,-20)
\put(104,-8){\vector(-1,2){2}} \put(70,13){\vector(-1,-1){2}} \put(51,-18){\vector(-1,-2){2}}

\put(130,-2){\mbox{$=$}}

\put(100,65){
\put(70,-60){\vector(1,0){40}}
\put(110,-70){\vector(-1,0){40}}
\put(120,-67){\mbox{$+\ \ \phi_1$}}
\put(155,-50){\vector(0,-1){30}}
\put(165,-80){\vector(0,1){30}}
}}

\put(0,-150){
\put(-17,20){\line(1,-1){17}}\put(-17,20){\vector(1,-1){14}}   \put(0,3){\vector(1,1){17}}
\put(0,-3){\vector(-1,-1){17}}   \put(17,-20){\line(-1,1){17}} \put(17,-20){\vector(-1,1){14}}
\put(-1,-3){\line(0,1){6}}  \put(1,-3){\line(0,1){6}}

\put(30,-2){\mbox{$:=$}}

\qbezier(50,20)(75,-40)(97,-4)  \qbezier(102,4)(105,9)(110,20)
\qbezier(50,-20)(55,-9)(58,-4) \qbezier(63,4)(85,40)(110,-20)
%\qbezier(50,20)(55,9)(58,4) \qbezier(63,-4)(85,-40)(110,20)
\put(56,8){\vector(1,-2){2}} \put(90,-13){\vector(1,1){2}} \put(109,18){\vector(1,2){2}}
%\qbezier(50,-20)(75,40)(97,4)  \qbezier(102,-4)(105,-9)(110,-20)
\put(105,-9){\vector(-1,2){2}} \put(70,13){\vector(-1,-1){2}} \put(51,-18){\vector(-1,-2){2}}

\put(130,-2){\mbox{$=$}}

\put(100,65){
\put(70,-60){\vector(1,0){40}}
\put(110,-70){\vector(-1,0){40}}
\put(120,-67){\mbox{$+\ \ \bar \phi_1$}}
\put(155,-50){\vector(0,-1){30}}
\put(165,-80){\vector(0,1){30}}
}}

\end{picture}
\caption{\footnotesize Vertical AP lock from \cite{ALM}.
Also shown is the ``opposite" of the vertical lock made from inverse vertices.
%Note that notation for black and white vertices is different from \cite{ALM}.
} \label{fig:pladeco}
\end{figure}

In~\cite{ALM}, we have suggested a remarkable extension of the ``Kauffman bracket'' \cite{Kauff,DM12,DM12-2} from
the Jones to HOMFLY polynomials --
but for a restricted set of knot/link diagrams $\Gamma$, made entirely from
the antiparallel (AP) lock tangles from Fig.\,\ref{fig:pladeco}. For a small review see also Section~\ref{sec:rep-[1]}.
The crucial feature of these tangles is the {\it planar decomposition}, see Fig.\,\ref{fig:pladeco},
which allows to represent the HOMFLY polynomial as a sum of trivial monomials,
selected by a topology of a diagram with $\Gamma_\bullet$ black and $\Gamma_\circ$ white vertices:
\be
H^\Gamma_\Box(A,q) = \ A^{-w} \Big<( \, = +\, \phi_1 \cdot || \,)^{\Gamma_\bullet}
(\, = +\, \bar \phi_1 \cdot || \,)^{\Gamma_\circ}
\Big>_\Gamma
= A^{-w} \sum_{(a,b,c)\in <\Gamma>_{[1]}} \phi_1^a{\bar \phi}_1^b D_{[1]}^c\,.
\label{pladec}
\ee
Here $w$ is the algebraic number of crossings also called the writhe number of a link diagram $\Gamma$, $D_{[1]}:=[N] = \frac{\{q^N\}}{\{q\}} = \frac{\{A\}}{\{q\}}$ (with $\{x\}:=x-x^{-1}$ and $A=q^N$)
is the quantum dimension of the fundamental representation,
and $c$ is the number of closed circles in the particular item of the planar decomposition $<\Gamma>_{[1]}$ of $\Gamma$. In this formula, $\langle \dots \rangle_\Gamma $ means that one replaces each crossing of a link diagram $\Gamma$ by its planar resolution depicted in Fig.\,\ref{fig:pladeco}.
All the coefficients in the sum (\ref{pladec}) are unities, and all the $A,q$ dependence is contained
in just three quantities $A=q^N$, $\phi_1=A\{q\}$ and $\bar \phi_1=-A^{-1}\{q\}$.
Since some planar contractions in $<\Gamma>_{[1]}$ provide the same values of $a,b,c$,
one can instead say that monomials come with some non-negative integer multiplicities.

As mentioned in \cite{ALM}, an additional bonus of this construction is immediate generalizability
to symmetric representations.
The analogue of (\ref{pladec}) for $R=[2]$, to be explained in detail in Section~\ref{sec:rep-[2]},
in somewhat symbolical notation is 
\begin{equation}\label{pladec2}
\begin{aligned}
    H^\Gamma_{[2]}(A,q) &=   (A^2 q^2)^{-w}\Big<\Big<
\left(\,\overset{\text{\large =}}{} \stackinset{r}{2.5pt}{b}{-4pt}{\text{\large =}}\; + \phi_2 \cdot ||\; || +\psi_2\ \cdot \Big|\horr\Big|\right)^{\Gamma_\bullet}
\left(\,\overset{\text{\large =}}{} \stackinset{r}{2.5pt}{b}{-4pt}{\text{\large =}}\; + \bar \phi_2\cdot ||\; || + \bar \psi_2\cdot \Big|\horr\Big| \right)^{\Gamma_\circ}
\Big>\Big>_\Gamma
= \\
&=(A^2 q^2)^{-w} \!\!\!\!\!
\sum_{(a,b,c,d,e)\in <\Gamma>_{[2]}}
\!\!\!\!\!\!\!\!
K_2^{a,b,c,d,e}(q)\,\phi_2^a{\phi_2^*}^b\psi_2^c{\psi_2^*}^d D_{[2]}^e
\end{aligned}
\end{equation}
where planar decomposition of the symmetric AP lock is shown in detail in Fig.\,\ref{fig:proje2lock} below.
New ingredients here are the $q$-dependent coefficients $K_2^{a,b,c,d,e}(q)$
formed by a collection of projectors to the representation $R=[2]$.
They are well defined and calculable from the first principles, see Section~\ref{sec:rep-[2]}.
Still it is a separate story, somewhat orthogonal or complementary to the beautiful (\ref{pladec}),
thus it was not included into \cite{ALM}.
The goal of the present paper is to present a detailed description of (\ref{pladec2})
and its further generalizations to $R=[r]$ (see Sections~\ref{sec:rep-[3]}, \ref{sec:rep-[r]}) --
and provide examples for some simple families of AP lock diagrams.
%including twist, double-braid, ???? (all arborescent)
%together with some??? non-arborescent examples like Kanenobu family and its immediate link extension.???
Cabling does not help -- the AP lock of multiple lines is not planar(ised).
However, {\bf  its projection to a symmetric representation possess planar decomposition}.
%(at least)
What is needed is a {\bf calculus of projectors}.
The main general formula for coefficients of the planar decomposition of the antiparallel lock in an arbitrary
symmetric representation $R=[r]$ is in (\ref{psi_r^(i)}).

%Bipartite evolution ??? Examples ???

\setcounter{equation}{0}
\section{Representation $[1]$}\label{sec:rep-[1]}

This section is a brief remainder of some of the results of our previous paper~\cite{ALM}. The presented statements are generalized to higher symmetric representations in subsequent sections.

In~\cite{ALM}, we have extended the Kauffman bracket in Fig.\,\ref{fig:Kauff}, valid for computing the fundamental Jones polynomial ($N=2$) for any link, to the case of the HOMFLY polynomial in the fundamental representation at an arbitrary $N$, see Fig.\,\ref{fig:pladeco}, but for a special class of bipartite links consisting entirely from antiparallel (AP) lock tangles. The planar decomposition of AP locks is a direct consequence of the HOMFLY skein relation.

\begin{figure}[h!]
\begin{picture}(300,130)(0,0)

\put(200,15){

\put(-40,38){\mbox{$:=$}}

\put(-70,40){
\put(-13,10){\vector(1,-1){10}}
\put(7,0){\vector(1,1){10}}
\put(17,-10){\vector(-1,1){10}}
\put(-3,0){\vector(-1,-1){10}}

\put(-3,0){\line(1,0){10}}
\put(0,5){\mbox{$n$}}
}

\put(0,100){
\qbezier(-3,0)(-13,-7)(-12,-10) \put(-12,-10){\vector(0,-1){2}}
\qbezier(7,0)(17,-7)(15,-10) \put(9,-2){\vector(-1,1){2}}
\put(-3,0){\line(1,0){10}}

\put(-13,10){\vector(1,-1){10}}
\put(7,0){\vector(1,1){10}}
}

\put(-4,80){\mbox{$\ldots$}}

\put(0,60){
\qbezier(-3,0)(-13,7)(-12,10) \put(-6,2){\vector(1,-1){2}}
\qbezier(7,0)(17,7)(15,10) \put(15,9){\vector(0,1){2}}
\put(-3,0){\line(1,0){10}}
}

\put(0,40){
\qbezier(-3,0)(-20,10)(-3,20) \put(-6,2){\vector(1,-1){2}}
\qbezier(7,0)(20,10)(7,20) \put(9,18){\vector(-1,1){2}}
\put(-3,0){\line(1,0){10}}
}

\put(0,20){
\qbezier(-3,0)(-20,10)(-3,20) \put(-6,2){\vector(1,-1){2}}
\qbezier(7,0)(20,10)(7,20) \put(9,18){\vector(-1,1){2}}
\put(-3,0){\line(1,0){10}}
}

\qbezier(-3,0)(-20,10)(-3,20) \put(-6,2){\vector(1,-1){2}}
\qbezier(7,0)(20,10)(7,20) \put(9,18){\vector(-1,1){2}}
\put(-3,0){\line(1,0){10}}

\put(-3,0){\vector(-1,-1){10}}
\put(17,-10){\vector(-1,1){10}}

\put(-20,105){
\put(50,-67){\mbox{$=$}}

\put(70,-50){\vector(0,-1){30}}
\put(80,-80){\vector(0,1){30}}

\put(90,-67){\mbox{$+\ \ \ \PPPhi_n^{[1]}$}}

\put(135,-60){\vector(1,0){40}}
\put(175,-70){\vector(-1,0){40}}
}

}
    
\end{picture}
    \caption{\footnotesize A vertical iteration of the horizontal AP lock tangle called a vertical chain operator and its planar decomposition in the fundamental representation.}
    \label{fig:chain-[1]}
\end{figure}
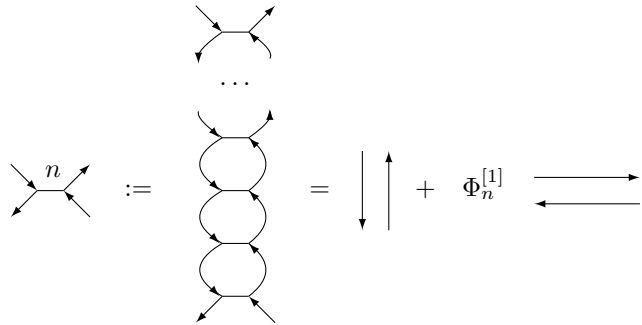

Then, we have provided examples of calculation of the fundamental HOMFLY polynomial for some bipartite links. The considered bipartite families included {\it evolution} parameters. The evolution in almost all families consisted of insertion of 2-strand braids (being an iteration of AP locks) instead of single crossings inside a link. Then, the numbers of crossings in this 2-strand braids become evolution parameters of the resulting knot family. The planar decomposition of this 2-braid is provided in Fig.\,\ref{fig:chain-[1]}. More sophisticated examples of the evolution of links and its consequences can be studied in~\cite{mironov2013evolution,dunin2013superpolynomials,mironov2016racah,mironov2015colored2,mironov2014colored,dunin2022evolution,anokhina2019nimble,willis2021khovanov,nakagane2019action,anokhina2018khovanov}.

In {\it ordinary evolution} technique, the colored HOMFLY polynomial for a link including 2-strand braids is expressed in terms of the $\cal R$-matrix eigenvalues to the powers of the numbers of crossings inside these evolutionary 2-braids. In the planar technique, the evolution is bipartite but still compatible with the evolution by the eigenvalues. In particular, in the case of the fundamental representation the {\it bipartite evolution} manifests through the $\PPPhi_n^{[1]}$ coefficient in front of the $=$ resolution: 
\begin{equation}\label{chain-[1]}
    \Phi_n^{[1]} = \frac{(1+\phi_1 D_{[1]})^n-1}{D_{[1]}}=A^{2n}\cdot D_{[1]}^{-1}\left(\lambda_{\varnothing}^{2n}-\lambda_{\rm adj}^{2n}\right)\,.
\end{equation}
Taking into account the framing factor $A^{-2}$ being assigned to each lock vertex to restore the topological invariance, we come to the following correspondence with the $\cal R$-matrix eigenvalues, and thus, with the ordinary evolution:
\begin{equation}\label{eigen-[1]}
\begin{aligned}
    \lambda_{\rm adj}^2 &= A^{-2}\,, \\
    \lambda_{\varnothing}^2 &= A^{-2}(1+\phi_1 D_{[1]})=1\,.
\end{aligned}
\end{equation}
The above formulas provide what we call the bipartite decomposition of squares of $\mathcal{R}$-matrices eigenvalues in the antiparallel channel. 

An interesting and simplifying fact is that there are the following relations on the planar resolution coefficients:
\begin{equation}\label{coef-rel-[1]}
\begin{aligned}
    \PPPhi_{n+m} &= \PPPhi_n+\PPPhi_m + \PPPhi_n\PPPhi_m D\,, \\
    0 &= \PPhi + \bar{\PPhi} +\PPhi\bar{\PPhi}\cdot D\,.
\end{aligned}
\end{equation}
The first formula can be derived if considering the vertical chain of $n+m$ locks, and the second one is the particular case of the first relation.

%All the discussed peculiarities have straightforward lift to other symmetric representations and are demonstrated in what follows.

As we will see below, all these peculiarities have straightforward lift to all symmetric representations.

\setcounter{equation}{0}
\section{Representation $[2]$}\label{sec:rep-[2]}

In this section, we lift the planar decomposition method for the first symmetric representation, briefly described for the Jones polynomial in Section~\ref{sec:colored-Kauff}, to the HOMFLY polynomial. The method include the cabling technique~\cite{AnoAnd} and further projection to the representation $[2]$. Thus, a projector calculus is needed, and it is described in different details in Section~\ref{sec:proj-calc-[2]}. The subsequent subsections are devoted to simple examples of calculation of the HOMFLY polynomials via our planar decomposition method.

%The corresponding colored HOMFLY polynomial is obtained by the cabling technique 

\subsection{Generalities}

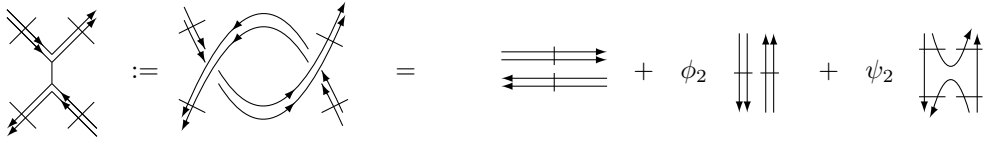
\begin{figure}[h]
\begin{picture}(100,80)(20,-40)

\put(100,0){
%\put(-20,17){\line(1,-1){17}}\put(-20,17){\vector(1,-1){14}}   \put(3,0){\vector(1,1){17}}
%\put(-3,0){\vector(-1,-1){17}}   \put(20,-17){\line(-1,1){17}} \put(20,-17){\vector(-1,1){14}}
%\put(-3,0){\line(1,0){6}}
%\put(-23,17){\line(1,-1){17}}\put(-23,17){\vector(1,-1){14}}   \put(6,0){\vector(1,1){17}}
%\put(-6,0){\vector(-1,-1){17}}   \put(23,-17){\line(-1,1){17}} \put(23,-17){\vector(-1,1){14}}
%\put(-19,6){\line(1,1){10}}  \put(-19,-6){\line(1,-1){10}}
%\put(19,6){\line(-1,1){10}}  \put(19,-6){\line(-1,-1){10}}

\put(-17,21){\line(1,-1){17}}\put(-17,24){\line(1,-1){17}}   \put(0,4){\vector(1,1){17}}
\put(-17,21){\vector(1,-1){14}} \put(-17,24){\vector(1,-1){14}}
\put(0,7){\vector(1,1){17}}
\put(0,-4){\vector(-1,-1){17}}   \put(17,-21){\line(-1,1){17}} \put(17,-21){\vector(-1,1){14}}
\put(0,-7){\vector(-1,-1){17}}   \put(17,-24){\line(-1,1){17}} \put(17,-24){\vector(-1,1){14}}
\put(0,4){\line(0,-1){8}}

\put(-16,11){\line(1,1){10}}  \put(-16,-11){\line(1,-1){10}}
\put(16,11){\line(-1,1){10}}  \put(16,-11){\line(-1,-1){10}}

\put(30,-2){\mbox{$:=$}}

\qbezier(50,20)(55,9)(58,4) \qbezier(63,-4)(85,-40)(110,20)
\put(56,8){\vector(1,-2){2}} \put(90,-13){\vector(1,1){2}} \put(109,18){\vector(1,2){2}}
\qbezier(50,-20)(75,40)(97,4)  \qbezier(102,-4)(105,-9)(110,-20)
\put(104,-8){\vector(-1,2){2}} \put(70,13){\vector(-1,-1){2}} \put(51,-18){\vector(-1,-2){2}}

\qbezier(50,25)(55,14)(58,9) \qbezier(63,1)(85,-35)(110,25)
\put(56,13){\vector(1,-2){2}} \put(90,-8){\vector(1,1){2}} \put(109,23){\vector(1,2){2}}
\qbezier(50,-15)(75,45)(97,9)  \qbezier(102,1)(105,-4)(110,-15)
\put(104,-3){\vector(-1,2){2}} \put(70,18){\vector(-1,-1){2}} \put(51,-13){\vector(-1,-2){2}}

\put(48,15){\line(2,1){10}}  \put(48,-10){\line(2,-1){10}}
\put(110,10){\line(-2,1){10}}  \put(112,-10){\line(-2,-1){10}}

\put(130,-2){\mbox{$=$}}

\put(100,65){
\put(70,-60){\vector(1,0){40}} \put(70,-57){\vector(1,0){40}}
\put(110,-70){\vector(-1,0){40}}  \put(110,-67){\vector(-1,0){40}}
\put(90,-55){\line(0,-1){7}}   \put(90,-65){\line(0,-1){7}}
\put(120,-67){\mbox{$+\ \ \ \phi_2$}}
\put(160,-50){\vector(0,-1){30}} \put(163,-50){\vector(0,-1){30}}
\put(170,-80){\vector(0,1){30}}  \put(173,-80){\vector(0,1){30}}
\put(158,-65){\line(1,0){7}}   \put(168,-65){\line(1,0){7}}
\put(190,-67){\mbox{$+\ \ \ \psi_2$}}
\put(230,-50){\vector(0,-1){30}}
\put(250,-80){\vector(0,1){30}}
\qbezier(233,-50)(240,-75)(247,-50)  \put(246,-52){\vector(1,2){2}}
\qbezier(233,-80)(240,-55)(247,-80)  \put(234,-78){\vector(-1,-2){2}}
\put(228,-56){\line(1,0){10}}   \put(242,-56){\line(1,0){10}}
\put(228,-74){\line(1,0){10}}   \put(242,-74){\line(1,0){10}}
}}

\end{picture}
\caption{\footnotesize  The lock tangle projected to the symmetric representation $[2]$
and its planar decomposition.
} \label{fig:proje2lock}
\end{figure}

The planar decomposition of the lock diagram is shown in Fig.\,\ref{fig:proje2lock}. It is a generalization of the Kauffman expanded 2-cabled single vertex (see Fig.\,\ref{fig:2-cabled-vert}) projected to the representation $[2]$. Note that after projection only three diagrams in the first line of Fig.\,\ref{fig:2-cabled-vert} survive, and they are exactly the diagrams from the r.h.s. of Fig.\,\ref{fig:proje2lock}. 
The main point is that there are  three items in the product
%???$[2]\times [2] = [4] + [3,1] + [2,2]$,
$[2]\otimes \overline{[2]} = [2]\otimes [2^{N-1}]=  [4,2^{N-2}] \oplus [2,1^{N-2}] \oplus \varnothing$,
thus three eigenvalues of $\cal R$-matrix,
thus three diagrams are sufficient for decomposition --
and there are exactly three planar diagrams left after the projection on the first symmetric representation.
Thus, we can look for the decomposition for the AP the lock tangle
\be
\tau_{[2]} = \left(\,\overset{\text{\large =}}{} \stackinset{r}{2.5pt}{b}{-4pt}{\text{\large =}}\; + \phi_2 \cdot ||\; || +\psi_2\ \cdot \Big|\horr\Big|\right)\cdot P^{\otimes 4}\,. 
\ee
Important is that projectors are present at all the external ends,
and they must be converted in the actual knot diagram.

%\newpage ???

At the classical level, the symmetric projector in Fig.\,\ref{fig:proje2} is
\be
P_{i  j}^{i'  j'} = \frac{1}{2}\left(\delta_i^{i'} \delta_j^{j'} + \delta_i^{j'} \delta_j^{i'}\right)\,.
\label{proj2}
\ee

\begin{figure}[h]
\begin{picture}(100,50)(-220,0)

\put(-40,13){\mbox{$P_{i  j}^{i'  j'} \ :=$}}
\put(20,0){\vector(0,1){30}}
\put(30,0){\vector(0,1){30}}
\put(12,-2){\mbox{$i$}} \put(32,-2){\mbox{$j$}}\put(12,32){\mbox{$i'$}}\put(32,32){\mbox{$j'$}}

\put(15,13){\line(1,0){20}}
%\put(22,15){\mbox{$S$}}

\end{picture}
\caption{\footnotesize  Pictorial image of the symmetric projector (\ref{proj2}).
} \label{fig:proje2}
\end{figure}

\begin{figure}[h!]
\begin{picture}(100,200)(40,-150)

\put(100,0){
\put(-20,17){\line(1,-1){17}}\put(-20,17){\vector(1,-1){14}}   \put(3,0){\vector(1,1){17}}
\put(-3,0){\vector(-1,-1){17}}   \put(20,-17){\line(-1,1){17}} \put(20,-17){\vector(-1,1){14}}
\put(-3,0){\line(1,0){6}}
\put(-23,17){\line(1,-1){17}}\put(-23,17){\vector(1,-1){14}}   \put(6,0){\vector(1,1){17}}
\put(-6,0){\vector(-1,-1){17}}   \put(23,-17){\line(-1,1){17}} \put(23,-17){\vector(-1,1){14}}
\put(-19,6){\line(1,1){10}}  \put(-19,-6){\line(1,-1){10}}
%\put(19,6){\line(-1,1){10}}  \put(19,-6){\line(-1,-1){10}}
\qbezier(-20,17)(-40,42)(0,42) \qbezier(20,17)(40,42)(0,42)
\qbezier(-20,-17)(-40,-42)(0,-42) \qbezier(20,-17)(40,-42)(0,-42)
\qbezier(-23,17)(-43,45)(0,45) \qbezier(23,17)(43,45)(0,45)
\qbezier(-23,-17)(-43,-45)(0,-45) \qbezier(23,-17)(43,-45)(0,-45)

\put(50,-2){\mbox{$=$}}

\put(100,0){\circle{36}} \put(100,0){\circle{30}}

\put(140,-2){\mbox{$+ \ \ \ \phi_2$}}

\put(200,20){\circle{36}} \put(200,20){\circle{30}}
\put(200,-20){\circle{36}} \put(200,-20){\circle{30}}
\put(200,32){\line(0,1){10}} \put(200,-32){\line(0,-1){10}}

\put(230,-2){\mbox{$+ \ \ \ \psi_2$}}

\put(-50,0){
\put(340,20){\circle{30}} \put(340,-20){\circle{30}}
\qbezier(320,20)(320,40)(340,40) \qbezier(360,20)(360,40)(340,40)
\qbezier(320,-20)(320,-40)(340,-40) \qbezier(360,-20)(360,-40)(340,-40)
\put(320,20){\line(0,-1){40}}  \put(360,-20){\line(0,1){40}}

}

%-----------------------------------

\put(0,-100){
\put(-20,17){\line(1,-1){17}}\put(-20,17){\vector(1,-1){14}}   \put(3,0){\vector(1,1){17}}
\put(-3,0){\vector(-1,-1){17}}   \put(20,-17){\line(-1,1){17}} \put(20,-17){\vector(-1,1){14}}
\put(-3,0){\line(1,0){6}}
\put(-23,17){\line(1,-1){17}}\put(-23,17){\vector(1,-1){14}}   \put(6,0){\vector(1,1){17}}
\put(-6,0){\vector(-1,-1){17}}   \put(23,-17){\line(-1,1){17}} \put(23,-17){\vector(-1,1){14}}
\put(-19,6){\line(1,1){10}}  %\put(-19,-6){\line(1,-1){10}}
\put(19,6){\line(-1,1){10}}  %\put(19,-6){\line(-1,-1){10}}
\qbezier(-20,17)(-40,42)(-40,0) \qbezier(-20,-17)(-40,-42)(-40,0)
\qbezier(-23,17)(-37,36)(-37,0) \qbezier(-23,-17)(-37,-36)(-37,0)
\qbezier(20,17)(40,42)(40,0) \qbezier(20,-17)(40,-42)(40,0)
\qbezier(23,17)(37,36)(37,0) \qbezier(23,-17)(37,-36)(37,0)

\put(50,-2){\mbox{$=$}}

\put(100,0){\circle{36}} \put(100,0){\circle{30}}
\put(150,0){\circle{36}} \put(150,0){\circle{30}}
\put(100,12){\line(0,1){10}} \put(150,12){\line(0,1){10}}

\put(190,-2){\mbox{$+ \ \ \ \phi_2$}}

\put(250,0){\circle{36}} \put(250,0){\circle{30}}
\put(250,12){\line(0,1){10}}

\put(280,-2){\mbox{$+ \ \ \ \psi_2$}}

\put(340,0){\circle{30}} \put(380,0){\circle{30}}
\qbezier(320,0)(320,20)(340,20) \qbezier(320,0)(320,-20)(340,-20)
\qbezier(400,0)(400,20)(380,20) \qbezier(400,0)(400,-20)(380,-20)
\put(340,20){\line(1,0){40}}  \put(340,-20){\line(1,0){40}}

\put(100,12){\line(0,1){10}}
\put(340,12){\line(0,1){10}} \put(380,12){\line(0,1){10}}

}

\put(290,32){\line(0,1){12}} \put(290,-32){\line(0,-1){12}}
}

\end{picture}
\caption{\footnotesize  The two closures of a single lock: the unknot in the first line
and the Hopf link in the second line. Note that the diagrams near $\psi_2$ are the same in the unknot and in the Hopf cases and contribute as $P^{i'j}_{ij} P^{ik}_{i'k}$ in the resulting HOMFLY polynomials in~\eqref{unknotHopf2}.
} \label{fig:singlelockclosures}
\end{figure}
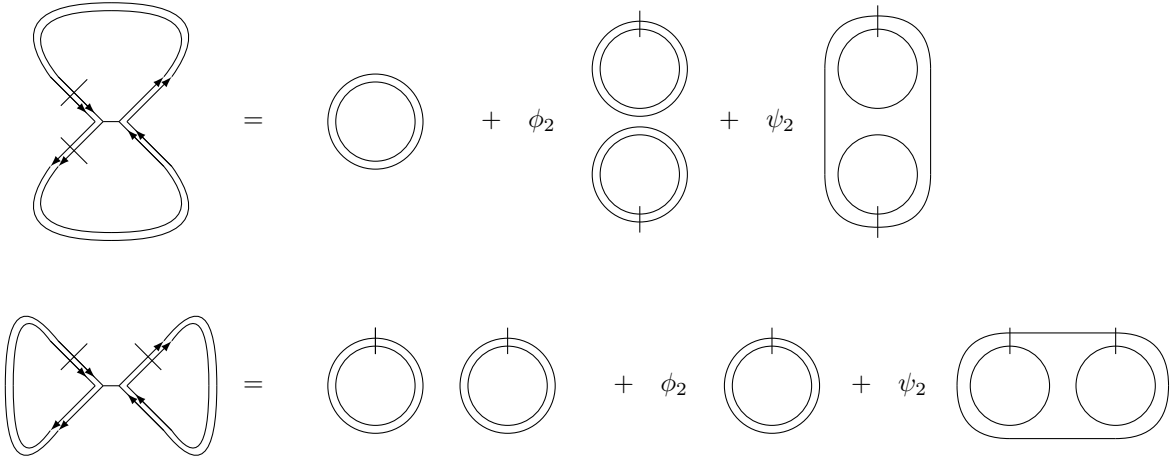

\noindent From a single lock we can make two diagrams -- one is an unknot, another one is   the Hopf link, see Fig.\,\ref{fig:singlelockclosures}.

This allows to find the two coefficients\footnote{Summation over repeated indices is implied in formula~\eqref{unknotHopf2}.} $\phi_2$ and $\psi_2$ because we know the corresponding HOMFLY polynomials:

\begin{equation}\label{unknotHopf2}
\begin{aligned}
    H^{\rm unknot}_{[2]} &=
D_{[2]} = \frac{1}{A^4q^4} \left(D_{[2]} + \phi_2 D_{[2]}^2 + \psi_2\, P^{i'j}_{ij} P^{ik}_{i'k} \right),
\\
H^{\rm Hopf}_{[2]}&=\frac{1}{A^4q^8}\left(q^{12} D_{[4]} + q^4 D_{[3,1]} + D_{[2,2]}\right)
=  1 + \frac{1}{A^2}\frac{\{Aq\}\{A/q\}}{\{q\}^2} + \frac{1}{q^4A^4}\frac{\{Aq^3\}\{A\}^2\{A/q\}}{\{q\}^2\{q^2\}^2}
= \\
&= \frac{1}{A^4q^4} \left(D_{[2]}^2 + \phi_2 D_{[2]} + \psi_2\, P^{i'j}_{ij} P^{ik}_{i'k} \right).
\end{aligned}
\end{equation}
%\tr (\hat P\otimes I) (I\otimes \hat P) =
%\left(P_{ij}^{i'j'}\delta_k^{k'} \cdot P_{i'k'}^{i''j''}\delta_{j'}^{j''}\right)\delta^{i}_{i''}\delta^j_{j''}\delta^k_{k''}
The first two expressions for the Hopf link correspond to its parallel and antiparallel realizations,
which are both possible.
Since classically $
P^{i'j}_{ij} P^{ik}_{i'k} = \frac{N(N+1)^2}{2^2}$,
%???it is clear 
one can guess that its quantization is just $\frac{[N][N+1]^2}{[2]^2}=\frac{\{A\}\{Aq\}^2}{\{q\}\{q^2\}^2}$. A fair computation in the quantum case is provided in Section~\ref{sec:quantum-proj}. 
Then, it follows that
\be\label{planar-coef-[2]}
\phi_2 = A^2q^3 \{q\}\{q^2\}\,, \ \ \ \ \ \ \psi_2 = Aq^2[2]\{q^2\}\,.
\label{phi2psi2}
\ee
For the inverse crossings, one can similarly compute by the knowledge that $H^{\bar{\cal L}}_R(q, A)=H^{\cal L}_R(q^{-1},A^{-1})$ (where $\bar{\cal L}$ is the link mirror to $\cal L$):
\be
\bar\phi_2 = \frac{1}{A^2q^3} \{q\}\{q^2\}\,, \ \ \ \ \ \ \bar\psi_2 = -\frac{1}{Aq^2}[2]\{q^2\}\,,
\label{oppophi2psi2}
\ee
what can be also obtained by the replacement $A\rightarrow A^{-1}$, $q\rightarrow q^{-1}$ in formula~\eqref{planar-coef-[2]}. Note that the parallel and antiparallel Hopf links are the same and their HOMFLY polynomials coincide --
but this is not the case for 2-strand torus links with more intersections.
For examples glued from several AP locks, more complicated combinations of projectors are needed, which deserve separate consideration, see the next section.

\subsection{Projector calculus}\label{sec:proj-calc-[2]}

In this subsection, we calculate contractions of projectors appearing in the planar decomposition of the two-strand braid tangle from Section~\ref{sec:chain} and in the computations of the HOMFLY polynomials for links from Sections~\ref{sec:2nunknots}--\ref{sec:DB-[2]}.

\subsubsection{Vertical chain of classical projectors}
%??????Projector combinations. Classical part????}

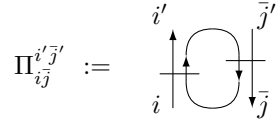
\begin{figure}[h!]
\begin{picture}(100,40)(-220,0)

\put(-40,13){\mbox{$\Pi_{i  \bar j}^{i'  \bar j'} \ :=$}}
\put(20,0){\vector(0,1){30}}
\put(25,10){\vector(0,1){10}}

\put(45,20){\vector(0,-1){10}}
\put(50,30){\vector(0,-1){30}}

\qbezier(25,20)(25,30)(35,30)
\qbezier(45,20)(45,30)(35,30)
\qbezier(25,10)(25,0)(35,0)
\qbezier(45,10)(45,0)(35,0)

\put(12,-2){\mbox{$i$}} \put(52,-2){\mbox{$\bar j$}}\put(12,32){\mbox{$i'$}}\put(52,32){\mbox{$\bar j'$}}

\put(15,13){\line(1,0){15}} \put(40,18){\line(1,0){15}}
%\put(22,15){\mbox{$S$}}

\end{picture}
\caption{\footnotesize  The contraction of two two projectors appearing in the planar decomposition of the double-lock tangle (\ref{Proj2}).
} \label{Proje2}
\end{figure}

For our examples, we need a few simple combinations of projectors. An important building block is shown in Fig.\,\ref{Proje2}:
\be
\Pi_{i\bar j}^{i'\bar j'} = \Pi_{ij'}^{i'j}:= \sum_{k,k'=1}^N P_{ik}^{i'k'}  P^{jk}_{j'k'}
= \frac{1}{4}\left((N+2)\delta_i^{i'} \delta^j_{j'} +  \delta_i^j \delta^{i'}_{j'}\right)
= \frac{1}{4}\left((N+2)\delta_i^{i'} \delta_{\bar j}^{\bar j'} +  \delta_{i \bar j} \delta^{i'\bar j'}\right).
\label{Proj2}
\ee
Note an ambiguity\footnote{This ambiguity with ordinary and bar indices is present only in the classical case as one gets rid of bar indices within the grading operator $\hat{\cal M}$ from equation~\eqref{M-op}, and in the case $q\rightarrow 1$ the grading operator turns to the identity matrix. In the quantum case, it is crucial to raise and lower indices by the quantum $\hat{\cal M}$-matrix, see Section~\ref{sec:quantum-proj}.} in notation: now we have differently directed arrows, and we denote the change of direction either
by bar or by raising/lowering indices.
Accordingly, we get two different notations $\delta_{i\bar j} = \delta_i^j$ for one and the same invariant tensor.

\noindent The obvious combination form a {\bf vertical chain} of projectors, see Fig.\,\ref{fig:vertchain2}\,:

%\begin{itemize}

%\item{Vertical chain, Fig.\ref{vertchain2}}
\be
(\hat \Pi^n)_{i\bar j}^{i'\bar j'} :=
\Pi_{i\bar j}^{i_1\bar j_1} \Pi_{i_1,\bar j_1}^{i_2,\bar j_2}  \ldots \Pi_{i_{n-1}\bar j_{n-1}}^{i'\bar j'}
= \frac{1}{4^n}\left(u_n\delta_i^{i'} \delta_{\bar j}^{\bar j'} + v_n \delta_{i \bar j} \delta^{i'\bar j'}\right).
\ee
There are simple recursions, of which one is easily solved:
\be
u_{n+1} = (N+2)u_n \ \Longrightarrow \ u_n = (N+2)^n, \ \ \ \ \ \ \ \ v_{n+1} = (N+2)v_n + u_n +Nv_n\,.
%\ \Longrightarrow \ v_n = ???
\ee
For expressions of $v_n$, see Table~\eqref{v-coef} below.
This implies that planar diagrams, contributing to the unknot closure of 2-strand braid of $(n-1)$ AP locks contain
\be
\Pi_n:=
(\tr\otimes \tr) \, \hat \Pi^n := \sum_{i,i'=1}^N (\hat \Pi^n)_{i\bar i}^{i'\bar i'} = \frac{1}{4^n}(N u_n + N^2 v_n) =  \frac{N(N+1)^n}{2^n}
\ \longrightarrow \nn \\
\ \longrightarrow \ \frac{[N][N+1]^n}{[2]^n} = D_{[2]} \left(\frac{[N+1]}{[2]}\right)^{n-1}.
\label{Pin}
\ee
%\vspace{-0.3cm}
Quantization in this case is obvious, and is straightforwardly derived in the next subsubsection. However, another closure of the same vertical chain, corresponding to the 2-strand double links,
is a little more sophisticated:
\be\label{TrPi^n}
\Tr \hat \Pi^n := \sum_{i,j =1}^N
 (\hat \Pi^n)_{i\bar j}^{i\bar j}   =
\frac{1}{4^n}(N^2 u_n + N v_n)\,,
\ee
and gives the following chain of expressions:

\begin{figure}[h!]
\begin{picture}(100,205)(-50,-20)

\put(-30,83){\mbox{$\hat \Pi^n  \ :=$}}
\put(20,0){\vector(0,1){170}}
\put(50,170){\vector(0,-1){170}}

\put(25,10){\vector(0,1){10}}
\put(45,20){\vector(0,-1){10}}
\qbezier(25,20)(25,30)(35,30)
\qbezier(45,20)(45,30)(35,30)
\qbezier(25,10)(25,0)(35,0)
\qbezier(45,10)(45,0)(35,0)
\put(15,13){\line(1,0){15}} \put(40,18){\line(1,0){15}}

\put(0,40){
\put(25,10){\vector(0,1){10}}
\put(45,20){\vector(0,-1){10}}
\qbezier(25,20)(25,30)(35,30)
\qbezier(45,20)(45,30)(35,30)
\qbezier(25,10)(25,0)(35,0)
\qbezier(45,10)(45,0)(35,0)
\put(15,13){\line(1,0){15}} \put(40,18){\line(1,0){15}}
}

\put(0,80){
\put(25,10){\vector(0,1){10}}
\put(45,20){\vector(0,-1){10}}
\qbezier(25,20)(25,30)(35,30)
\qbezier(45,20)(45,30)(35,30)
\qbezier(25,10)(25,0)(35,0)
\qbezier(45,10)(45,0)(35,0)
\put(15,13){\line(1,0){15}} \put(40,18){\line(1,0){15}}
}

\put(30,120){\mbox{$\ldots$}}

\put(0,140){
\put(25,10){\vector(0,1){10}}
\put(45,20){\vector(0,-1){10}}
\qbezier(25,20)(25,30)(35,30)
\qbezier(45,20)(45,30)(35,30)
\qbezier(25,10)(25,0)(35,0)
\qbezier(45,10)(45,0)(35,0)
\put(15,13){\line(1,0){15}} \put(40,18){\line(1,0){15}}
}

\put(12,-2){\mbox{$i$}} \put(52,-2){\mbox{$\bar j$}}\put(12,172){\mbox{$i'$}}\put(52,172){\mbox{$\bar j'$}}

\put(15,13){\line(1,0){15}} \put(40,18){\line(1,0){15}}
%\put(22,15){\mbox{$S$}}

\put(20,30){\vector(0,1){2}}  \put(50,140){\vector(0,-1){2}}

%-------------------------------------------------

\put(180,0){
\put(-90,83){\mbox{$\Pi_n:=(\tr\otimes \tr)\, \hat \Pi^n   \ :=$}}
\put(20,15){\vector(0,1){140}}
\put(50,150){\vector(0,-1){140}}

\qbezier(20,150)(20,175)(35,175)
\qbezier(50,150)(50,175)(35,175)
\qbezier(20,15)(20,-5)(35,-5)
\qbezier(50,15)(50,-5)(35,-5)

\put(25,10){\vector(0,1){10}}
\put(45,20){\vector(0,-1){10}}
\qbezier(25,20)(25,30)(35,30)
\qbezier(45,20)(45,30)(35,30)
\qbezier(25,10)(25,0)(35,0)
\qbezier(45,10)(45,0)(35,0)
\put(15,13){\line(1,0){15}} \put(40,18){\line(1,0){15}}

\put(0,40){
\put(25,10){\vector(0,1){10}}
\put(45,20){\vector(0,-1){10}}
\qbezier(25,20)(25,30)(35,30)
\qbezier(45,20)(45,30)(35,30)
\qbezier(25,10)(25,0)(35,0)
\qbezier(45,10)(45,0)(35,0)
\put(15,13){\line(1,0){15}} \put(40,18){\line(1,0){15}}
}

\put(0,80){
\put(25,10){\vector(0,1){10}}
\put(45,20){\vector(0,-1){10}}
\qbezier(25,20)(25,30)(35,30)
\qbezier(45,20)(45,30)(35,30)
\qbezier(25,10)(25,0)(35,0)
\qbezier(45,10)(45,0)(35,0)
\put(15,13){\line(1,0){15}} \put(40,18){\line(1,0){15}}
}

\put(30,120){\mbox{$\ldots$}}

\put(0,140){
\put(25,10){\vector(0,1){10}}
\put(45,20){\vector(0,-1){10}}
\qbezier(25,20)(25,30)(35,30)
\qbezier(45,20)(45,30)(35,30)
\qbezier(25,10)(25,0)(35,0)
\qbezier(45,10)(45,0)(35,0)
\put(15,8){\line(1,0){15}} \put(40,18){\line(1,0){15}}
}

%\put(12,-2){\mbox{$i$}} \put(52,-2){\mbox{$\bar j$}}\put(12,172){\mbox{$i'$}}\put(52,172){\mbox{$\bar j'$}}

%\put(15,13){\line(1,0){15}} \put(40,18){\line(1,0){15}}

%\put(20,30){\vector(0,1){2}}  \put(50,140){\vector(0,-1){2}}
}

%-------------------------------------------------

\put(330,0){

\put(-60,83){\mbox{$\Tr  \hat\Pi^n   \ :=$}}
\put(20,0){\vector(0,1){170}}
\put(50,170){\vector(0,-1){170}}

\qbezier(20,170)(20,185)(5,185)
\qbezier(50,170)(50,185)(65,185)
\qbezier(-10,170)(-10,185)(5,185)
\qbezier(80,170)(80,185)(65,185)
\qbezier(20,0)(20,-20)(5,-20)
\qbezier(50,0)(50,-20)(65,-20)
\qbezier(-10,0)(-10,-20)(5,-20)
\qbezier(80,0)(80,-20)(65,-20)

\put(80,0){\vector(0,1){170}}
\put(-10,170){\vector(0,-1){170}}

\put(25,10){\vector(0,1){10}}
\put(45,20){\vector(0,-1){10}}
\qbezier(25,20)(25,30)(35,30)
\qbezier(45,20)(45,30)(35,30)
\qbezier(25,10)(25,0)(35,0)
\qbezier(45,10)(45,0)(35,0)
\put(15,13){\line(1,0){15}} \put(40,18){\line(1,0){15}}

\put(0,40){
\put(25,10){\vector(0,1){10}}
\put(45,20){\vector(0,-1){10}}
\qbezier(25,20)(25,30)(35,30)
\qbezier(45,20)(45,30)(35,30)
\qbezier(25,10)(25,0)(35,0)
\qbezier(45,10)(45,0)(35,0)
\put(15,13){\line(1,0){15}} \put(40,18){\line(1,0){15}}
}

\put(0,80){
\put(25,10){\vector(0,1){10}}
\put(45,20){\vector(0,-1){10}}
\qbezier(25,20)(25,30)(35,30)
\qbezier(45,20)(45,30)(35,30)
\qbezier(25,10)(25,0)(35,0)
\qbezier(45,10)(45,0)(35,0)
\put(15,13){\line(1,0){15}} \put(40,18){\line(1,0){15}}
}

\put(30,120){\mbox{$\ldots$}}

\put(0,140){
\put(25,10){\vector(0,1){10}}
\put(45,20){\vector(0,-1){10}}
\qbezier(25,20)(25,30)(35,30)
\qbezier(45,20)(45,30)(35,30)
\qbezier(25,10)(25,0)(35,0)
\qbezier(45,10)(45,0)(35,0)
\put(15,13){\line(1,0){15}} \put(40,18){\line(1,0){15}}
}

%\put(12,-2){\mbox{$i$}} \put(52,-2){\mbox{$\bar j$}}\put(12,172){\mbox{$i'$}}\put(52,172){\mbox{$\bar j'$}}

\put(15,13){\line(1,0){15}} \put(40,18){\line(1,0){15}}
%\put(10,-5){\line(1,0){50}}

\put(20,3){\vector(0,1){2}}  \put(50,140){\vector(0,-1){2}}
\put(80,3){\vector(0,1){2}}  \put(-10,140){\vector(0,-1){2}}

%\put(5,220){\line(0,1){15}}
%\put(65,220){\line(0,1){15}}

%\qbezier(25,10)(25,0)(35,0)
%\qbezier(45,10)(45,0)(35,0)
%\put(15,13){\line(1,0){15}} \put(40,18){\line(1,0){15}}
}

\end{picture}
\caption{\footnotesize  "Vertical" chain of projectors.
Shown are its closures, which are relevant for the unknot and 2-strand link
examples in Sections \ref{sec:2nunknots}, \ref{sec:2links} and in a derivation of the chain planar decomposition in Section~\ref{sec:chain}.
$n$ is the number of $\hat\Pi$, which is the same as the number of small cycles in the picture.
The lock vertices would stand in between the cycles, thus there would be $(n-1)$ of them in $\Pi_n$
and $(n+1)$ ones in $\Tr \hat{\Pi}^n$. %--
%this is the reason why there is a shift by one in the indices of $\Pi_{l+1}$
%in the second line of (\ref{2unknot}) and no such shift in (\ref{???}).
The picture for $\Tr \hat{\Pi}^n$ can be also redrawn as a wheel with $(n+1)$ small circles between the two
concentric rings, see Fig.\,\ref{fig:circles}.
Note that the pictures in the {\it present} subsection concern only projectors,
$\cal R$-matrices and planar decompositions do not yet appear.
} \label{fig:vertchain2}
\end{figure}
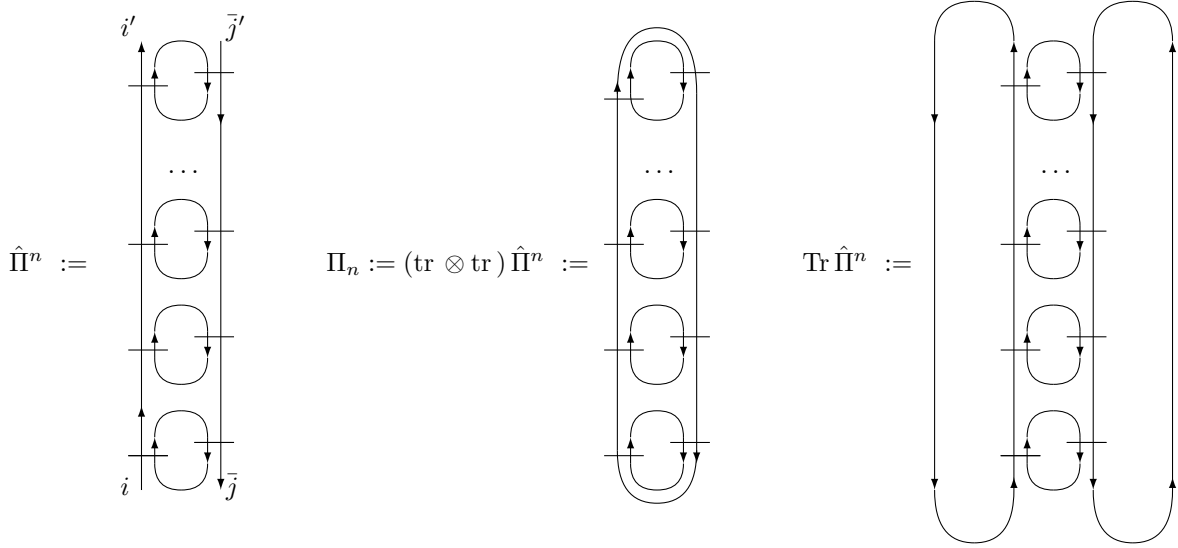

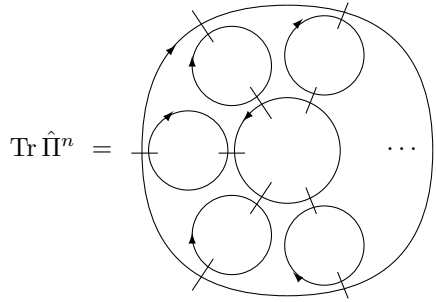
\begin{figure}[h!]
\begin{picture}(100,105)(-250,-50)

\put(-105,-2){\mbox{$\Tr\hat \Pi^n \ =$}}

\put(0,0){\circle{40}}
\qbezier(0,-55)(-55,-55)(-55,0)
\qbezier(0,55)(-55,55)(-55,0)
\qbezier(0,-55)(55,-55)(55,0)
\qbezier(0,55)(55,55)(55,0)

\put(-37.5,0){\circle{30}}
\put(-21,32){\circle{30}}
\put(-21,-32){\circle{30}}
\put(14,36){\circle{30}}
\put(14,-36){\circle{30}}

\put(37.5,0){\mbox{$\ldots$}}

\put(-44,38){\vector(1,1){2}}
\put(-15,13){\vector(-1,-1){2}}
\put(4,48){\vector(1,1){2}}
\put(-36,34){\vector(0,1){2}}
\put(-45,13){\vector(1,1){2}}
\put(-36,-33){\vector(0,1){2}}
\put(4,-48){\vector(-1,1){2}}

\put(23,56){\line(-2,-5){4}}
\put(11,24){\line(-2,-5){4}}
\put(-36,53){\line(2,-3){8}}
\put(-14,24){\line(2,-3){8}}
\put(-59,-1){\line(1,0){10}}
\put(-26,-1){\line(1,0){10}}
\put(-36,-53){\line(2,3){8}}
\put(-14,-24){\line(2,3){8}}
\put(23,-56){\line(-2,5){4}}
\put(11,-24){\line(-2,5){4}}

\end{picture}
\caption{\footnotesize  Alternative, more symmetric, picture for $\Tr\hat \Pi^n$
} \label{fig:circles}
\end{figure}

\be\label{v-coef}
\begin{array}{c|c||c  }
n & v_n &  \Tr \hat \Pi^n
%& {\rm quantization} ???
\\
\hline
1 & 1 &  \frac{N(N+1)^2 }{4}
%& \frac{[N][N+1]^2}{[2]^2} =D_{[2]}\frac{[N+1]}{[2]}
 {\phantom{\int^{\int^{\int^\int}}}} \\
2& 3N+4 & \frac{N(N+1)(N^2+3N+4)}{16}
%& D_{[2]}\frac{[N+1][N+2] + [2]}{[2]^3}
 {\phantom{\int^{\int^{\int^\int}}}} \\
3& 7N^2+18N+12 & \frac{N(N+1)(N^3+5N^2+14N+12)}{64}
%& ??? D_{[2]}\frac{[N+1][N+2]^2+[2](2[N+2]+[N])}{[2]^5}
 {\phantom{\int^{\int^{\int^\int}}}} \\
4& (3N+4)(5N^2+12N+8) & \frac{N(N+1)(N^4+7N^3+32N^2+56N+32)}{256}
%& ??? D_{[2]}\frac{[N+1][N+2]^3+???14N^2+36N+24}{[2]^7}
 {\phantom{\int^{\int^{\int^\int}}}} \\
5&31N^4 +150N^3+280N^2+240N+80 & \frac{N(N+1)(N^5+9N^4+62N^3+168N^2+192N+80)}{1024}
%& ??? D_{[2]}\frac{[N+1][N+2]^4+???30N^3+112N^2+144N+64}{[2]^7}
 {\phantom{\int^{\int^{\int^\int}}}} \\
\ldots &&\ldots
%&\ldots
\\
&&  \frac{N(N+1)\big((N+1)(N+2)^{n-1}+2v_{n-1}\big)}{4^n}
%&  D_{[2]}\frac{[N+1][N+2]^{n-1}+[2]V_{n-1}}{[2]^{2n-1}}
\end{array}
\ee

\noindent Note that $\Tr \hat \Pi = \Pi_2\,$, what can be easily seen from their pictures. Formula~\eqref{TrPi^n} implies that the quantization of $\Tr \hat \Pi^n$ can be reduced to a quantization rule for $v_n\longrightarrow V_n\,$:
\be
\Tr_q\, \hat \Pi^n = D_{[2]}\frac{[N+1][N+2]^{n-1}+[2]V_{n-1}}{[2]^{2n-1}}
\ee
which is considerably simpler to invent:
\be
\begin{array}{c|c|c}
n & v_n & V_n \\
\hline
1 & 1 & 1 \\
2 &3N+4 & 2[N+2]+[N] = [2][N+1]+[N+2] \\
3 & 7N^2+18N+12 &  [2]^2[N+1]^2+[2][N+1][N+2]+[N+2]^2 \\
4 & 15N^3 +56N^2+72N + 32 &  \\
5 & 31N^4 +150N^3+280N^2+240N+80 &  \\
\ldots & \ldots & \ldots \\
n &  \sum_{k=0}^n \frac{n!}{k!(n-k)!} (2^{n}-2^k)N^{n-k-1}
& \sum_{k=0}^{n-1} \big([2][N+1]\big)^{n-k-1}[N+2]^k  \\
& =\frac{(2N+2)^n-(N+2)^n}{N} & = \frac{\big([2][N+1]\big)^n-[N+2]^n}{[N]}
\end{array}
\ee
As we will see in Section~\ref{sec:quantum-proj}, this is indeed the case, see~\eqref{TrPi^n-quant}.

\subsubsection{Combinations of quantum projectors}\label{sec:quantum-proj}

In what follows, $\Tr$ mean ordinary traces, and $\Tr_q$ are quantum traces including the non-trivial grading operators $\hat{\cal M}$ and their inverse, see Section~\ref{sec:RT-CS}. Quantum projector~\cite{AnoAnd} is made out of the ${\cal R}$-matrix:
\be\label{proj-[2]}
\hat P = \frac{{\cal R}+q^{-1}}{[2]}\,.
\ee
It satisfies $\hat P^2 = \hat P$ because of the skein relation $({\cal R}-q)({\cal R}+q^{-1})=0$.
${\cal R}$, and thus, $\hat{P}$ are not diagonal, e.g. for $N=2$ the fundamental ${\cal R}$-matrix is
\begin{equation}
    {\cal R} = \left(\begin{array}{cccc} q & 0 & 0 & 0 \\ 0 & \{q\} & 1 & 0 \\ 0 & 1 & 0 & 0 \\ 0 & 0 & 0 & q \end{array}\right),\quad \text{and} \quad \hat P = \left(\begin{array}{cccc}
 1 & 0 & 0 & 0 \\\vspace{0.1cm} 0 & \frac{q}{[2]} & \frac{1}{[2]} & 0 \\ 0 & \frac{1}{[2]} & \frac{q^{-1}}{[2]} & 0 \\ 0 & 0 & 0 & 1 \end{array}\right).
\end{equation}
For generic $N$, the projector on the representation $[2]$ is
\be
P_{ij}^{i'j'} = \sum_{k=1}^N \delta_{ik}\delta_{jk}\delta^{i'}_k\delta^{j'}_k
+ \frac{1}{[2]}\sum_{k\neq l}^N \delta_{ik}\delta_{jl}\delta^{i'}_l\delta^{j'}_k
+ \frac{q}{[2]} \sum_{k<l}^N \delta_{ik}\delta_{jl}\delta^{i'}_k\delta^{j'}_l
+ \frac{q^{-1}}{[2]} \sum_{k>l}^N \delta_{ik}\delta_{jl}\delta^{i'}_k\delta^{j'}_l\,.
\ee
%Introduce a grading $N\times N$ diagonal matrix???
%\be
%\hat M:= {\rm diag} \Big(q^{2i-N-1},\ \  i = 1,\ldots, N\Big)
%\ee
%and its "conjugate"???
%\be
%\hat {\bar M}:= {\rm diag} \Big(q^{N+1-2i},\ \  i = 1,\ldots, N\Big)
%\ee
Note that for $q\neq 1$, $P_{ij}^{i'j'}\neq P_{i'j'}^{ij}$, so it is important to keep indices order in the quantum case. Traces of operators, as being closures, inevitably include critical points, and thus, the grading operators $\hat{\cal M}$, $\hat{\cal M}^{-1}$, see Fig.\,\ref{fig:R-M} and expression~\eqref{M-op}.
Then, one can use these quantities to define various traces.
To begin with,
\begin{equation}
\begin{aligned}
    D_{[1]} &= \Tr \hat {\cal M} = [N] = \frac{\{q^N\}}{\{q\}}\,, \\
D_{[2]} &= \Tr  \hat{\cal M}^{-1} \hat P\hat {\cal M} =
%\sum_{i_1,i_2,j_1,j_2}
\bar {\cal M}^{i_1}_{i_2}P_{i_1j_1}^{i_2j_2}{\cal M}^{j_1}_{j_2} = \frac{[N][N+1]}{[2]}\,.
\end{aligned}
\end{equation}
The quantization of our operator $\hat \Pi$ becomes
\be
\hat\Pi:=  \hat{P} \hat {\cal M} \hat{\bar P},  \ \  \ \ \ {\rm i.e.} \  \ \ \ \
\Pi_{i_1\bar j_1}^{i_2\bar j_2} := P_{i_1k_1}^{i_2k_2} {\cal M}^{k_1 \bar{j}_1}\delta_{k_2 \bar{j}_2} P_{\bar j_1 \bar k_1 }^{\bar j_2\bar l_2 }\,.
\ee
%Note that for $q\neq 1$ $P_{\bar j_1 \bar k_1 }^{\bar j_2\bar l_2 }\neq P_{\bar k_1\bar j_1  }^{\bar l_2\bar j_2}$
%and it is the first version, which should be used in the downarrow??? projector.
Now, we can compute the quantum version of $\Pi_n$ from Fig.\,\ref{fig:vertchain2}\,:
\begin{equation}
\begin{aligned}
    \Pi_1&=(\tr\otimes \tr) \hat {\cal M}\hat \Pi := {\cal M}^{i \bar j}\delta_{i_1\bar{j}_1} \Pi_{i\bar j}^{i_1\bar j_1} = \frac{[N][N+1]}{[2]}\,,
\\
\Pi_2&=(\tr\otimes \tr) \hat {\cal M}\hat \Pi^2 :=   {\cal M}^{i \bar j}\delta_{i_2\bar{j}_2} \Pi_{i\bar j}^{i_1\bar j_1}\Pi_{i_1\bar j_1}^{i_2\bar j_2}
= \frac{[N][N+1]^2}{[2]^2}\,,   \\
\Pi_3&=(\tr\otimes \tr) \hat {\cal M}\hat \Pi^3 :=   {\cal M}^{i \bar j}\delta_{i_3\bar{j}_3} \Pi_{i\bar i}^{i_1\bar j_1}\Pi_{i_1\bar j_1}^{i_2\bar j_2}
\Pi_{i_2\bar j_2}^{i_3\bar j_3}
= \frac{[N][N+1]^3}{[2]^3}\,,  \\
&\ldots \\
\Pi_n&=(\tr\otimes \tr) \hat {\cal M}\hat \Pi^n=\frac{[N][N+1]^n}{[2]^n} = D_{[2]} \left(\frac{[N+1]}{[2]}\right)^{n-1}.
\end{aligned}
\end{equation}
as anticipated in (\ref{Pin}). Then, quantization of $\Tr \hat{\Pi}^n$ from Fig.\,\ref{fig:vertchain2} is
{\small \begin{equation}\label{TrPi^n-quant}
\begin{aligned}
    \Tr_q\, \hat \Pi &:= \Tr \hat {\cal M}^{-1} \hat \Pi  \hat {\cal M} = \bar{\cal M}^{i_1}_{i_2}\Pi_{i_1\bar j_1}^{i_2\bar j_2} {\cal M}^{\bar j_1}_{\bar j_2}
= \frac{[N][N+1]^2}{[2]^2}\,,  \\
\Tr_q\, \hat \Pi^2 &:= \Tr \hat {\cal M}^{-1} \hat \Pi^2  \hat {\cal M} = \frac{[N][N+1]}{[2]^4}\Big( [N+1][N+2] +[2] \Big)\,, \\
\Tr_q\, \hat \Pi^3 &:= \Tr \hat {\cal M}^{-1} \hat \Pi^3  \hat {\cal M} = \frac{[N][N+1]}{[2]^6}\Big( [N+1][N+2]^2+[2]\big([2][N+1]+[N+2]\big)\Big)\,, \\
\Tr_q\, \hat \Pi^4 &:= \Tr \hat {\cal M}^{-1} \hat \Pi^4  \hat {\cal M} = \frac{[N][N+1]}{[2]^8}\Big( [N+1][N+2]^3
+[2]\big([2]^2[N+1]^2+[2][N+1][N+2]+[N+2]^2\big)\Big)\,, \\
%\Tr_q\, \hat \Pi^5 &:= \Tr \hat M\hat \Pi^5 \hat {\bar M}   = \frac{[N][N+1]}{[2]^{10}}\Big( [N+1][N+2]^4
%+[2]\big([2]^3[N+1]^3+[2]^2[N+1]^2[N+2]+[2][N+1][N+2]^2+[N+2]^3\big)\Big)
&\ldots \\
\Tr_q\, \hat \Pi^n &:= \Tr \hat {\cal M}^{-1} \hat \Pi^n  \hat {\cal M}  = \frac{[N][N+1]}{[2]^{2n}}
\left([N+1][N+2]^{n-1} +  [2] \sum_{i=0}^{n-2} [2]^{n-2-i}[N+1]^{n-2-i}[N+2]^i\right)\,.
\end{aligned}
\end{equation}}

\subsubsection{Projector combinations from planar decomposition}\label{sec:proj-prop-[2]}

Another approach to projector calculus relies on intermediate configurations
which explicitly admit planar decomposition.
%It neither rely on naive quantization, nor on explicit computation of the quantum contraction of $\Pi$ and $M$ in each case.
It is based on the following statements, which can be derived from the definitions of projectors:
%Explicit form of the projectors and  $M$-matrices leads to the planar decompositions
\begin{figure}[h!]
\centering
\includegraphics[width=6cm]{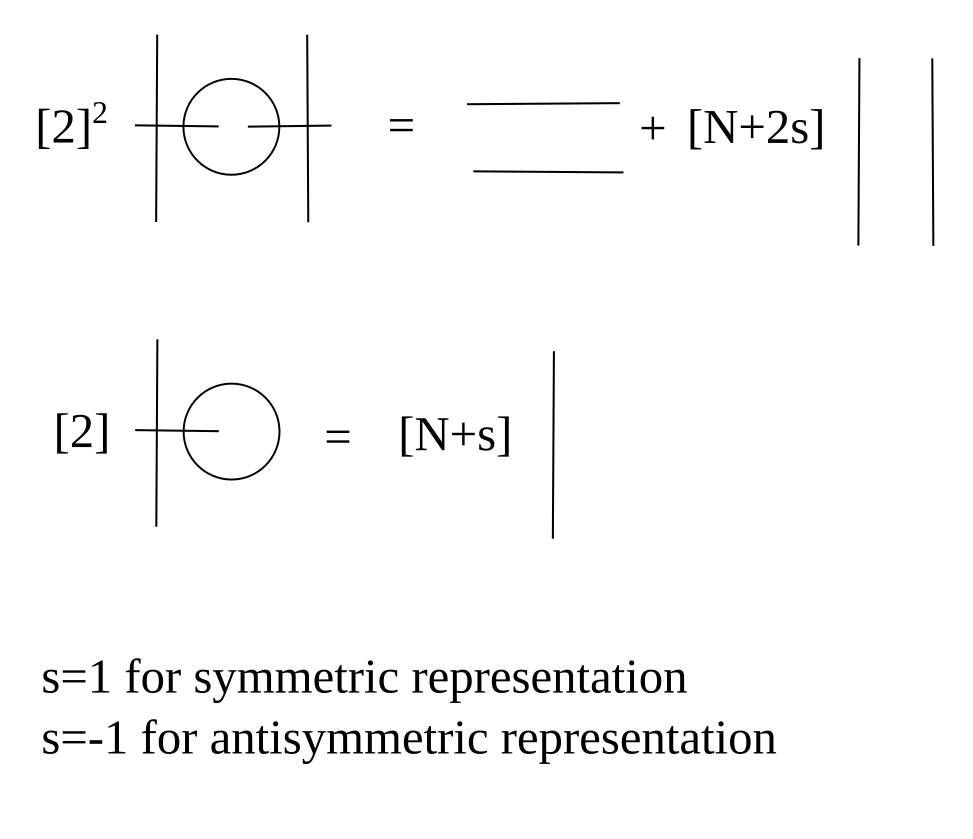}
\caption{\footnotesize Erasure of single circles -- as implied by projector properties. At the l.h.s. we have contractions of projectors, while at the r.h.s. there are only single lines carrying fundamental representations.}\label{fig:proj-prop}
\end{figure}

Note that the r.h.s. of these formulas contain only single lines --
i.e. only fundamental representations,
and in the first line we obtain their antiparallel combinations. Actually, it is {\it not} the true planar decomposition in representation $[2]$
shown in Fig.\,\ref{fig:proje2lock}. Quantum projectors are made up from $\cal R$-matrices and their eigenvalues, see the example of the projector on representation $[2]$ in~\eqref{proj-[2]}, and thus, contain non-planar crossings. Still, this refers to open tangles,
while a closed $r$-cabled bipartite link possesses a truly planar decomposition. First, one needs to decompose cabled lock vertices to $(r+1)$ diagrams (as shown in Fig.\,\ref{fig:proje2lock}, for example). Second, one erases cycles inside resolved diagrams (as dictated by rules similar to Fig.\,\ref{fig:proj-prop}) and ends up with planarized (cabled) cycles decompositions giving $D_{[1]},\dots,D_{[r]}$ contributions to the resulting HOMFLY polynomial. 

%In other words 
%Still, it can be useful.
The decomposition in Fig.\,\ref{fig:proje2lock} is also useful to provide answers for certain projectors contractions without tedious calculations. For example, this decomposition immediately provides recurrent relations for $\Pi_n$ and $\Tr \hat \Pi_n$,
which can be easily solved. Namely,
\be
\begin{array}{c}
\includegraphics[width=10cm]{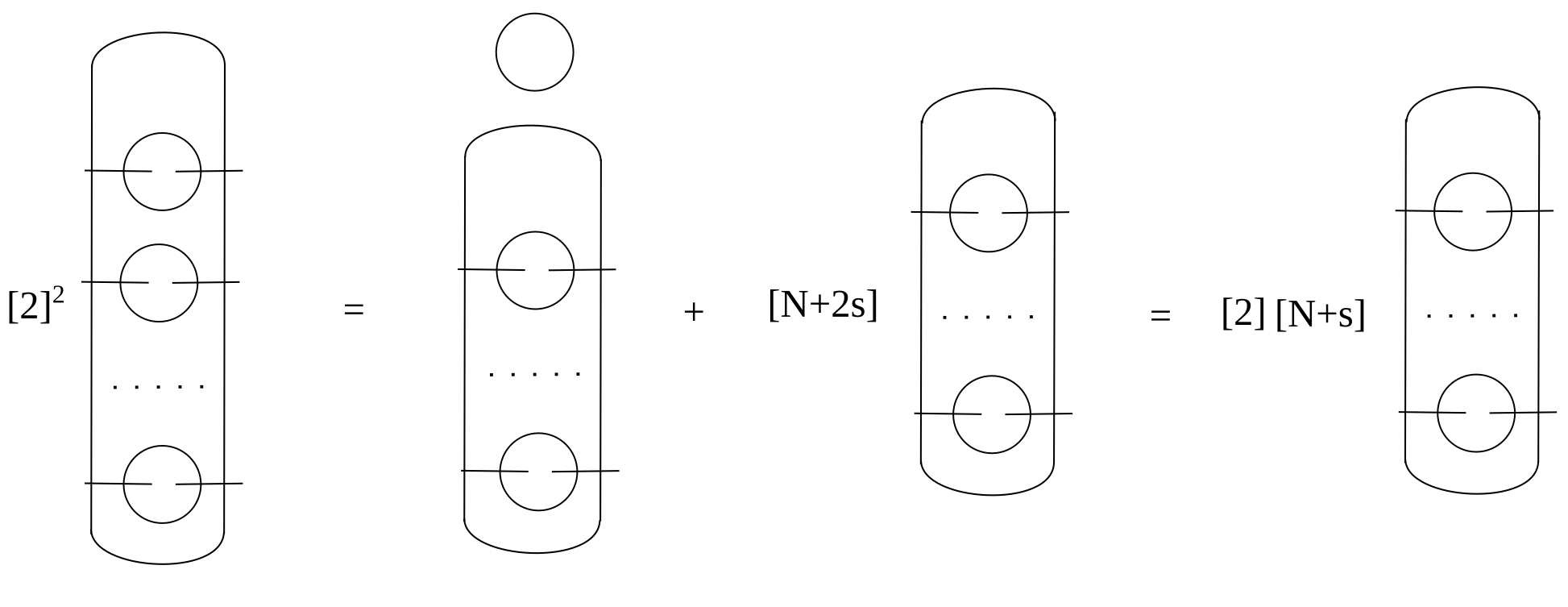}\\
%\Pi_n=\frac{[N]+[N+2s]}{[2]^2}=\frac{[N+s]}{[2]}\Pi_n,\\\\
\Pi_n= \frac{[N]+[N+2s]}{[2]^2}\,\Pi_{n-1} =\frac{[N+s]}{[2]}\, \Pi_{n-1}\,,\\\\
\Pi_1=D_{[2]}\qquad\Rightarrow\qquad
\Pi_n=D_{[2]}\left(\frac{[N+s]}{[2]}\right)^{n-1}.
\end{array}
\ee
and
\be
\includegraphics[width=8cm]{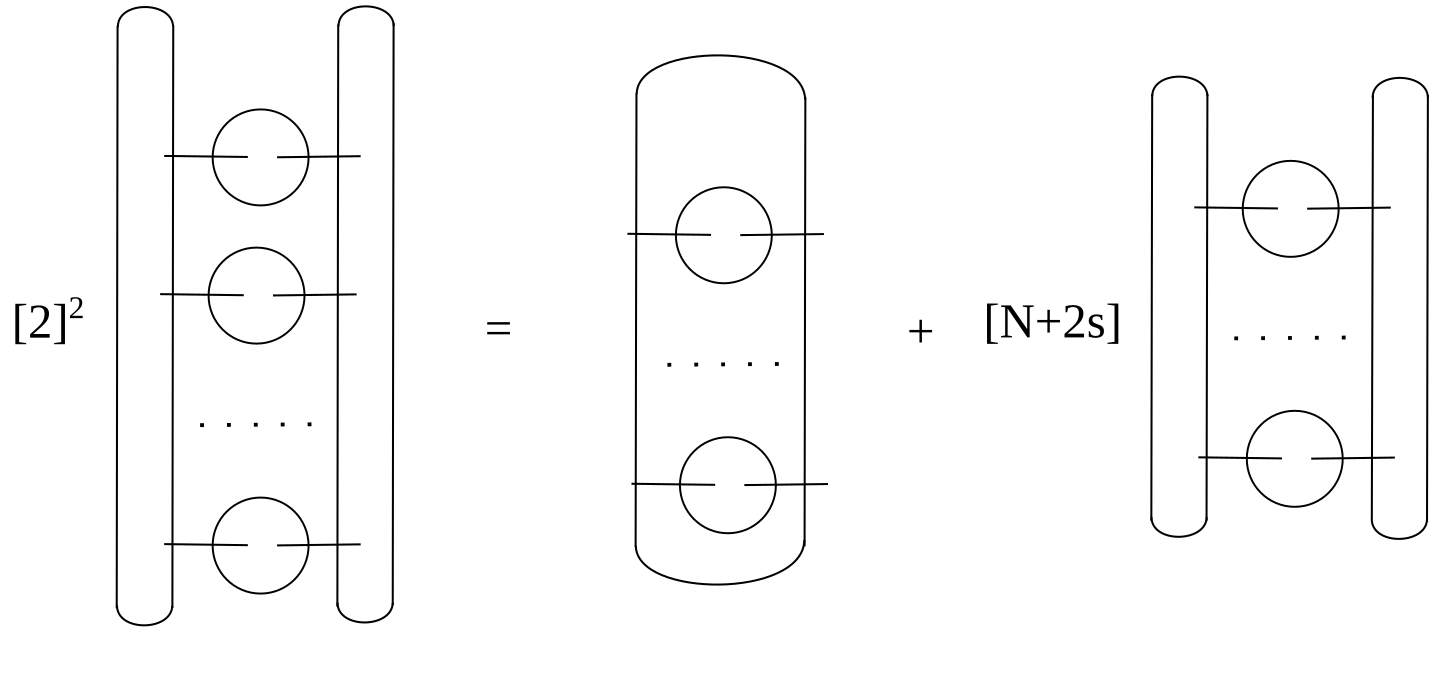}
\ee
\be
\begin{array}{c}
\Tr_q \,\hat\Pi^n=\frac{1}{[2]^2}\Pi_{n-1}+\frac{[N+2s]}{[2]^2}\Tr_q \, \hat\Pi^{n-1}\,, \\\\
\Tr_q \,\hat\Pi^1=\Pi_2\qquad \Rightarrow\ \\ \Tr_q \,\hat\Pi^n=\frac{D_{[2]}}{[2]^{2n-1}}\left([N+2s]^{n-1}[N+s]+[2]\sum\limits_{u=0}^{n-2} [2]^u[N+2s]^u[N+s]^{n-2-u}\right).
\end{array}
\ee	
Note that here we have provided calculations in both the first symmetric and antisymmetric representations. The answers transform into each other by the substitution $q\rightarrow q^{-1}$ due to the symmetry of the colored HOMFLY polynomial $H_{R^{T}}^{\cal L}(q,A)=H_R^{\cal L}(q^{-1},A)$ where we denote by $T$ the transposition of a Young diagram. 

%This observation is a one more way to compute the above projector combinations.

%???\newpage

\subsection{Chain of antiparallel lock tangles}\label{sec:chain}

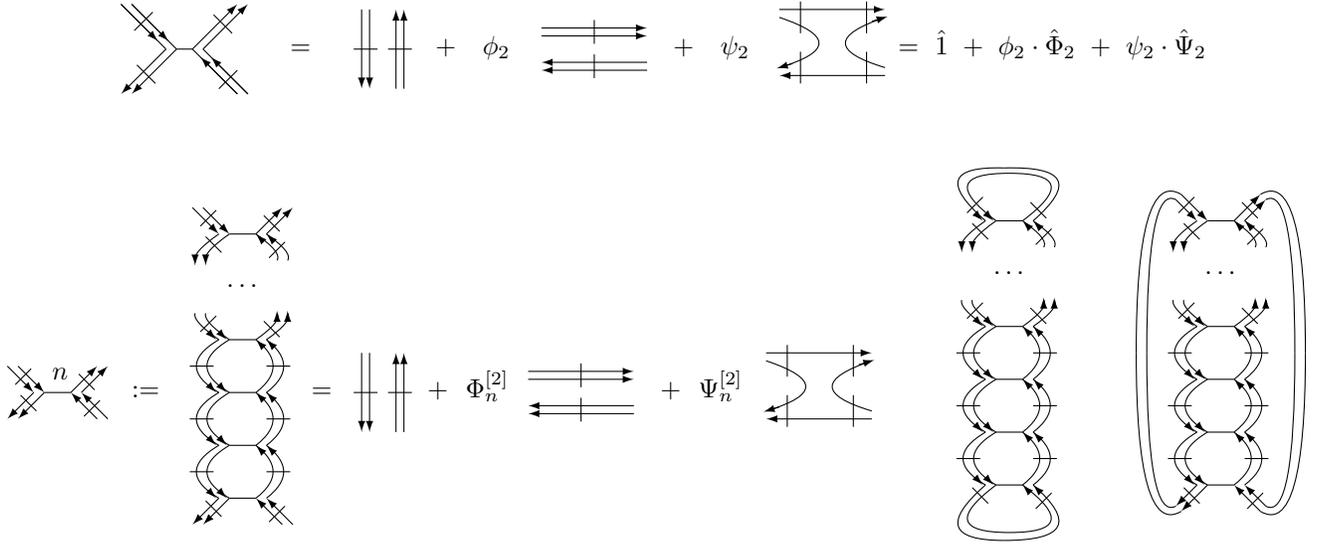
\begin{figure}[h!]

\begin{picture}(100,220)(-320,-190)

\put(-250,0){

\put(0,0){

\put(-19,7){\line(1,1){7}}
\put(13,-15){\line(1,1){7}}
\put(11,13){\line(1,-1){7}}
\put(-18,-6){\line(1,-1){7}}
\put(-24,17){\line(1,-1){17}}
\put(-24,17){\vector(1,-1){14}}
\put(24,-17){\line(-1,1){17}}
\put(24,-17){\vector(-1,1){14}}
\put(7,0){\vector(1,1){17}}
 \put(-7,0){\vector(-1,-1){17}}

 \put(-20,17){\line(1,-1){17}}\put(-20,17){\vector(1,-1){14}}   \put(3,0){\vector(1,1){17}}
 \put(-3,0){\vector(-1,-1){17}}   \put(20,-17){\line(-1,1){17}} \put(20,-17){\vector(-1,1){14}}
 \put(-3,0){\line(1,0){6}}
 }

\put(0,0){
\put(40,-2){\mbox{$=$}}

\put(0,65){

\put(77,-65){\line(1,0){9}}
\put(64,-65){\line(1,0){9}}
\put(70,-50){\vector(0,-1){30}}
\put(80,-80){\vector(0,1){30}}
\put(67,-50){\vector(0,-1){30}}
\put(83,-80){\vector(0,1){30}}

\put(95,-67){\mbox{$+\ \ \ \phi_2$}}

\put(155,-63){\line(0,1){9}}
\put(155,-76){\line(0,1){9}}
\put(135,-60){\vector(1,0){40}}
\put(175,-70){\vector(-1,0){40}}
\put(135,-57){\vector(1,0){40}}
\put(175,-73){\vector(-1,0){40}}

\put(185,-67){\mbox{$+\ \ \ \psi_2$}}

\put(225,-50){\vector(1,0){40}}
\put(265,-75){\vector(-1,0){40}}

\qbezier(225,-53)(255,-63)(225,-72)
\qbezier(265,-53)(235,-63)(265,-72)
\put(230,-70.5){\vector(-1,-0.3){6}}
\put(260,-54.5){\vector(1,0.3){6}}

\put(258,-59){\line(0,1){12}}
\put(258,-78){\line(0,1){12}}
\put(233,-59){\line(0,1){12}}
\put(233,-78){\line(0,1){12}}

\put(270,-67){\mbox{$ = \ \hat{1} \ + \ \phi_2 \cdot \hat{\Phi}_2 \ + \ \psi_2 \cdot \hat{\Psi}_2 $}}

}
}
}

%--------------------------------------
%vertical chain operator

\put(-230,-170){

\put(-40,38){\mbox{$:=$}}

\put(-70,40){

\put(-14,3){\line(1,1){6}}
\put(12,-9){\line(1,1){6}}
\put(16,1){\line(-1,1){6}}
\put(-6,-7){\line(-1,1){6}}

\put(-17,10){\vector(1,-1){10}}
\put(-13,10){\vector(1,-1){10}}
\put(11,0){\vector(1,1){10}}
\put(7,0){\vector(1,1){10}}
\put(21,-10){\vector(-1,1){10}}
\put(17,-10){\vector(-1,1){10}}
\put(-3,0){\vector(-1,-1){10}}
\put(-7,0){\vector(-1,-1){10}}

\put(-3,0){\line(1,0){10}}
\put(0,5){\mbox{$n$}}
}

\put(0,100){

\put(-14,3){\line(1,1){6}}
\put(12,-9){\line(1,1){6}}
\put(16,1){\line(-1,1){6}}
\put(-6,-7){\line(-1,1){6}}

\qbezier(-7,0)(-17,-7)(-16,-10)
\put(-16,-10){\vector(0,-1){2}}
\qbezier(-3,0)(-13,-7)(-12,-10)
\put(-12,-10){\vector(0,-1){2}}

\qbezier(11,0)(21,-7)(19,-10)
\put(13,-2){\vector(-1,1){2}}
\qbezier(7,0)(17,-7)(15,-10)
\put(9,-2){\vector(-1,1){2}}
\put(-3,0){\line(1,0){10}}

\put(-13,10){\vector(1,-1){10}}
\put(7,0){\vector(1,1){10}}
\put(-17,10){\vector(1,-1){10}}
\put(11,0){\vector(1,1){10}}
}

\put(-4,80){\mbox{$\ldots$}}

\put(0,60){

\put(-14,3){\line(1,1){6}}
\put(16,1){\line(-1,1){6}}

\put(-18,-10){\line(1,0){9}}
\put(11,-10){\line(1,0){9}}

\qbezier(-7,0)(-17,7)(-16,10)
\put(-10,2){\vector(1,-1){2}}
\qbezier(-3,0)(-13,7)(-12,10)
\put(-6,2){\vector(1,-1){2}}

\qbezier(11,0)(21,7)(19,10)
\put(19,9){\vector(0,1){2}}
\qbezier(7,0)(17,7)(15,10)
\put(15,9){\vector(0,1){2}}
\put(-3,0){\line(1,0){10}}

}

\put(0,40){

\qbezier(-7,0)(-24,10)(-7,20)
\put(-10,2){\vector(1,-1){2}}
\qbezier(-3,0)(-20,10)(-3,20)
\put(-6,2){\vector(1,-1){2}}

\put(-18,-10){\line(1,0){9}}
\put(11,-10){\line(1,0){9}}

\qbezier(11,0)(24,10)(11,20)
\put(13,18){\vector(-1,1){2}}
\qbezier(7,0)(20,10)(7,20)
\put(9,18){\vector(-1,1){2}}
\put(-3,0){\line(1,0){10}}

}

\put(0,20){

\qbezier(-3,0)(-20,10)(-3,20)
\put(-6,2){\vector(1,-1){2}}
\qbezier(-7,0)(-24,10)(-7,20)
\put(-10,2){\vector(1,-1){2}}

\put(-18,-10){\line(1,0){9}}
\put(11,-10){\line(1,0){9}}

\qbezier(7,0)(20,10)(7,20)
\put(9,18){\vector(-1,1){2}}
\qbezier(11,0)(24,10)(11,20)
\put(13,18){\vector(-1,1){2}}
\put(-3,0){\line(1,0){10}}

}

\qbezier(-3,0)(-20,10)(-3,20)
\put(-6,2){\vector(1,-1){2}}
\qbezier(-7,0)(-24,10)(-7,20)
\put(-10,2){\vector(1,-1){2}}

\qbezier(7,0)(20,10)(7,20)
\put(9,18){\vector(-1,1){2}}
\qbezier(11,0)(24,10)(11,20)
\put(13,18){\vector(-1,1){2}}
\put(-3,0){\line(1,0){10}}

\put(-3,0){\vector(-1,-1){10}}
\put(17,-10){\vector(-1,1){10}}
\put(-7,0){\vector(-1,-1){10}}
\put(21,-10){\vector(-1,1){10}}

\put(12,-9){\line(1,1){6}}
\put(-6,-7){\line(-1,1){6}}

\put(-20,105){
\put(48,-67){\mbox{$=$}}

\put(64,-65){\line(1,0){9}}
\put(77,-65){\line(1,0){9}}

\put(70,-50){\vector(0,-1){30}}
\put(80,-80){\vector(0,1){30}}
\put(67,-50){\vector(0,-1){30}}
\put(83,-80){\vector(0,1){30}}

\put(92,-67){\mbox{$+\ \ \Phi_n^{[2]}$}}

\put(130,-60){\vector(1,0){40}}
\put(170,-70){\vector(-1,0){40}}
\put(130,-57){\vector(1,0){40}}
\put(170,-73){\vector(-1,0){40}}

\put(150,-63){\line(0,1){9}}
\put(150,-76){\line(0,1){9}}

}

\put(-25,105){

\put(185,-67){\mbox{$+\ \ \Psi_n^{[2]}$}}

\put(225,-50){\vector(1,0){40}}
\put(265,-75){\vector(-1,0){40}}

\qbezier(225,-53)(255,-63)(225,-72)
\qbezier(265,-53)(235,-63)(265,-72)
\put(230,-70.5){\vector(-1,-0.3){6}}
\put(260,-54.5){\vector(1,0.3){6}}

\put(258,-59){\line(0,1){12}}
\put(258,-78){\line(0,1){12}}
\put(233,-59){\line(0,1){12}}
\put(233,-78){\line(0,1){12}}

}
}

%--------------------------------------
%unknot

\put(60,-165){

\put(0,100){

\put(-14,3){\line(1,1){6}}
\put(12,-9){\line(1,1){6}}
\put(16,1){\line(-1,1){6}}
\put(-6,-7){\line(-1,1){6}}

\qbezier(-3,0)(-13,-7)(-12,-10)
\put(-12,-10){\vector(0,-1){2}}
\qbezier(-7,0)(-17,-7)(-16,-10)
\put(-16,-10){\vector(0,-1){2}}

\qbezier(7,0)(17,-7)(15,-10)
\put(9,-2){\vector(-1,1){2}}
\qbezier(11,0)(21,-7)(19,-10)
\put(13,-2){\vector(-1,1){2}}
\put(-3,0){\line(1,0){10}}

\qbezier(-3,0)(-27,18)(2,18)
\qbezier(7,0)(31,18)(2,18)
 \put(-6,2){\vector(1,-1){2}}
 \qbezier(-7,0)(-31,21)(0,20)
\qbezier(11,0)(35,21)(0,20)
 \put(-10,2){\vector(1,-1){2}}
}

\put(-4,80){\mbox{$\ldots$}}

\put(0,60){

\put(-14,3){\line(1,1){6}}
\put(16,1){\line(-1,1){6}}

\put(-18,-10){\line(1,0){9}}
\put(11,-10){\line(1,0){9}}

\qbezier(-3,0)(-13,7)(-12,10)
\put(-6,2){\vector(1,-1){2}}
\qbezier(-7,0)(-17,7)(-16,10)
\put(-10,2){\vector(1,-1){2}}

\qbezier(7,0)(17,7)(15,10)
\put(15,9){\vector(0,1){2}}
\qbezier(11,0)(21,7)(19,10)
\put(19,9){\vector(0,1){2}}
\put(-3,0){\line(1,0){10}}
}

\put(0,40){

\put(-18,-10){\line(1,0){9}}
\put(11,-10){\line(1,0){9}}

\qbezier(-3,0)(-20,10)(-3,20)
\put(-6,2){\vector(1,-1){2}}
\qbezier(-7,0)(-24,10)(-7,20)
\put(-10,2){\vector(1,-1){2}}

\qbezier(7,0)(20,10)(7,20)
\put(9,18){\vector(-1,1){2}}
\qbezier(11,0)(24,10)(11,20)
\put(13,18){\vector(-1,1){2}}
\put(-3,0){\line(1,0){10}}
}

\put(0,20){

\put(-18,-10){\line(1,0){9}}
\put(11,-10){\line(1,0){9}}

\qbezier(-3,0)(-20,10)(-3,20)
\put(-6,2){\vector(1,-1){2}}
\qbezier(-7,0)(-24,10)(-7,20)
\put(-10,2){\vector(1,-1){2}}

\qbezier(7,0)(20,10)(7,20)
\put(9,18){\vector(-1,1){2}}
\qbezier(11,0)(24,10)(11,20)
\put(13,18){\vector(-1,1){2}}
\put(-3,0){\line(1,0){10}}
}

\put(12,-9){\line(1,1){6}}
\put(-6,-7){\line(-1,1){6}}

\qbezier(-3,0)(-20,10)(-3,20)
\put(-6,2){\vector(1,-1){2}}
\qbezier(-7,0)(-24,10)(-7,20)
\put(-10,2){\vector(1,-1){2}}

\qbezier(7,0)(20,10)(7,20)
\put(9,18){\vector(-1,1){2}}
\qbezier(11,0)(24,10)(11,20)
\put(13,18){\vector(-1,1){2}}
\put(-3,0){\line(1,0){10}}

\qbezier(-3,0)(-27,-18)(2,-18)
\qbezier(7,0)(31,-18)(2,-18)
\put(9,-2){\vector(-1,1){2}}
\qbezier(-7,0)(-31,-21)(0,-21)
\qbezier(11,0)(35,-21)(0,-21)
\put(13,-2){\vector(-1,1){2}}
}

%--------------------------------------
%APT link

\put(140,-165){

\put(0,100){

\put(-14,3){\line(1,1){6}}
\put(12,-9){\line(1,1){6}}
\put(16,1){\line(-1,1){6}}
\put(-6,-7){\line(-1,1){6}}

\qbezier(-3,0)(-13,-7)(-12,-10)
\put(-12,-10){\vector(0,-1){2}}
\qbezier(-7,0)(-17,-7)(-16,-10)
\put(-16,-10){\vector(0,-1){2}}

\qbezier(7,0)(17,-7)(15,-10)
\put(9,-2){\vector(-1,1){2}}
\qbezier(11,0)(21,-7)(19,-10)
\put(13,-2){\vector(-1,1){2}}
\put(-3,0){\line(1,0){10}}

\put(-13,10){\vector(1,-1){10}}
\put(7,0){\vector(1,1){10}}
\put(-15,8){\vector(1,-1){8}}
\put(11,0){\vector(1,1){8}}
}

\put(-4,80){\mbox{$\ldots$}}

\put(0,60){

\put(-18,-10){\line(1,0){9}}
\put(11,-10){\line(1,0){9}}

\put(-14,3){\line(1,1){6}}
\put(16,1){\line(-1,1){6}}

\qbezier(-3,0)(-13,7)(-12,10)
\put(-6,2){\vector(1,-1){2}}
\qbezier(-7,0)(-17,7)(-16,10)
\put(-10,2){\vector(1,-1){2}}

\qbezier(7,0)(17,7)(15,10)
\put(15,9){\vector(0,1){2}}
\qbezier(11,0)(21,7)(19,10)
\put(19,9){\vector(0,1){2}}
\put(-3,0){\line(1,0){10}}
}

\put(0,40){

\put(-18,-10){\line(1,0){9}}
\put(11,-10){\line(1,0){9}}

\qbezier(-3,0)(-20,10)(-3,20)
\put(-6,2){\vector(1,-1){2}}
\qbezier(-7,0)(-24,10)(-7,20)
\put(-10,2){\vector(1,-1){2}}

\qbezier(7,0)(20,10)(7,20)
\put(9,18){\vector(-1,1){2}}
\qbezier(11,0)(24,10)(11,20)
\put(13,18){\vector(-1,1){2}}
\put(-3,0){\line(1,0){10}}
}

\put(0,20){

\put(-18,-10){\line(1,0){9}}
\put(11,-10){\line(1,0){9}}

\qbezier(-3,0)(-20,10)(-3,20)
\put(-6,2){\vector(1,-1){2}}
\qbezier(-7,0)(-24,10)(-7,20)
\put(-10,2){\vector(1,-1){2}}

\qbezier(7,0)(20,10)(7,20)
\put(9,18){\vector(-1,1){2}}
\qbezier(11,0)(24,10)(11,20)
\put(13,18){\vector(-1,1){2}}
\put(-3,0){\line(1,0){10}}
}

\put(12,-9){\line(1,1){6}}
\put(-6,-7){\line(-1,1){6}}

\qbezier(-3,0)(-20,10)(-3,20)
\put(-6,2){\vector(1,-1){2}}
\qbezier(-7,0)(-24,10)(-7,20)
\put(-10,2){\vector(1,-1){2}}

\qbezier(7,0)(20,10)(7,20)
\put(9,18){\vector(-1,1){2}}
\qbezier(11,0)(24,10)(11,20)
\put(13,18){\vector(-1,1){2}}
\put(-3,0){\line(1,0){10}}

\put(-3,0){\vector(-1,-1){10}}
\put(17,-10){\vector(-1,1){10}}
\put(-7,0){\vector(-1,-1){8}}
\put(19,-8){\vector(-1,1){8}}

\qbezier(-13,110)(-30,120)(-30,50)
\qbezier(-13,-10)(-30,-20)(-30,50)
\qbezier(-15,108)(-26,114)(-26,50)
\qbezier(-15,-8)(-26,-14)(-26,50)

\qbezier(19,108)(30,114)(30,50)
\qbezier(19,-8)(30,-14)(30,50)
\qbezier(17,110)(34,120)(34,50)
\qbezier(17,-10)(34,-20)(34,50)
}

\end{picture}
\caption{\footnotesize
%??? add different orientation of chain ??? 
The horizontal AP lock tangle and the {\bf vertical chain operator}, obtained by vertical iteration of this lock.
Since it appear in further examples, we also introduce a special abbreviated notation.
A chain tangle can be closed in two different ways giving rise to the unknot and to the 2-component 2-strand AP torus links
$APT[2,2n]$. 
%, see Fig.\,\ref{fig:chain-[3]-closures}.
With the exception of Hopf at $n=1$, they are different from the more familiar parallel torus links,
which are more similar to torus 2-strand {\it knots} with odd number of intersections.
%Two-strand torus {\it knots} are also bipartite (as being rational, see Theorem 2 in Section~\ref{sec:bipknots}) -- but depicted by more sophisticated diagrams.
Since the decomposition of the chain is just the same as for the single AP lock
(only $\phi_2$ is changed for $\Phi_n^{[2]}$ and $\psi_2$ is changed for $\Psi_n^{[2]}$),
it is especially simple to make further compositions. %beginning from the double braids.
}\label{fig:vertchain}
\end{figure}

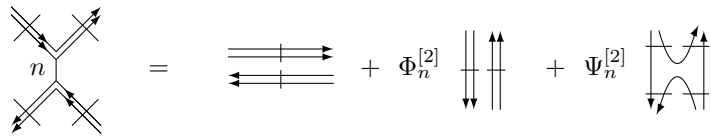
\begin{figure}[h!]
\begin{picture}(100,50)(20,-30)

\put(100,0){

\put(40,-2){\mbox{$n$}}

\put(50,0){
\put(-17,21){\line(1,-1){17}}\put(-17,24){\line(1,-1){17}}   \put(0,4){\vector(1,1){17}}
\put(-17,21){\vector(1,-1){14}} \put(-17,24){\vector(1,-1){14}}
\put(0,7){\vector(1,1){17}}
\put(0,-4){\vector(-1,-1){17}}   \put(17,-21){\line(-1,1){17}} \put(17,-21){\vector(-1,1){14}}
\put(0,-7){\vector(-1,-1){17}}   \put(17,-24){\line(-1,1){17}} \put(17,-24){\vector(-1,1){14}}
\put(0,4){\line(0,-1){8}}

\put(-16,11){\line(1,1){10}}  \put(-16,-11){\line(1,-1){10}}
\put(16,11){\line(-1,1){10}}  \put(16,-11){\line(-1,-1){10}}
}

\put(85,-2){\mbox{$=$}}

\put(45,65){
\put(70,-60){\vector(1,0){40}} \put(70,-57){\vector(1,0){40}}
\put(110,-70){\vector(-1,0){40}}  \put(110,-67){\vector(-1,0){40}}
\put(90,-55){\line(0,-1){7}}   \put(90,-65){\line(0,-1){7}}
\put(120,-67){\mbox{$+\ \ \Phi^{[2]}_n$}}
\put(160,-50){\vector(0,-1){30}} \put(163,-50){\vector(0,-1){30}}
\put(170,-80){\vector(0,1){30}}  \put(173,-80){\vector(0,1){30}}
\put(158,-65){\line(1,0){7}}   \put(168,-65){\line(1,0){7}}
\put(190,-67){\mbox{$+\ \ \Psi^{[2]}_n$}}
\put(230,-50){\vector(0,-1){30}}
\put(250,-80){\vector(0,1){30}}
\qbezier(233,-50)(240,-75)(247,-50)  \put(246,-52){\vector(1,2){2}}
\qbezier(233,-80)(240,-55)(247,-80)  \put(234,-78){\vector(-1,-2){2}}
\put(228,-56){\line(1,0){10}}   \put(242,-56){\line(1,0){10}}
\put(228,-74){\line(1,0){10}}   \put(242,-74){\line(1,0){10}}
}
}

\end{picture}
\caption{\footnotesize  The horizontal chain tangle, made from vertical locks from Fig.\,\ref{fig:proje2lock}, projected to the symmetric representation $[2]$
and its planar decomposition.
} \label{fig:proje2lock-chain}
\end{figure}

\begin{figure}[h!]
$
\begin{array}{cccc}
\begin{array}{c}\includegraphics[width=1cm]{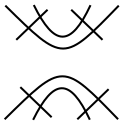}\end{array}&\hat 1&D_{[2]}&D_{[2]}^2\\
\begin{array}{c}\includegraphics[width=1cm]{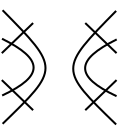}\end{array}&\hat\Phi_2&D_{[2]}^2&D_{[2]}\\
\begin{array}{c}\includegraphics[width=1cm]{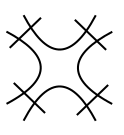}\end{array}&\hat\Psi_2&\Pi_2&\Tr\hat\Pi_1=\Pi_2\\
\begin{array}{c}\includegraphics[width=2cm]{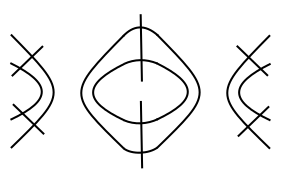}\end{array}&\hat\Phi_2\hat\Phi_2&D_{[2]}^3&D_{[2]}^2\\
\begin{array}{c}\includegraphics[width=2cm]{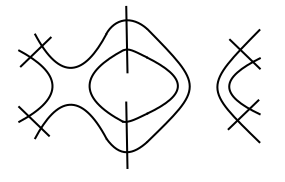}\end{array}&\hat\Psi_2\hat\Phi_2&D_{[2]}\Pi_2&\Pi_2\\
\begin{array}{c}\includegraphics[width=2cm]{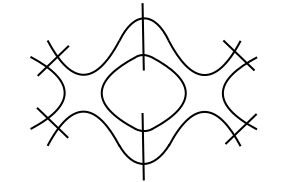}\end{array}&\hat\Psi_2\hat\Psi_2&\Pi_3&\Tr\hat\Pi_2\\
\end{array}
$
$
\begin{array}{cccc}
\begin{array}{c}\includegraphics[width=3cm]{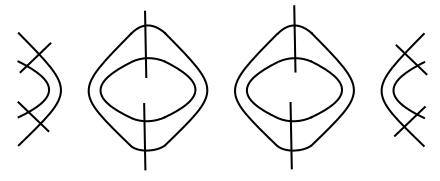}\end{array}&\hat\Phi_2\hat\Phi_2\hat\Phi_2&D_{[2]}^4&D_{[2]}^3\\
\begin{array}{c}\includegraphics[width=3cm]{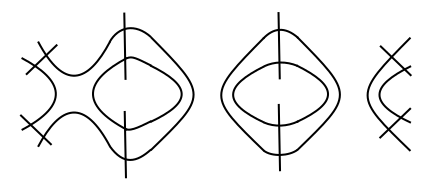}\end{array}&\hat\Psi_2\hat\Phi_2\hat\Phi_2&D_{[2]}^2\Pi_2&D_{[2]}\Pi_2\\
\begin{array}{c}\includegraphics[width=3cm]{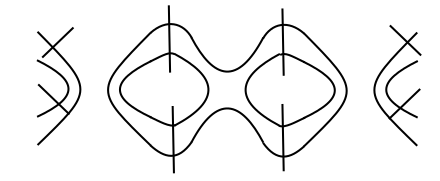}\end{array}&\hat\Phi_2\hat\Psi_2\hat\Phi_2&D_{[2]}^2\Pi_2&D_{[2]}\Pi_2\\
\begin{array}{c}\includegraphics[width=3cm]{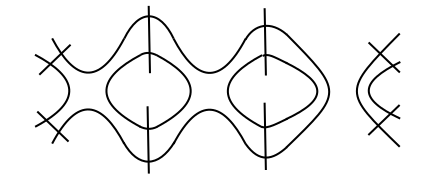}\end{array}&\hat\Psi_2\hat\Psi_2\hat\Phi_2&D_{[2]}\Pi_3&\Pi_3\\
\begin{array}{c}\includegraphics[width=3cm]{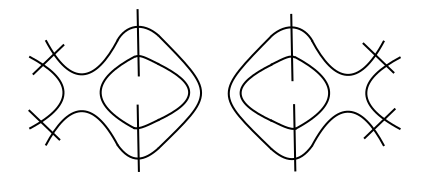}\end{array}&\hat\Psi_2\hat\Phi_2\hat\Psi_2&\Pi_2^2&\Pi_3\\
\begin{array}{c}\includegraphics[width=3cm]{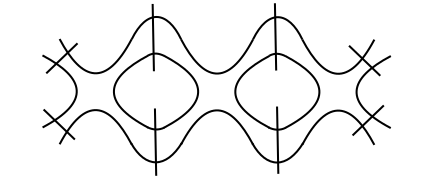}\end{array}&\hat\Psi_2\hat\Psi_2\hat\Psi_2&\Pi_4&\Tr\hat\Pi_3\\
\end{array}
$
\caption{\footnotesize %??? expand products of operators, add Fig. from SVALKA ??? 
Different terms in the planar expansion of the horizontal chain from Fig\,\ref{fig:proje2lock-chain} and of the diagrams for unknots
in Section~\ref{sec:2nunknots}, AP links in Section~\ref{sec:2links}.
They are drawn for another orientation as compared to Fig.\,\ref{fig:vertchain}: for horizontal chain made from vertical blocks,
shown in Fig.\,\ref{fig:proje2lock}.
From left to right: a picture of the operator, expression for the operator, its contribution to the unknot closure,
its contribution to the AP link closure.
Note a striking difference for two different closures.
}
\label{fig:PhiPsi}
\end{figure}

\begin{figure}[h!]
$
\begin{array}{cccc}
\begin{array}{c}\includegraphics[width=3cm]{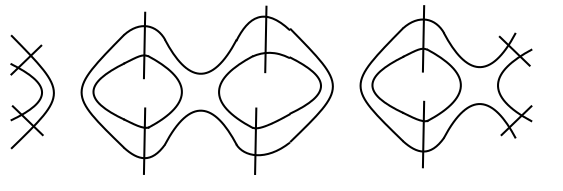}\end{array}&\hat \Phi_2\hat \Psi_2\hat \Phi_2\hat \Psi_2&\Pi_2^2D_{[2]} & \Pi_2^2 \\
\begin{array}{c}\includegraphics[width=3cm]{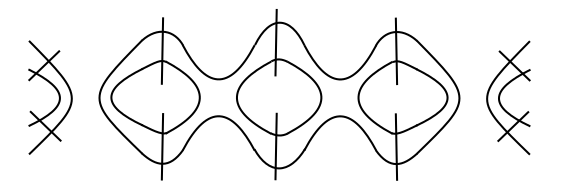}\end{array}&\hat \Phi_2\hat \Psi_2\hat \Psi_2\hat \Phi_2&\Pi_3D_{[2]}^2 & \Pi_3D_{[2]} \\
\begin{array}{c}\includegraphics[width=3cm]{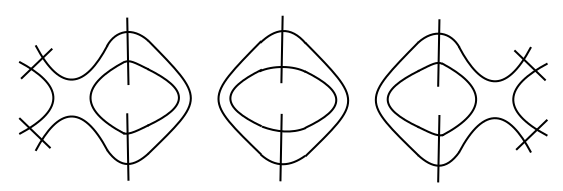}\end{array}&\hat \Psi_2\hat \Phi_2\hat \Phi_2\hat \Psi_2&\Pi_2^2D_{[2]} & \Pi_3D_{[2]} \\
\begin{array}{c}\includegraphics[width=3cm]{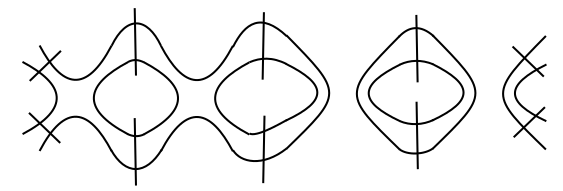}\end{array}&\hat \Psi_2\hat \Psi_2\hat \Phi_2\hat \Phi_2&\Pi_3D_{[2]}^2 & \Pi_3D_{[2]} \\
\end{array}
$
$
\begin{array}{cccc}
\begin{array}{c}\includegraphics[width=3cm]{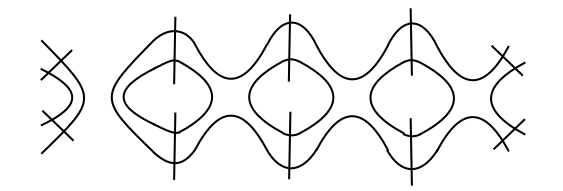}\end{array}&\hat \Phi_2\hat \Psi_2\hat \Psi_2\hat \Psi_2&\Pi_4D_{[2]} & \Pi_4 \\
\begin{array}{c}\includegraphics[width=3cm]{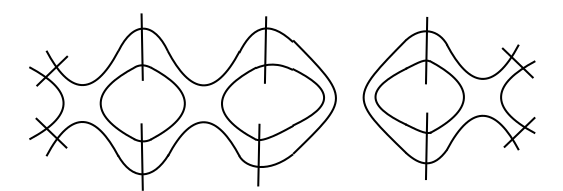}\end{array}&\hat \Psi_2\hat \Psi_2\hat \Phi_2\hat \Psi_2&\Pi_3\Pi_2 & \Pi_4 \\
\end{array}
$
\caption{\footnotesize 
Further examples of monomials in the expansion of~\eqref{chain-n}. From left to right: a picture of the operator, expression for the operator, its contribution to the unknot closure,
its contribution to the AP link closure.}\label{fig:PhiPsi4}
\end{figure}

\noindent An antiparallel {\it vertical} chain is a vertical iteration of the horizontal lock tangle, see Fig.\,\ref{fig:vertchain}. Its planar resolution is obtained from
\begin{equation}\label{chain-n}
    (\hat{1} + \phi_2 \cdot \hat{\Phi}_2 + \psi_2 \cdot \hat{\Psi}_2)^n.
\end{equation}
Expansion of this expression contains monomial diagrams, some of them are shown in Figs.\,\ref{fig:PhiPsi},\,\ref{fig:PhiPsi4}.
To save space, they are drawn in another orientation: for horizontal chain made from vertical blocks, see Fig.\,\ref{fig:proje2lock-chain}.

 It is clear that for $\sum_i k_i \neq 0$
\begin{equation}\label{mon-1}
\hat\Phi_2^{k_1}\hat\Psi_2^{l_1}\hat\Phi_2^{k_2}\hat\Psi_2^{l_2}\hat\Phi_2^{k_3} \ldots
\hat\Phi_2^{k_m} \hat\Psi_2^{l_{m}}=\hat\Phi_2^{\sum_i k_i}\hat\Psi_2^{\sum_i l_i}\sim \hat\Phi_2
\end{equation}
because a double circle gives just $D_{[2]}$ factor and single circles can be factorised due to projector property in Fig.\,\ref{fig:proj-prop} giving just $\frac{[N+1]}{[2]}$ multiplier. The rest monomials are of the form
\begin{equation}\label{mon-2}
    \hat{\Psi}_2^l
\end{equation}
%in the case of $l\neq 0$
with $\hat{\Psi}_2^0=\hat{1}$. In the case of $l\neq 0$, $\hat{\Psi}_2^l$ is reduced to a linear combination of $\hat{\Psi}_2$ and $\hat{\Phi}_2$ diagrams. In other words, a connected diagram with projectors may eventually contain disconnected contributions that must be attributed to $\Phi^{[2]}_n$. In total, we get\footnote{For the representation $[1]$, the similar formula was~\eqref{chain-[1]}.}%\cite{ALM}
%$$\Phi_n^{[1]} = \frac{(1+\phi_1 D_{[1]})^n-1}{D_{[1]}}$$}
\begin{equation}\label{Phi_n^[2]}
\begin{aligned}
    \Phi_n^{[2]}&=\underbrace{D_{[2]}^{-1}\left\{\left(1+\phi_2D_{[2]} +\psi_2\,\frac{[N+1]}{[2]}\right)^n-\left(1 +\psi_2\,\frac{[N+1]}{[2]}\right)^n\right\}}_{\text{contribution of~\eqref{mon-1}}}+\\
    &+\underbrace{D_{[2]}^{-1}\left\{\left(1 +\psi_2\,\frac{[N+1]}{[2]}\right)^n-\frac{[2][N+1]}{[N+2]}\left(1 +\psi_2\,\frac{[N+2]}{[2]^2}\right)^n+\frac{[N]}{[N+2]}\right\}}_{\text{contribution of~\eqref{mon-2} for }l\,\neq\, 0}=\\
    &=\frac{1}{D_{[2]}}\left(\left\{\left(1+\phi_2D_{[2]} +\psi_2\,\frac{[N+1]}{[2]}\right)^n-1\right\}
    - \frac{[2][N+1]}{[N+2]}\left\{\left(1 +\psi_2\,\frac{[N+2]}{[2]^2}\right)^n-1\right\}\right)= \\
    &=(A^2 q^2)^{2n}\cdot D_{[2]}^{-1}\left\{\lambda_\varnothing^{2n}-\frac{[2][N+1]}{[N+2]}\lambda_{\rm adj}^{2n}+\frac{[N]}{[N+2]}\lambda_{[4,2^{N-2}]}^{2n}\right\},
\end{aligned}
\end{equation}
and
\begin{equation}\label{Psi_n^[2]}
    \Psi_n^{[2]}=\frac{[2]^2}{[N+2]}\left(1 +\psi_2\,\frac{[N+2]}{[2]^2}\right)^n-\frac{[2]^2}{[N+2]}
    = (A^2 q^2)^{2n}\cdot \frac{[2]^2}{[N+2]}\left\{\lambda_{\rm adj}^{2n}-\lambda_{[4,2^{N-2}]}^{2n}\right\}.
\end{equation}
Here we introduce the eigenvalues relying {\it bipartite} and {\it ordinary evolutions}, see discussion in Section~\ref{sec:rep-[1]}\,:
\begin{equation}\label{eigen}
\begin{aligned}
    \lambda_{[4,2^{N-2}]}^2&=A^{-4} q^{-4}=\left(A^{-2}q^{-2}\right)^2\,, \\
    \lambda_{\rm adj}^2&=A^{-4} q^{-4}\left(1 +\psi_2\,\frac{[N+2]}{[2]^2}\right)=A^{-2}\,, \\
    \lambda_\varnothing^2 &= A^{-4} q^{-4} \left(1+\phi_2D_{[2]} +\psi_2\,\frac{[N+1]}{[2]}\right)=1\,.
\end{aligned}
\end{equation}
These two types of evolutions are still almost the same and are just written in different bases/within different objects. In the bipartite evolution, parameters are hidden in coefficients in front of planar resolutions while in the ordinary evolution, parameters stay in powers of the exponents being the $\cal R$-matrix eigenvalues.

Note that the last term in the middle of~\eqref{Psi_n^[2]} is present in order to cancel the contribution proportional to $\psi_2^0$ because $\hat{\Psi}_2^0=\hat{1}$ and must not contribute to the coefficient $\Psi_n^{[2]}$. It is also worth noting that $\Psi_n^{[2]}$ coefficient is independent on $\phi_2$ as operators including $\hat{\Phi}_2$ diagrams~\eqref{mon-1} contribute only in $\Phi_n^{[2]}$ coefficient, see~\eqref{Phi_n^[2]}. We remind that the quantum numbers are $[n]:=\frac{\{q^n\}}{\{q\}}$, so that for example $[N+2]=\frac{\{Aq^2\}}{\{q\}}$.

%???  ???
%(??? term evolution for chains, bipartite evolution vs ordinary evolution; /objects but  ???)

Considering the chain with $(n+m)$ lock elements, one can easily prove that
\begin{equation}\label{rel-(m+n)}
\begin{aligned}
    \Phi_{n+m}^{[2]}&=\Phi_n^{[2]}+\Phi_m^{[2]}+\Phi_n^{[2]}\Phi_m^{[2]}D_{[2]}+\frac{1}{[2]^2}\Psi_n^{[2]}\Psi_m^{[2]}+\frac{[N+1]}{[2]}\left(\Phi_n^{[2]}\Psi_m^{[2]}+\Psi_n^{[2]}\Phi_m^{[2]}\right)\,, \\
    \Psi_{n+m}^{[2]}&=\Psi_n^{[2]}+\Psi_m^{[2]}+\frac{[N+2]}{[2]^2}\Psi_n^{[2]}\Psi_m^{[2]}\,.
\end{aligned}
\end{equation}
In the particular case of $n=1$, $m=-1$
\begin{equation}\label{rel-(m+n)-part}
\begin{aligned}
    \Phi_1^{[2]}&=\phi_2\,,\quad \Phi_0^{[2]}=0\,,\quad \Phi_{-1}^{[2]}=\bar{\phi}_2\,, \\
    \Psi_1^{[2]}&=\psi_2\,,\quad \Psi_0^{[2]}=0,\quad \Psi_{-1}^{[2]}=\bar{\psi}_2\,,
\end{aligned}
\end{equation}
and from~\eqref{rel-(m+n)}, we get the relations
\begin{equation}
\begin{aligned}
    0&=\phi_2+\bar{\phi}_2 + \phi_2 \bar{\phi}_2 D_{[2]}+\frac{1}{[2]^2}\psi_2 \bar{\psi}_2 + \frac{[N+1]}{[2]}\left(\psi_2 \bar{\phi}_2+\phi_2 \bar{\psi}_2\right)\,, \\
    0&=\psi_2 + \bar{\psi}_2 + \frac{[N+2]}{[2]^2}\psi_2 \bar{\psi}_2\,.
\end{aligned}
\end{equation}
Coefficients of planar decomposition of a mirror chain of $n$ locks is obtained from ones by the change $A\rightarrow A^{-1}$, $q\rightarrow q^{-1}$, or equivalently, by $n\rightarrow -n$.

\subsection{Unknots from 2-strand braids}\label{sec:2nunknots}

We already know from (\ref{unknotHopf2}) that
\be
H^{\rm unknot}_{[2]}  = \frac{1}{A^4q^4} \left(D_{[2]} + \phi_2 D_{[2]}^2 + \psi_2\Pi_2 \right) =
D_{[2]}\,.
\ee
Now we can calculate another realization of the same unknot constructed from any number of lock vertices. The general expression for the unknot closure of a chain of $n$ horizontal AP locks can be written from Fig.\,\ref{fig:vertchain}\,:
\begin{equation}
    H_{[2]}^{\rm unknot}=(A^2q^2)^{-2n}\left(D_{[2]}+\Phi_n^{[2]}D_{[2]}^2+\Psi_n^{[2]}\Pi_2\right)=(A^2q^2)^{-2n}\cdot D_{[2]}\left(1+\phi_2D_{[2]} +\psi_2\,\frac{[N+1]}{[2]}\right)^n= D_{[2]}\,.
\end{equation}

\subsection{Antiparallel torus links from 2-strand braids}\label{sec:2links}

The answer for the HOMFLY polynomial for anti-parallel 2-strand links is already known from the celebrated Rosso--Jones formula~\cite{rosso1993invariants,lin2010hecke,tierz2004soft,brini2012torus}:
\be\label{AP(2,2n)-answer}
H_{[2]}^{APT[2,2n]} = 1 + \frac{1}{A^{2n}}\frac{\{Aq\}\{A/q\}}{\{q\}^2} + \frac{1}{(q^2A^2)^{2n}}\frac{\{Aq^3\}\{A\}^2\{A/q\}}{\{q\}^2\{q^2\}^2}\,.
\ee
The terms in the r.h.s. of this formula correspond to the irreducible representations from the tensor product $[2]\otimes \overline{[2]}=\varnothing \, \oplus \, {\rm adj}\, \oplus \, [4,2^{N-2}]$. The structure of planar calculus expansion is very different -- it is a cycle decomposition. For example, for the $APT[2,4]$ 2-component link, we have
\be
\begin{aligned}
    &H_{[2]}^{APT[2,4]} = 1 + \frac{1}{A^4}\frac{\{Aq\}\{A/q\}}{\{q\}^2} + \frac{1}{q^8A^8}\frac{\{Aq^3\}\{A\}^2\{A/q\}}{\{q\}^2\{q^2\}^2}= \\
&= \frac{1}{(A^4q^4)^2}\left(D_{[2]}^2 + 2\phi_2D_{[2]} + \phi_2^2D_{[2]}^2 + 2\psi_2(1+\phi_2)\cdot\Pi_2+ \psi_2^2\cdot\Pi_3
\right) = \\
&= \frac{1}{(A^4q^4)^2}\left(D_{[2]}^2 + 2\phi_2D_{[2]} + \phi_2^2D_{[2]}^2 + 2\psi_2(1+\phi_2)\frac{\{Aq\}^2\{A\}}{\{q^2\}^2\{q\}}
%+\right. \nn \\ \left.
+ \psi_2^2\, \frac{\{A\}\{Aq\}\Big(\{Aq^2\}\{Aq\} + \{q^2\}\{q\}\Big)}{\{q^2\}^4}
\right).
\end{aligned}
\ee
Note that in the classical case $\Pi_3=N^2+3N+4$. In general, the way to quantize a classical answer is not unique. However, in this special case we know from equation (59) of \cite{DM3} the right quantization rule $N^2+3N+4 \ \longrightarrow\ [N+1][N+2]+[2]$ which exactly leads to the quantum answer for $\Pi_3$. Hopefully, this new application to calculation of the colored HOMFLY polynomials for bipartite knots/links
and this new interpretation in terms of projectors will attract new attention
to the quantization puzzles raised in \cite{DM3,AnoM}.

It is tempting to calculate only classical answers and obtain quantum answers just by proper quantization rules. However, already in the $APT[2,6]$ case:
\be
\begin{aligned}
    &H_{[2]}^{APT[2,6]} = 1 + \frac{1}{A^6}\frac{\{Aq\}\{A/q\}}{\{q\}^2}
+ \frac{1}{q^{12}A^{12}}\frac{\{Aq^3\}\{A\}^2\{A/q\}}{\{q\}^2\{q^2\}^2}
= \\
&= \frac{1}{(A^4q^4)^3}\left(D_{[2]}^2 + 3\phi_2D_{[2]} + 3\phi_2^2D_{[2]}^2 +\phi_2^3D_{[2]}^3
+ 3\psi_2(2\phi_2+\phi_2^2D_{[2]})\Pi_2 + 3\psi_2^2\phi_2 \Pi_3
+ 3\psi_2 \Pi_2 + 3\psi_2^2\Pi_3 + \psi_2^3 \Pi_4\right).
\end{aligned}
\ee
In the classical case $\Pi_4=\frac{N(N+1)}{64}(N^3+5 N^2+14 N+12)$, and we do not know a proper way to quantize it. However, from our quantum planar technique, we can obtain the answer for $APT[2,2n]$ for generic $n$ (see Fig.\,\ref{fig:vertchain}):
\begin{equation}
\begin{aligned}
    &H_{[2]}^{APT[2,2n]} = \frac{1}{(A^2q^2)^{2n}}\cdot \Tr (1+ \phi_2\cdot \hat\Phi_2+\psi_2\cdot \hat \Psi_2)^n=\frac{1}{(A^2q^2)^{2n}}\left(D_{[2]}^2+\Phi_n^{[2]}D_{[2]}+\Psi_n^{[2]}\Pi_2\right)=\\
    &=(A^2q^2)^{-2n}\left\{\left(1+\phi_2D_{[2]} +\psi_2\,\frac{[N+1]}{[2]}\right)^n+[N+1][N-1]\left(1 +\psi_2\,\frac{[N+2]}{[2]^2}\right)^n+D_{[2]}^2-D_{[1]}^2\right\}
\end{aligned}
\end{equation}
which after the substitutions gets exactly the form~\eqref{AP(2,2n)-answer}.

\subsection{Twist knots -- the lock-closure of 2-strand braids}

A twist knot with $n$ full-twists is shown in Fig.\,\ref{fig:twist-knot} for $m=1$. The resolution of the chain part leads to two diagrams we have already calculated (see Fig.\,\ref{fig:singlelockclosures}), and the third diagram is the only one to be computed via the rule in Fig.\,\ref{fig:proje2lock}. In total, for the HOMFLY polynomial we get
\begin{equation}\label{H-twist}
\begin{aligned}
    H^{{\rm Tw}_{2n}}_{[2]}&=(A^2q^2)^{-2n-2}\left((D_{[2]}+\phi_2 D_{[2]}^2+\psi_2 \Pi_2)+\Phi_n^{[2]}\cdot (D_{[2]}^2+\phi_2 D_{[2]}+\psi_2 \Pi_2)+\Psi_n^{[2]}\cdot (\Pi_2(1+\phi_2)+\psi_2 \Tr \hat{\Pi}^2)\right)= \\
    &=\lambda_\varnothing^{2n} \cdot \tau_\varnothing + \lambda_{\rm adj}^{2n} \cdot \tau_{\rm adj} + \lambda_{[4,2^{N-2}]}^{2n} \cdot \tau_{[4,2^{N-2}]}
\end{aligned}
\end{equation}
with
\begin{equation}
\begin{aligned}
    \tau_{[4,2^{N-2}]} &= A^{-4} q^{-4}\left((D_{[2]}+\phi_2 D_{[2]}^2+\psi_2 \Pi_2) + \frac{[N]}{[N+2]}D_{[2]}^{-1}(D_{[2]}^2+\phi_2 D_{[2]}+\psi_2 \Pi_2)-\frac{[2]^2}{[N+2]}(\Pi_2(1+\phi_2)+\psi_2 \Tr \hat{\Pi}^2)\right)\\
    &=A^{-2}q^{-1}\frac{\{A\}^2\{A/q\}\{A q^3\}}{\{q^2\}\{q\}}\,, \\
    \tau_{\rm adj} &= A^{-4} q^{-4}\left(\frac{[2]^2}{[N+2]}\cdot (\Pi_2(1+\phi_2)+\psi_2 \Tr \hat{\Pi}^2)-\frac{[2][N+1]}{[N+2]}D_{[2]}^{-1}(D_{[2]}^2+\phi_2 D_{[2]}+\psi_2 \Pi_2)\right)= \\
    &=\frac{\{A q\}\{A/q\}}{\{q\}^2}(-A^{-4} q^{-4}+ A^{-2} q^{-4}+ q^2 A^{-2}-A^{-2} q^{-2})\,, \\
    \tau_\varnothing &= A^{-4} q^{-4} D_{[2]}^{-1}\cdot (D_{[2]}^2+\phi_2 D_{[2]}+\psi_2 \Pi_2)= \\
    &=\frac{A^4 q^{12}-A^4 q^{10}-A^4 q^8+2 A^4 q^6-A^4 q^2+A^4-A^2 q^8+A^2 q^4-A^2 q^2-A^2+q^2}{A^6 q^7 \{q^2\}\{q\}}\,,
\end{aligned}
\end{equation}
and the eigenvalues are given by~\eqref{eigen}.

\subsection{Double braid knots}\label{sec:DB-[2]}

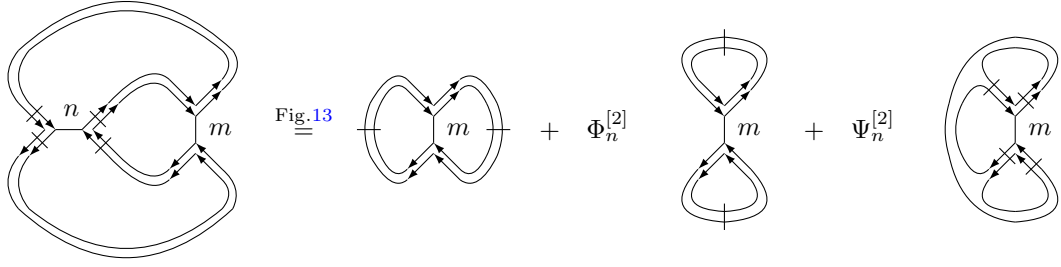
\begin{figure}[h!]
\begin{picture}(100,100)(-70,-50)

\put(-14,3){\line(1,1){6}}
\put(12,-9){\line(1,1){6}}
\put(16,1){\line(-1,1){6}}
\put(-6,-7){\line(-1,1){6}}

\qbezier(21,10)(33,22)(40,15)
\qbezier(17,10)(32,26)(40,19)

\qbezier(21,-10)(33,-22)(40,-15)
\qbezier(17,-10)(32,-26)(40,-19)

\qbezier(-17,10)(-25,20)(-17,30)
\qbezier(-13,10)(-21,20)(-13,30)

\qbezier(-17,-30)(-25,-20)(-17,-10)
\qbezier(-13,-30)(-21,-20)(-13,-10)

\qbezier(60,19)(65,25)(60,30)
\qbezier(60,15)(69,23)(64,30)

\qbezier(-13,30)(24,60)(60,30)
\qbezier(-17,30)(24,67)(64,30)

\qbezier(60,-30)(65,-24)(60,-19)
\qbezier(64,-30)(68,-22)(60,-15)

\qbezier(-13,-30)(24,-60)(60,-30)
\qbezier(-17,-30)(24,-67)(64,-30)

\put(-17,10){\vector(1,-1){10}}
\put(-13,10){\vector(1,-1){10}}
\put(11,0){\vector(1,1){10}}
\put(7,0){\vector(1,1){10}}
\put(21,-10){\vector(-1,1){10}}
\put(17,-10){\vector(-1,1){10}}
\put(-3,0){\vector(-1,-1){10}}
\put(-7,0){\vector(-1,-1){10}}

\put(-3,0){\line(1,0){10}}
\put(0,5){\mbox{$n$}}

\put(55,-2){\mbox{$m$}}

\put(50,-5){\line(0,1){10}}

\put(50,-5){\vector(-1,-1){10}}
\put(60,-15){\vector(-1,1){10}}

\put(60,-19){\vector(-1,1){10}}
\put(50,-9){\vector(-1,-1){10}}

\put(40,15){\vector(1,-1){10}}
\put(50,5){\vector(1,1){10}}

\put(40,19){\vector(1,-1){10}}
\put(50,9){\vector(1,1){10}}

\put(80,-3){\mbox{$\overset{\rm Fig.\ref{fig:vertchain}}{=}$}}

\put(90,0){

\put(55,-2){\mbox{$m$}}

\put(50,-5){\line(0,1){10}}

\put(21,0){\line(1,0){9}}
\put(70,0){\line(1,0){9}}

\put(50,-5){\vector(-1,-1){10}}
\put(60,-15){\vector(-1,1){10}}

\put(60,-19){\vector(-1,1){10}}
\put(50,-9){\vector(-1,-1){10}}

\put(40,15){\vector(1,-1){10}}
\put(50,5){\vector(1,1){10}}

\put(40,19){\vector(1,-1){10}}
\put(50,9){\vector(1,1){10}}

\qbezier(30,10)(35,20)(40,15)
\qbezier(26,10)(34,24)(40,19)

\qbezier(60,15)(65,20)(70,10)
\qbezier(60,19)(67,24)(74,10)

\qbezier(30,-10)(35,-20)(40,-15)
\qbezier(26,-10)(34,-24)(40,-19)

\qbezier(60,-15)(65,-20)(70,-10)
\qbezier(60,-19)(67,-24)(74,-10)

\qbezier(30,-10)(25,0)(30,10)
\qbezier(26,-10)(22,0)(26,10)

\qbezier(70,-10)(75,0)(70,10)
\qbezier(74,-10)(78,0)(74,10)

}

\put(180,-3){\mbox{$+ \ \ \ \Phi_n^{[2]}$}}

\put(200,0){

\put(55,-2){\mbox{$m$}}

\put(50,-5){\line(0,1){10}}

\put(50,28){\line(0,1){10}}
\put(50,-38){\line(0,1){10}}

\put(50,-5){\vector(-1,-1){10}}
\put(60,-15){\vector(-1,1){10}}

\put(60,-19){\vector(-1,1){10}}
\put(50,-9){\vector(-1,-1){10}}

\put(40,15){\vector(1,-1){10}}
\put(50,5){\vector(1,1){10}}

\put(40,19){\vector(1,-1){10}}
\put(50,9){\vector(1,1){10}}

\qbezier(40,19)(32,30)(50,31)
\qbezier(40,15)(24,33)(50,35)

\qbezier(60,19)(68,30)(50,31)
\qbezier(60,15)(76,33)(50,35)

\qbezier(40,-19)(32,-30)(50,-31)
\qbezier(40,-15)(24,-33)(50,-35)

\qbezier(60,-19)(68,-30)(50,-31)
\qbezier(60,-15)(76,-33)(50,-35)

}

\put(280,-3){\mbox{$+ \ \ \ \Psi_n^{[2]}$}}

\put(310,0){

\put(55,-2){\mbox{$m$}}

\put(50,-5){\line(0,1){10}}

\put(54,-17){\line(1,1){6}}
\put(38,13){\line(1,1){6}}

\put(44,-7){\line(1,-1){6}}
\put(51,14){\line(1,-1){6}}

\put(50,-5){\vector(-1,-1){10}}
\put(60,-15){\vector(-1,1){10}}

\put(60,-19){\vector(-1,1){10}}
\put(50,-9){\vector(-1,-1){10}}

\put(40,15){\vector(1,-1){10}}
\put(50,5){\vector(1,1){10}}

\put(40,19){\vector(1,-1){10}}
\put(50,9){\vector(1,1){10}}

\qbezier(40,19)(32,30)(50,31)
\qbezier(60,19)(68,30)(50,31)

\qbezier(40,-19)(32,-30)(50,-31)
\qbezier(60,-19)(68,-30)(50,-31)

\qbezier(30,-10)(25,0)(30,10)
\qbezier(25,-10)(22,0)(25,10)

\qbezier(30,-10)(35,-20)(40,-15)
\qbezier(30,10)(35,20)(40,15)

\qbezier(60,15)(76,33)(50,35)
\qbezier(60,-15)(76,-33)(50,-35)

\qbezier(50,35)(30,33)(25,10)
\qbezier(50,-35)(30,-33)(25,-10)

}

\end{picture}
    \caption{\footnotesize A double braid knot in our schematic notations and the resolution of the vertical chain inside it. For $m=1$ this double braid knot reduces to a twist knot.}
    \label{fig:twist-knot}
\end{figure}

\noindent A double braid knot is obtained from a twist knot by changing the vertical lock element to a horizontal chain with $m$ vertical locks, see Fig.\,\ref{fig:twist-knot}. Thus, the expression for the HOMFLY polynomial is obtained from one for a twist knot~\eqref{H-twist} by the change $\psi_2\rightarrow \Psi_m^{[2]}$, $\phi_2\rightarrow \Phi_m^{[2]}$ and the framing factor $(A^2q^2)^{-2n-2} \rightarrow (A^2q^2)^{-2n-2m}$:

{\footnotesize \begin{equation}\label{H-twist}
\begin{aligned}
    &H^{DB(2n,2m)}_{[2]}=\frac{1}{(A^2q^2)^{2n+2m}}\left((D_{[2]}+\Phi_m^{[2]} D_{[2]}^2+\Psi_m^{[2]} \Pi_2)+\Phi_n^{[2]} (D_{[2]}^2+\Phi_m^{[2]} D_{[2]}+\Psi_m^{[2]} \Pi_2)+\Psi_n^{[2]} (\Pi_2(1+\Phi_m^{[2]})+\Psi_m^{[2]} \Tr \hat{\Pi}^2)\right)\overset{\eqref{rel-(m+n)}}{=} \\
    &=\frac{1}{(A^2q^2)^{2n+2m}}\left(D_{[2]}+D_{[2]}\Phi_{n+m}^{[2]}+\Pi_2 \Psi_{n+m}^{[2]}+D_{[2]}(D_{[2]}-1)\left(\Phi_n^{[2]}+\Phi_m^{[2]}-\Phi_n^{[2]}\Phi_m^{[2]}\right)\right)= \\
    &\tau_{\varnothing}^\varnothing + \lambda_{\rm adj}^{2n+2m} \tau_{\rm adj}^{\rm adj} + \lambda_{[4,2^{N-2}]}^{2n+2m} \tau_{[4,2^{N-2}]}^{[4,2^{N-2}]}+(\lambda_{\rm adj}^{2m}+\lambda_{\rm adj}^{2n})\tau_{\rm adj}^{\varnothing}+(\lambda_{[4,2^{N-2}]}^{2m}+\lambda_{[4,2^{N-2}]}^{2n})\tau_{[4,2^{N-2}]}^{\varnothing}+(\lambda_{[4,2^{N-2}]}^{2m}\lambda_{\rm adj}^{2n}+\lambda_{\rm adj}^{2m}\lambda_{[4,2^{N-2}]}^{2n})\tau_{[4,2^{N-2}]}^{\rm adj}
\end{aligned}
\end{equation}}
where the eigenvalues are the same as in~\eqref{eigen} and
{\small \begin{equation}
\begin{aligned}
    &\tau_{[4,2^{N-2}]}^{[4,2^{N-2}]}=D_{[2]}+\frac{[N]}{[N+2]}-\Pi_2\frac{[2]^2}{[N+2]}+2(D_{[2]}-1)\frac{[N]}{[N+2]}-\frac{D_{[2]}-1}{D_{[2]}}\frac{[N]^2}{[N+2]^2}=D_{[2]}\frac{\{A q^3\}\{A\}\{A/q\}}{\{A q\}^2 \{A q^2\}}\,, \\
    &\tau_\varnothing^\varnothing=1-\frac{D_{[2]}-1}{D_{[2]}}=\frac{\{q^2\}\{q\}}{\{A\}\{A q\}}\,, \\
    &\tau_{\rm adj}^{\rm adj}=\Pi_2 \frac{[2]^2}{[N+2]}-\frac{[2][N+1]}{[N+2]}-\frac{D_{[2]}-1}{D_{[2]}}\left(\frac{[2][N+1]}{[N+2]}\right)^2=\frac{\{A/q\}\{A q\}}{\{A\}\{A q^2\}\{q\}^2}\left(A^2 q^2+A^{-2} q^{-2}-q^4-q^{-4}-q^2-q^{-2}+2\right)\,,\\
    &\tau_{[4,2^{N-2}]}^{\rm adj}=(D_{[2]}-1)\frac{[2][N+1]}{[N+2]}\left(D_{[2]}^{-1}\frac{[N]}{[N+2]}-1\right)=-[2]D_{[2]}\cdot \frac{\{A q^3\}\{A/q\}}{\{A q\}\{A q^2\}}\,, \\
    &\tau_{[4,2^{N-2}]}^{\varnothing}= (D_{[2]}-1)\left(1-D_{[2]}^{-1}\frac{[N]}{[N+2]}\right)=D_{[2]}\frac{\{A q^3\}\{A/q\}}{\{A q\}^2}\,, \\
    &\tau_{\rm adj}^{\varnothing}=\frac{D_{[2]}-1}{D_{[2]}}\cdot \frac{[2][N+1]}{[N+2]}=[2]\frac{\{A/q\}}{\{A\}}\,.
\end{aligned}
\end{equation}}

\setcounter{equation}{0}
\section{Representation $[3]$}\label{sec:rep-[3]}

\begin{figure}[H]
\begin{picture}(100,150)(20,-40)

\put(-40,70){
%\put(50,0){
%\put(-20,17){\line(1,-1){17}}\put(-20,17){\vector(1,-1){14}}   \put(3,0){\vector(1,1){17}}
%\put(-3,0){\vector(-1,-1){17}}   \put(20,-17){\line(-1,1){17}} \put(20,-17){\vector(-1,1){14}}
%\put(-3,0){\line(1,0){6}}
%\put(-23,17){\line(1,-1){17}}\put(-23,17){\vector(1,-1){14}}   \put(6,0){\vector(1,1){17}}
%\put(-6,0){\vector(-1,-1){17}}   \put(23,-17){\line(-1,1){17}} \put(23,-17){\vector(-1,1){14}}
%\put(-26,17){\line(1,-1){17}}\put(-26,17){\vector(1,-1){14}}   \put(9,0){\vector(1,1){17}}
%\put(-9,0){\vector(-1,-1){17}}   \put(26,-17){\line(-1,1){17}} \put(26,-17){\vector(-1,1){14}}
%\put(-20,4){\line(1,1){10}}  \put(-20,-4){\line(1,-1){10}}
%\put(20,4){\line(-1,1){10}}  \put(20,-4){\line(-1,-1){10}}
%}

\put(100,0){
\put(-17,21){\line(1,-1){17}}\put(-17,24){\line(1,-1){17}} \put(-17,27){\line(1,-1){17}}
\put(-17,21){\vector(1,-1){14}} \put(-17,24){\vector(1,-1){14}}   \put(-17,27){\vector(1,-1){14}}
\put(0,4){\vector(1,1){17}} \put(0,7){\vector(1,1){17}} \put(0,10){\vector(1,1){17}}
\put(0,-4){\vector(-1,-1){17}}   \put(17,-21){\line(-1,1){17}} \put(17,-21){\vector(-1,1){14}}
\put(0,-7){\vector(-1,-1){17}}   \put(17,-24){\line(-1,1){17}} \put(17,-24){\vector(-1,1){14}}
\put(0,-10){\vector(-1,-1){17}}   \put(17,-27){\line(-1,1){17}} \put(17,-27){\vector(-1,1){14}}
\put(0,4){\line(0,-1){8}}

\put(-15,11){\line(1,1){10}}  \put(-15,-11){\line(1,-1){10}}
\put(15,11){\line(-1,1){10}}  \put(15,-11){\line(-1,-1){10}}

\put(30,-2){\mbox{$:=$}}

\qbezier(50,20)(55,9)(58,4) \qbezier(63,-4)(85,-40)(110,20)
\put(56,8){\vector(1,-2){2}} \put(90,-13){\vector(1,1){2}} \put(109,18){\vector(1,2){2}}
\qbezier(50,-20)(75,40)(97,4)  \qbezier(102,-4)(105,-9)(110,-20)
\put(104,-8){\vector(-1,2){2}} \put(70,13){\vector(-1,-1){2}} \put(51,-18){\vector(-1,-2){2}}

\qbezier(50,25)(55,14)(58,9) \qbezier(63,1)(85,-35)(110,25)
\put(56,13){\vector(1,-2){2}} \put(90,-8){\vector(1,1){2}} \put(109,23){\vector(1,2){2}}
\qbezier(50,-15)(75,45)(97,9)  \qbezier(102,1)(105,-4)(110,-15)
\put(104,-3){\vector(-1,2){2}} \put(70,18){\vector(-1,-1){2}} \put(51,-13){\vector(-1,-2){2}}

\qbezier(50,30)(55,19)(58,14) \qbezier(63,6)(85,-30)(110,30)
\put(56,18){\vector(1,-2){2}} \put(90,-3){\vector(1,1){2}} \put(109,28){\vector(1,2){2}}
\qbezier(50,-10)(75,50)(97,14)  \qbezier(102,6)(105,1)(110,-10)
\put(104,2){\vector(-1,2){2}} \put(70,23){\vector(-1,-1){2}} \put(51,-8){\vector(-1,-2){2}}

\put(48,15){\line(2,1){10}}  \put(48,-10){\line(2,-1){10}}
\put(110,10){\line(-2,1){10}}  \put(112,-10){\line(-2,-1){10}}

\put(130,-2){\mbox{$=$}}

\put(100,65){
\put(70,-60){\vector(1,0){40}} \put(70,-57){\vector(1,0){40}}  \put(70,-54){\vector(1,0){40}}
\put(110,-70){\vector(-1,0){40}}  \put(110,-73){\vector(-1,0){40}} \put(110,-76){\vector(-1,0){40}}
\put(90,-52){\line(0,-1){10}}   \put(90,-68){\line(0,-1){10}}

\put(-70,0){
\put(190,-67){\mbox{$+\ \ \ \psi_3$}}
\put(230,-50){\vector(0,-1){30}}
\put(250,-80){\vector(0,1){30}}
\qbezier(233,-50)(240,-75)(247,-50)  \put(246,-52){\vector(1,2){2}}
\qbezier(236,-50)(240,-69)(244,-50)  \put(243,-52){\vector(1,2){2}}
\qbezier(236,-80)(240,-61)(244,-80)  \put(237,-78){\vector(-1,-2){2}}
\qbezier(233,-80)(240,-55)(247,-80)  \put(234,-78){\vector(-1,-2){2}}
\put(228,-56){\line(1,0){10}}   \put(242,-56){\line(1,0){10}}
\put(228,-74){\line(1,0){10}}   \put(242,-74){\line(1,0){10}}
}

\put(190,-67){\mbox{$+\ \ \ \chi_3$}}
\put(230,-50){\vector(0,-1){30}}  \put(227,-50){\vector(0,-1){30}}
\put(250,-80){\vector(0,1){30}}   \put(253,-80){\vector(0,1){30}}
\qbezier(233,-50)(240,-75)(247,-50)  \put(246,-52){\vector(1,2){2}}
\qbezier(233,-80)(240,-55)(247,-80)  \put(234,-78){\vector(-1,-2){2}}
\put(225,-56){\line(1,0){13}}   \put(242,-56){\line(1,0){13}}
\put(225,-74){\line(1,0){13}}   \put(242,-74){\line(1,0){13}}

\put(140,0){
\put(120,-67){\mbox{$+\ \ \ \phi_3$}}
\put(160,-50){\vector(0,-1){30}} \put(163,-50){\vector(0,-1){30}} \put(166,-50){\vector(0,-1){30}}
\put(173,-80){\vector(0,1){30}}  \put(176,-80){\vector(0,1){30}} \put(179,-80){\vector(0,1){30}}
\put(158,-65){\line(1,0){10}}   \put(171,-65){\line(1,0){10}}
}
}}
}

%----------------------------------------------------

\put(100,0){
%\put(-20,17){\line(1,-1){17}}\put(-20,17){\vector(1,-1){14}}   \put(3,0){\vector(1,1){17}}
%\put(-3,0){\vector(-1,-1){17}}   \put(20,-17){\line(-1,1){17}} \put(20,-17){\vector(-1,1){14}}
%\put(-3,0){\line(1,0){6}}
%\put(-23,17){\line(1,-1){17}}\put(-23,17){\vector(1,-1){14}}   \put(6,0){\vector(1,1){17}}
%\put(-6,0){\vector(-1,-1){17}}   \put(23,-17){\line(-1,1){17}} \put(23,-17){\vector(-1,1){14}}
%\put(-19,6){\line(1,1){10}}  \put(-19,-6){\line(1,-1){10}}
%\put(19,6){\line(-1,1){10}}  \put(19,-6){\line(-1,-1){10}}

\put(-17,21){\line(1,-1){17}}\put(-17,24){\line(1,-1){17}}   \put(0,4){\vector(1,1){17}}
\put(-17,21){\vector(1,-1){14}} \put(-17,24){\vector(1,-1){14}}
\put(0,7){\vector(1,1){17}}
\put(0,-4){\vector(-1,-1){17}}   \put(17,-21){\line(-1,1){17}} \put(17,-21){\vector(-1,1){14}}
\put(0,-7){\vector(-1,-1){17}}   \put(17,-24){\line(-1,1){17}} \put(17,-24){\vector(-1,1){14}}
\put(0,4){\line(0,-1){8}}

\put(-16,11){\line(1,1){10}}  \put(-16,-11){\line(1,-1){10}}
\put(16,11){\line(-1,1){10}}  \put(16,-11){\line(-1,-1){10}}

\put(30,-2){\mbox{$:=$}}

\qbezier(50,20)(55,9)(58,4) \qbezier(63,-4)(85,-40)(110,20)
\put(56,8){\vector(1,-2){2}} \put(90,-13){\vector(1,1){2}} \put(109,18){\vector(1,2){2}}
\qbezier(50,-20)(75,40)(97,4)  \qbezier(102,-4)(105,-9)(110,-20)
\put(104,-8){\vector(-1,2){2}} \put(70,13){\vector(-1,-1){2}} \put(51,-18){\vector(-1,-2){2}}

\qbezier(50,25)(55,14)(58,9) \qbezier(63,1)(85,-35)(110,25)
\put(56,13){\vector(1,-2){2}} \put(90,-8){\vector(1,1){2}} \put(109,23){\vector(1,2){2}}
\qbezier(50,-15)(75,45)(97,9)  \qbezier(102,1)(105,-4)(110,-15)
\put(104,-3){\vector(-1,2){2}} \put(70,18){\vector(-1,-1){2}} \put(51,-13){\vector(-1,-2){2}}

\put(48,15){\line(2,1){10}}  \put(48,-10){\line(2,-1){10}}
\put(110,10){\line(-2,1){10}}  \put(112,-10){\line(-2,-1){10}}

\put(130,-2){\mbox{$=$}}

\put(100,65){
\put(70,-60){\vector(1,0){40}} \put(70,-57){\vector(1,0){40}}
\put(110,-70){\vector(-1,0){40}}  \put(110,-67){\vector(-1,0){40}}
\put(90,-55){\line(0,-1){7}}   \put(90,-65){\line(0,-1){7}}

\put(-70,0){
\put(190,-67){\mbox{$+\ \ \ \psi_2$}}
\put(230,-50){\vector(0,-1){30}}
\put(250,-80){\vector(0,1){30}}
\qbezier(233,-50)(240,-75)(247,-50)  \put(246,-52){\vector(1,2){2}}
\qbezier(233,-80)(240,-55)(247,-80)  \put(234,-78){\vector(-1,-2){2}}
\put(228,-56){\line(1,0){10}}   \put(242,-56){\line(1,0){10}}
\put(228,-74){\line(1,0){10}}   \put(242,-74){\line(1,0){10}}
}

\put(70,0){
\put(120,-67){\mbox{$+\ \ \ \phi_2$}}
\put(160,-50){\vector(0,-1){30}} \put(163,-50){\vector(0,-1){30}}
\put(170,-80){\vector(0,1){30}}  \put(173,-80){\vector(0,1){30}}
\put(158,-65){\line(1,0){7}}   \put(168,-65){\line(1,0){7}}
}
}}

\end{picture}
\caption{\footnotesize  The lock tangle projected to the symmetric representation $[3]$
and its planar decomposition.
It is a direct generalization of Fig.\,\ref{fig:proje2lock}, which is reproduced/repeated in the second line.
Together, they make apparent the structure of planar decomposition for an arbitrary representation $R=[r]$.
For an antisymmetric representation $R=[1^r]$, one can consider the same picture but take $q$ inverted in formulas
for the coefficients and contractions of projectors.
For more sophisticated representations, even rectangular,
existence of a planar decomposition is still a question.
} \label{fig:proje3lock}
\end{figure}
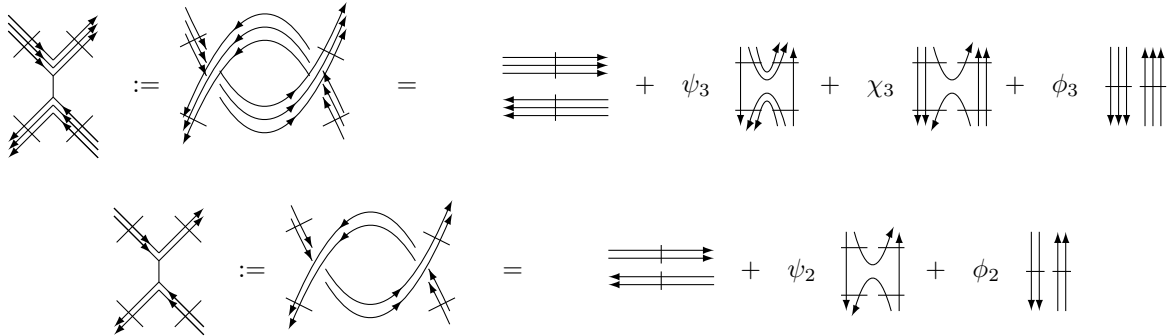

In this section, we develop the planar decomposition method for the HOMFLY polynomial for bipartite links in the representation $[3]$. Now, we consider $3$-cabled lock vertex with each strand carrying the fundamental representation. We then take the projection on the representation $[3]$ in order to obtain the corresponding colored HOMFLY polynomials. Thus, only four planar diagrams remain in the decomposition of the lock tangle, see Fig.\,\ref{fig:proje3lock}.

We do not need to repeat for the representation $[3]$ the introductory pieces of the previous Section~\ref{sec:rep-[2]},
and proceed directly to the analogue of Section~\ref{sec:proj-prop-[2]}.

\subsection{Projector calculus}\label{sec:proj-calc-[3]}

In this section we derive properties of the quantum projector on the representation $[3]$ which are in parallel with those ones of Section~\ref{sec:proj-prop-[2]}. From (4.16) of \cite{AnoAnd} we get the expression for the quantum projector
through two $\cal R$-matrices ${\cal R}_1={\cal R}\otimes \mathds{1}$ and ${\cal R}_2=\mathds{1}\otimes {\cal R}\,$:
\be\label{P[3]}
P_{[3]} = \frac{1+q {\cal R}_1+q {\cal R}_2+q^2 {\cal R}_1 {\cal R}_2+q^2 {\cal R}_2 {\cal R}_1 + q^3 {\cal R}_1 {\cal R}_2 {\cal R}_1}{q^3[2][3]}\,.
\ee
Even in two-strand braids examples in the case of representation $[3]$, there are many more different contractions of projectors than in the case of the first symmetric representation. Thus, it is more relevant not to calculate these combinations of projectors separately but use rules which reduce them recurrently to the simplest ones. The central identity is shown in Fig.\,\ref{fig:proj-prop-2} and can be derived by a straightforward contraction of two projectors~\eqref{P[3]} and the turning point operator~\eqref{M-op} at the l.h.s. 

\begin{figure}[h!]
\begin{picture}(300,60)(-120,-10)

\put(-26,17){\mbox{$[3]^2$}}

\put(0,0){\line(0,1){40}}
\put(3,0){\line(0,1){40}}
\put(20,20){\circle{25}}
\put(37.5,0){\line(0,1){40}}
\put(40.5,0){\line(0,1){40}}
\put(-4,20){\line(1,0){16}}
\put(28,20){\line(1,0){16}}

\put(55,17){\mbox{$= \ \ \, [2]^2$}}

\put(5,0){
\put(96,0){\line(0,1){40}}
\qbezier(100,40)(110,10)(120,40)
\put(124,0){\line(0,1){40}}
\qbezier(100,0)(110,30)(120,0)
\put(93,35){\line(1,0){13}}
\put(114,35){\line(1,0){13}}
\put(93,5){\line(1,0){13}}
\put(114,5){\line(1,0){13}}
}

\put(140,17){\mbox{$+ \ \ \, [N+4]$}}

\put(202,0){\line(0,1){40}}
\put(205,0){\line(0,1){40}}
\put(198,20){\line(1,0){11}}

\put(28,0){
\put(202,0){\line(0,1){40}}
\put(205,0){\line(0,1){40}}
\put(198,20){\line(1,0){11}}
}
    
\end{picture}
\caption{\footnotesize Erasure of a single circle -- as implied by projector properties. Note that the l.h.s contains projectors on the representation $[3]$, while there are only projectors on the representation $[2]$ at the r.h.s.}\label{fig:proj-prop-2}
\end{figure}

\begin{figure}[h!]
\begin{picture}(300,90)(-75,-85)

    \put(9,-16){\mbox{any diagram}}
    
    \put(0,0){\line(1,0){70}}
    \put(0,0){\line(0,-1){30}}
    \put(70,0){\line(0,-1){30}}
    \put(0,-30){\line(1,0){70}}

    \qbezier(70,0)(110,0)(110,-15)
    \qbezier(70,-3)(107,-3)(107,-15)
    \qbezier(70,-30)(110,-30)(110,-15)
    \qbezier(70,-27)(107,-27)(107,-15)
    \put(90,-15){\circle{16}}

    \put(90,-12){\line(0,1){15}}
    \put(90,-33){\line(0,1){15}}

    \put(125,-18){\mbox{$= \ \ \ \cfrac{[N+2]}{[3]}$}}

    \put(200,0){

    \put(9,-16){\mbox{any diagram}}

    \put(0,0){\line(1,0){70}}
    \put(0,0){\line(0,-1){30}}
    \put(70,0){\line(0,-1){30}}
    \put(0,-30){\line(1,0){70}}

    \qbezier(70,0)(110,0)(110,-15)
    \qbezier(70,-3)(107,-3)(107,-15)
    \qbezier(70,-30)(110,-30)(110,-15)
    \qbezier(70,-27)(107,-27)(107,-15)

    \put(90,-8){\line(0,1){10}}
    \put(90,-32){\line(0,1){10}}
    
    }

    \put(0,-50){

    \put(9,-16){\mbox{any diagram}}
    
    \put(0,0){\line(1,0){70}}
    \put(0,0){\line(0,-1){30}}
    \put(70,0){\line(0,-1){30}}
    \put(0,-30){\line(1,0){70}}

    \qbezier(70,0)(110,0)(110,-15)
    \qbezier(70,-30)(110,-30)(110,-15)
    \put(90,-15){\circle{16}}
    \put(90,-15){\circle{22}}

    \put(90,-12){\line(0,1){15}}
    \put(90,-33){\line(0,1){15}}

    \put(125,-18){\mbox{$= \ \ \ \cfrac{[N+2]}{[3]}\,\cdot \cfrac{[N+1]}{[2]}$}}
    
    }

    \put(240,-50){

    \put(9,-16){\mbox{any diagram}}

    \put(0,0){\line(1,0){70}}
    \put(0,0){\line(0,-1){30}}
    \put(70,0){\line(0,-1){30}}
    \put(0,-30){\line(1,0){70}}

    \qbezier(70,0)(110,0)(110,-15)
    \qbezier(70,-30)(110,-30)(110,-15)
    
    }
    
\end{picture}
    \caption{\footnotesize Erasure of cycles at a closed end of a diagram.}
    \label{fig:red-cycle-[3]}
\end{figure}
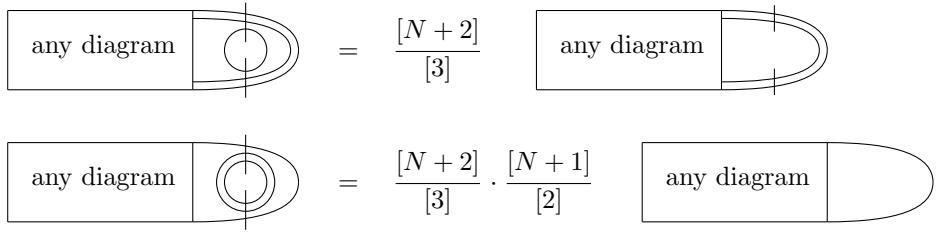

Next widely used in the subsequent examples identities are erasures of cycles at a closed end of a diagram, see Fig.\,\ref{fig:red-cycle-[3]}. These reductions are simple consequences of the rule in Fig.\,\ref{fig:proj-prop-2}.

\noindent For example, within the reduction rules from Fig.\,\ref{fig:red-cycle-[3]}, it is easy to calculate the quantities $K_3^{(1)}$, $K_3^{(2)}$ from Fig.\,\ref{fig:singlelockclosures3}:
\begin{equation}
\begin{aligned}
    K_2^{(1)} &:= \Pi_2 = \frac{[N][N+1]^2}{[2]^2}\,, \\
K_3^{(1)} &= \frac{[N][N+1][N+2]^2}{[2][3]^2}\,, \ \ \ \ \  K_3^{(2)} &= \frac{[N][N+1]^2[N+2]^2}{[2]^2[3]^2}\,,
\end{aligned}
\end{equation}
where we have also added the quantity $K_2^{(1)}$ appearing in computations of the bipartite HOMFLY polynomial for the representation $[2]$, see Section~\ref{sec:rep-[2]}.

\subsection{Warm-up examples}

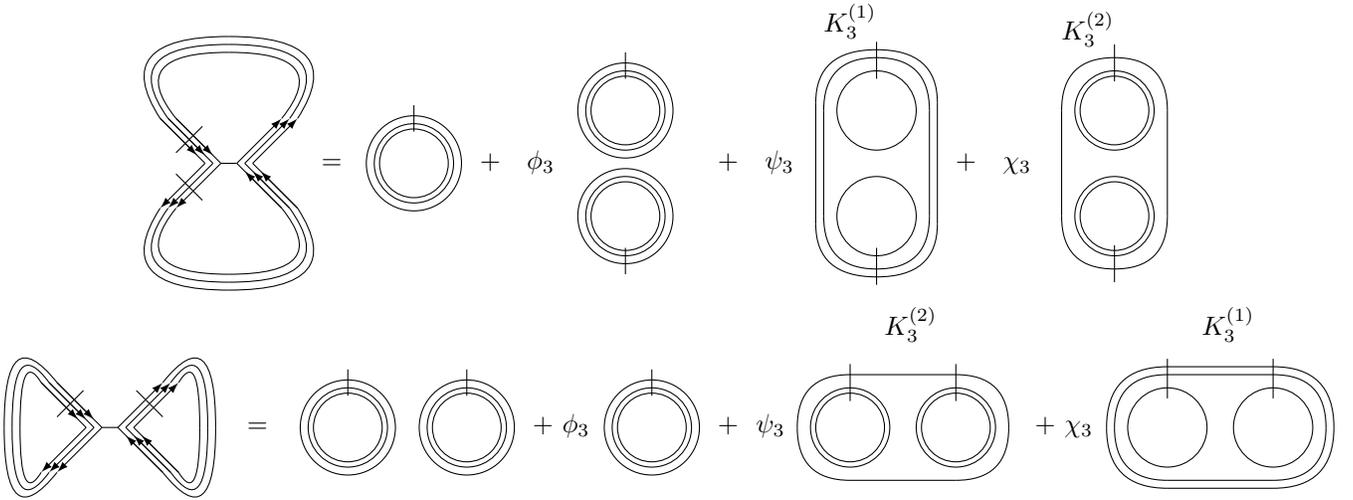
\begin{figure}[h!]
\begin{picture}(100,200)(32,-140)

\put(100,0){

\put(15,0){
\put(-20,17){\line(1,-1){17}}\put(-20,17){\vector(1,-1){14}}   \put(3,0){\vector(1,1){17}}
\put(-3,0){\vector(-1,-1){17}}   \put(20,-17){\line(-1,1){17}} \put(20,-17){\vector(-1,1){14}}
\put(-3,0){\line(1,0){6}}
\put(-23,17){\line(1,-1){17}}\put(-23,17){\vector(1,-1){14}}   \put(6,0){\vector(1,1){17}}
\put(-6,0){\vector(-1,-1){17}}   \put(23,-17){\line(-1,1){17}} \put(23,-17){\vector(-1,1){14}}
\put(-26,17){\line(1,-1){17}}\put(-26,17){\vector(1,-1){14}}   \put(9,0){\vector(1,1){17}}
\put(-9,0){\vector(-1,-1){17}}   \put(26,-17){\line(-1,1){17}} \put(26,-17){\vector(-1,1){14}}
\put(-20,4){\line(1,1){10}}  \put(-20,-4){\line(1,-1){10}}
%\put(19,6){\line(-1,1){10}}  \put(19,-6){\line(-1,-1){10}}
\qbezier(-20,17)(-40,42)(0,42) \qbezier(20,17)(40,42)(0,42)
\qbezier(-20,-17)(-40,-42)(0,-42) \qbezier(20,-17)(40,-42)(0,-42)
\qbezier(-23,17)(-43,45)(0,45) \qbezier(23,17)(43,45)(0,45)
\qbezier(-23,-17)(-43,-45)(0,-45) \qbezier(23,-17)(43,-45)(0,-45)
\qbezier(-26,17)(-46,48)(0,48) \qbezier(26,17)(46,48)(0,48)
\qbezier(-26,-17)(-46,-48)(0,-48) \qbezier(26,-17)(46,-48)(0,-48)
}

\put(50,-2){\mbox{$=$}}

\put(85,0){\circle{36}} \put(85,0){\circle{30}} \put(85,0){\circle{26}}
\put(85,12){\line(0,1){10}}

\put(110,-2){\mbox{$+ \ \ \ \phi_3$}}

\put(165,20){\circle{36}} \put(165,20){\circle{30}} \put(165,20){\circle{26}}
\put(165,-20){\circle{36}} \put(165,-20){\circle{30}}\put(165,-20){\circle{26}}
\put(165,32){\line(0,1){10}} \put(165,-32){\line(0,-1){10}}

\put(-30,0){

\put(230,-2){\mbox{$+ \ \ \ \psi_3$}}

\put(-50,0){
\put(340,20){\circle{30}} \put(340,-20){\circle{30}}
\qbezier(320,20)(320,40)(340,40) \qbezier(360,20)(360,40)(340,40)
\qbezier(320,-20)(320,-40)(340,-40) \qbezier(360,-20)(360,-40)(340,-40)
\put(320,20){\line(0,-1){40}}  \put(360,-20){\line(0,1){40}}
\qbezier(317,23)(317,43)(340,43) \qbezier(363,23)(363,43)(340,43)
\qbezier(317,-23)(317,-43)(340,-43) \qbezier(363,-23)(363,-43)(340,-43)
\put(317,23){\line(0,-1){46}}  \put(363,-23){\line(0,1){46}}

\put(320,50){\mbox{$K^{(1)}_3$}}

}

\put(320,-2){\mbox{$+ \ \ \ \chi_3$}}

\put(40,0){
\put(340,20){\circle{30}} \put(340,20){\circle{26}} \put(340,-20){\circle{30}} \put(340,-20){\circle{26}}
\qbezier(320,20)(320,40)(340,40) \qbezier(360,20)(360,40)(340,40)
\qbezier(320,-20)(320,-40)(340,-40) \qbezier(360,-20)(360,-40)(340,-40)
\put(320,20){\line(0,-1){40}}  \put(360,-20){\line(0,1){40}}

\put(320,47){\mbox{$K^{(2)}_3$}}
}

\put(290,32){\line(0,1){14}} \put(290,-32){\line(0,-1){14}}
\put(380,31){\line(0,1){14}} \put(380,-31){\line(0,-1){14}}
}
}

%-----------------------------------

\put(70,-100){

\put(-20,17){\line(1,-1){17}}\put(-20,17){\vector(1,-1){14}}   \put(3,0){\vector(1,1){17}}
\put(-3,0){\vector(-1,-1){17}}   \put(20,-17){\line(-1,1){17}} \put(20,-17){\vector(-1,1){14}}
\put(-3,0){\line(1,0){6}}
\put(-23,17){\line(1,-1){17}}\put(-23,17){\vector(1,-1){14}}   \put(6,0){\vector(1,1){17}}
\put(-6,0){\vector(-1,-1){17}}   \put(23,-17){\line(-1,1){17}} \put(23,-17){\vector(-1,1){14}}
\put(-26,17){\line(1,-1){17}}\put(-26,17){\vector(1,-1){14}}   \put(9,0){\vector(1,1){17}}
\put(-9,0){\vector(-1,-1){17}}   \put(26,-17){\line(-1,1){17}} \put(26,-17){\vector(-1,1){14}}

\put(-20,4){\line(1,1){10}}  %\put(-19,-6){\line(1,-1){10}}
\put(20,4){\line(-1,1){10}}  %\put(19,-6){\line(-1,-1){10}}

\qbezier(-20,17)(-40,42)(-40,0) \qbezier(-20,-17)(-40,-42)(-40,0)
\qbezier(-23,17)(-37,36)(-37,0) \qbezier(-23,-17)(-37,-36)(-37,0)
\qbezier(-26,17)(-34,30)(-34,0) \qbezier(-26,-17)(-34,-30)(-34,0)
\qbezier(20,17)(40,42)(40,0) \qbezier(20,-17)(40,-42)(40,0)
\qbezier(23,17)(37,36)(37,0) \qbezier(23,-17)(37,-36)(37,0)
\qbezier(26,17)(34,30)(34,0) \qbezier(26,-17)(34,-30)(34,0)

\put(52,-2){\mbox{$=$}}

\put(90,0){\circle{36}} \put(90,0){\circle{30}} \put(90,0){\circle{26}}
\put(135,0){\circle{36}} \put(135,0){\circle{30}} \put(135,0){\circle{26}}
\put(90,12){\line(0,1){10}} \put(135,12){\line(0,1){10}}

\put(160,-2){\mbox{$+ \  \phi_3$}}

\put(205,0){\circle{36}} \put(205,0){\circle{30}} \put(205,0){\circle{26}}
\put(205,12){\line(0,1){10}}

\put(-10,0){
\put(240,-2){\mbox{$+ \ \  \psi_3$}}

\put(-50,0){
\put(340,0){\circle{30}} \put(340,0){\circle{26}} \put(380,0){\circle{30}} \put(380,0){\circle{26}}
\qbezier(320,0)(320,20)(340,20) \qbezier(320,0)(320,-20)(340,-20)
\qbezier(400,0)(400,20)(380,20) \qbezier(400,0)(400,-20)(380,-20)
\put(340,20){\line(1,0){40}}  \put(340,-20){\line(1,0){40}}

\put(353,35){\mbox{$K^{(2)}_3$}}
}

\put(70,0){
\put(290,-2){\mbox{$+ \   \chi_3$}}

\put(340,0){\circle{30}} \put(380,0){\circle{30}}
\qbezier(320,0)(320,20)(340,20) \qbezier(320,0)(320,-20)(340,-20)
\qbezier(400,0)(400,20)(380,20) \qbezier(400,0)(400,-20)(380,-20)
\put(340,20){\line(1,0){40}}  \put(340,-20){\line(1,0){40}}
\qbezier(317,0)(317,23)(340,23) \qbezier(317,0)(317,-23)(340,-23)
\qbezier(403,0)(403,23)(380,23) \qbezier(403,0)(403,-23)(380,-23)
\put(340,23){\line(1,0){40}}  \put(340,-23){\line(1,0){40}}

\put(353,35){\mbox{$K^{(1)}_3$}}
}
\put(290,10){\line(0,1){14}} \put(330,10){\line(0,1){14}}
\put(410,11){\line(0,1){15}} \put(450,11){\line(0,1){15}}
}
}

\end{picture}
\caption{\footnotesize  The two closures of a single lock: the unknot in the first line
and the Hopf link in the second line.
Now in representation $R=[3]$, transverse lines denote projectors $P_{[3]}$.
} \label{fig:singlelockclosures3}
\end{figure}

\noindent As usual, we begin from the unknot and the Hopf link obtained by respectively the horizontal and vertical closures
of the horizontal lock, see Fig.\,\ref{fig:singlelockclosures3}:
\begin{equation}\label{unknot-[3]}
\begin{aligned}
    H^{\rm unknot}_{[1]} &=
D_{[1]} = \frac{1}{A^2} \left(D_{[1]} + \phi_1 D_{[1]}^2   \right),
\\
H^{\rm unknot}_{[2]} &=
D_{[2]} = \frac{1}{q^4A^4} \left(D_{[2]} + \phi_2 D_{[2]}^2 + \psi_2K_2^{(1)} \right),
\\
H^{\rm unknot}_{[3]} &=
D_{[3]} = \frac{1}{q^{12}A^6} \left(D_{[3]} + \phi_3 D_{[3]}^2 + \psi_3K^{(1)}_3 + \chi_3K^{(2)}_3 \right),
\end{aligned}    
\end{equation}
and
\begin{equation}\label{unknotHopf3}
\begin{aligned}
    H^{\rm Hopf}_{[1]} &= \frac{1}{q^2A^2}\left(q^{4} D_{[2]}   + D_{[1,1]}\right)
=  1 + \frac{1}{A^2}\frac{\{Aq\}\{A/q\}}{\{q\}^2}
=  \frac{1}{A^2} \left(D_{[1]}^2 + \phi_1 D_{[1]} \right),
\\
H^{\rm Hopf}_{[2]}&=\frac{1}{q^8A^4}\left(q^{12} D_{[4]} + q^4 D_{[3,1]} + D_{[2,2]}\right) =  \\
&=  1 + \frac{1}{A^2}\frac{\{Aq\}\{A/q\}}{\{q\}^2} + \frac{1}{q^4A^4}\frac{\{Aq^3\}\{A\}^2\{A/q\}}{\{q\}^2\{q^2\}^2}
=  \frac{1}{q^4A^4} \left(D_{[2]}^2 + \phi_2 D_{[2]} + \psi_2K_2^{(1)} \right),
 \\
H^{\rm Hopf}_{[3]}&=\frac{1}{q^{18}A^6}\left(q^{24} D_{[6]} + q^{12} D_{[5,1]} + q^{4}D_{[4,2]} +  D_{[3,3]}\right) = \\
&=  1 + \frac{1}{A^2}\frac{\{Aq\}\{A/q\}}{\{q\}^2} + \frac{1}{q^4A^4}\frac{\{Aq^3\}\{A\}^2\{A/q\}}{\{q\}^2\{q^2\}^2}
+  \frac{1}{q^{12}A^6}\frac{\{Aq^5\}\{Aq\}^2\{A\}^2\{A/q\}}{\{q\}^2\{q^2\}^2\{q^3\}^2}
= \\
&=  \frac{1}{q^{12}A^6} \left(D_3^2 + \phi_3 D_3 + \psi_3 K_3^{(2)}+\chi_3 K^{(1)}_3\right).
\end{aligned}
\end{equation}
These formulas are in parallel with (\ref{unknotHopf2}) for representation $R=[2]$. For the sake of convenience we list here all the quantities, including those ones for lower symmetric representations. The decomposition coefficients are
\be
\phi_1 = A \{q\}\,, \nn \\
\phi_2 = A^2 q^3[2] \{q\}^2, \ \ \ \ \ \ \psi_2 = Aq^2[2]^2\{q\}\,, \nn \\
\phi_3 =A^3 q^9[2][3] \{q\}^3 , \ \ \ \ \chi_3=A^2q^7[2][3]^2\{q\}^2, \ \ \ \ \psi_3 =  Aq^4[3]^2 \{q\}\,.
\label{phi3psi3chi3}
\ee
For the mirror lock tangle
\be
\bar\phi_1 = -\frac{1}{A} \{q\}\,, \nn \\
\bar\phi_2 = \frac{1}{A^2q^3}[2] \{q\}^2, \ \ \ \ \ \ \bar\psi_2 = -\frac{1}{Aq^2}[2]\{q^2\}\,, \nn \\
\bar\phi_3 = -\frac{1}{A^3q^9}[2][3] \{q\}^3,
\ \ \ \ \bar\chi_3 =\frac{1}{A^2q^7}[2][3]^2\{q\}^2, \ \ \ \ \bar\psi_3 = -\frac{1}{A q^4 }[3]^2 \{q\}\,.
\label{oppophi3psi3chi3}
\ee
%\be
%K_3^{(1)} = \frac{[N][N+1][N+2]^2}{[2][3]^2}, \ \ \ \ \  K_3^{(2)} = \frac{[N][N+1]^2[N+3]^2}{[[2]^2[3]^2}
%\ee
For antisymmetric representations one should invert everywhere just $q$, leaving $A$ intact.

\subsection{Chain of antiparallel lock tangles}\label{sec:chain-[3]}

\begin{figure}[h!]

\begin{picture}(100,190)(-320,-165)

\put(-295,0){

\put(0,0){

\put(-22,5){\line(1,1){10}}
\put(13,-16){\line(1,1){10}}
\put(10,13){\line(1,-1){10}}
\put(-20,-3){\line(1,-1){10}}

\put(-28,17){\line(1,-1){17}}
\put(-28,17){\vector(1,-1){14}}
\put(28,-17){\line(-1,1){17}}
\put(28,-17){\vector(-1,1){14}}
\put(11,0){\vector(1,1){17}}
 \put(-11,0){\vector(-1,-1){17}}

\put(-24,17){\line(1,-1){17}}
\put(-24,17){\vector(1,-1){14}}
\put(24,-17){\line(-1,1){17}}
\put(24,-17){\vector(-1,1){14}}
\put(7,0){\vector(1,1){17}}
 \put(-7,0){\vector(-1,-1){17}}

 \put(-20,17){\line(1,-1){17}}
 \put(-20,17){\vector(1,-1){14}}   
 \put(3,0){\vector(1,1){17}}
 \put(-3,0){\vector(-1,-1){17}}   
 \put(20,-17){\line(-1,1){17}} 
 \put(20,-17){\vector(-1,1){14}}
 \put(-3,0){\line(1,0){6}}
 }

\put(0,0){
\put(32,-2){\mbox{$=$}}

\put(0,65){

\put(-15,0){
\put(77,-65){\line(1,0){12}}
\put(61,-65){\line(1,0){12}}

\put(70,-50){\vector(0,-1){30}}
\put(67,-50){\vector(0,-1){30}}
\put(64,-50){\vector(0,-1){30}}

\put(80,-80){\vector(0,1){30}}
\put(83,-80){\vector(0,1){30}}
\put(86,-80){\vector(0,1){30}}
}

\put(82,-67){\mbox{$+\ \ \phi_3$}}

\put(-20,0){
\put(155,-63){\line(0,1){12}}
\put(155,-79){\line(0,1){12}}

\put(135,-60){\vector(1,0){40}}
\put(135,-57){\vector(1,0){40}}
\put(135,-54){\vector(1,0){40}}

\put(175,-70){\vector(-1,0){40}}
\put(175,-73){\vector(-1,0){40}}
\put(175,-76){\vector(-1,0){40}}
}

\put(162,-67){\mbox{$+\ \ \psi_3$}}

\put(-30,0){
\put(225,-50){\vector(1,0){40}}
\put(265,-75){\vector(-1,0){40}}

\qbezier(225,-56)(254,-63)(225,-69)
\qbezier(265,-56)(236,-63)(265,-69)
\put(230,-68){\vector(-1,-0.25){6}}
\put(260,-57){\vector(1,0.25){6}}

\qbezier(225,-53)(260,-63)(225,-72)
\qbezier(265,-53)(230,-63)(265,-72)
\put(230,-70.5){\vector(-1,-0.3){6}}
\put(260,-54.5){\vector(1,0.3){6}}

\put(256,-61){\line(0,1){15}}
\put(256,-79){\line(0,1){15}}
\put(232,-61){\line(0,1){15}}
\put(232,-79){\line(0,1){15}}
}

\put(242,-67){\mbox{$+\ \ \chi_3$}}

\put(50,0){
\put(225,-50){\vector(1,0){40}}
\put(225,-53){\vector(1,0){40}}
\put(265,-75){\vector(-1,0){40}}
\put(265,-72){\vector(-1,0){40}}

\qbezier(225,-56)(254,-63)(225,-69)
\qbezier(265,-56)(236,-63)(265,-69)
\put(230,-68){\vector(-1,-0.25){6}}
\put(260,-57){\vector(1,0.27){6}}

\put(256,-61){\line(0,1){15}}
\put(256,-79){\line(0,1){15}}
\put(232,-61){\line(0,1){15}}
\put(232,-79){\line(0,1){15}}
}

\put(322,-67){\mbox{$ = \, \hat{1} \, + \, \phi_3 \cdot \hat{\Phi}_3 \, + \, \psi_3 \cdot \hat{\Psi}_3 \, + \, \chi_3 \cdot \hat{\cal X}_3$}}

}
}
}

%--------------------------------------
%vertical chain operator

\put(-225,-150){

\put(-30,38){\mbox{$:=$}}

\put(-70,40){

\put(-18,3){\line(1,1){9}}
\put(13,-12){\line(1,1){9}}
\put(20,1){\line(-1,1){9}}
\put(-7,-10){\line(-1,1){9}}

\put(-23,12){\vector(1,-1){12.5}}
\put(-19,12){\vector(1,-1){12.5}}
\put(-15,12){\vector(1,-1){12.5}}

\put(15,0){\vector(1,1){12.5}}
\put(11,0){\vector(1,1){12.5}}
\put(7,0){\vector(1,1){12.5}}

\put(27,-12){\vector(-1,1){12.5}}
\put(23,-12){\vector(-1,1){12.5}}
\put(19,-12){\vector(-1,1){12.5}}

\put(-3,0){\vector(-1,-1){12.5}}
\put(-7,0){\vector(-1,-1){12.5}}
\put(-11,0){\vector(-1,-1){12.5}}

\put(-3,0){\line(1,0){10}}
\put(0,5){\mbox{$n$}}
}

\put(15,0){
\put(0,100){

\put(-18,3){\line(1.5,1){10}}
\put(12,-9){\line(1.5,1){10}}
\put(19,0){\line(-1.5,1){10}}
\put(-7,-7){\line(-1.5,1){10}}

\qbezier(-3,0)(-13,-7)(-12,-10)
\qbezier(-7,0)(-17,-7)(-16,-10)
\qbezier(-11,0)(-21,-7)(-20,-10)

\put(-20,-10){\vector(0,-1){2}}
\put(-16,-10){\vector(0,-1){2}}
\put(-12,-10){\vector(0,-1){2}}

\qbezier(15,0)(25,-7)(23,-10)
\qbezier(11,0)(21,-7)(19,-10)
\qbezier(7,0)(17,-7)(15,-10)

\put(17,-2){\vector(-1,1){2}}
\put(13,-2){\vector(-1,1){2}}
\put(9,-2){\vector(-1,1){2}}

\put(-3,0){\line(1,0){10}}

\put(-13,10){\vector(1,-1){10}}
\put(-17,10){\vector(1,-1){10}}
\put(-21,10){\vector(1,-1){10}}

\put(7,0){\vector(1,1){10}}
\put(11,0){\vector(1,1){10}}
\put(15,0){\vector(1,1){10}}
}

\put(-4,80){\mbox{$\ldots$}}

\put(0,60){

\put(-18.5,3){\line(1.5,1){10}}
\put(19.5,0){\line(-1.5,1){10}}

\put(-22,-10){\line(1,0){13}}
\put(11,-10){\line(1,0){13}}

\qbezier(-11,0)(-21,7)(-20,10)
\qbezier(-7,0)(-17,7)(-16,10)
\qbezier(-3,0)(-13,7)(-12,10)

\put(-14,2){\vector(1,-1){2}}
\put(-10,2){\vector(1,-1){2}}
\put(-6,2){\vector(1,-1){2}}

\qbezier(15,0)(25,7)(23,10)
\qbezier(11,0)(21,7)(19,10)
\qbezier(7,0)(17,7)(15,10)

\put(23,9){\vector(0,1){2}}
\put(19,9){\vector(0,1){2}}
\put(15,9){\vector(0,1){2}}

\put(-3,0){\line(1,0){10}}
}

\put(0,40){

\qbezier(-11,0)(-28,10)(-11,20)
\qbezier(-7,0)(-24,10)(-7,20)
\qbezier(-3,0)(-20,10)(-3,20)

\put(-14,2){\vector(1,-1){2}}
\put(-10,2){\vector(1,-1){2}}
\put(-6,2){\vector(1,-1){2}}

\put(-22,-10){\line(1,0){13}}
\put(11,-10){\line(1,0){13}}

\qbezier(15,0)(28,10)(15,20)
\qbezier(11,0)(24,10)(11,20)
\qbezier(7,0)(20,10)(7,20)

\put(17,18){\vector(-1,1){2}}
\put(13,18){\vector(-1,1){2}}
\put(9,18){\vector(-1,1){2}}

\put(-3,0){\line(1,0){10}}

}

\put(0,20){

\qbezier(-11,0)(-28,10)(-11,20)
\qbezier(-7,0)(-24,10)(-7,20)
\qbezier(-3,0)(-20,10)(-3,20)

\put(-14,2){\vector(1,-1){2}}
\put(-10,2){\vector(1,-1){2}}
\put(-6,2){\vector(1,-1){2}}

\put(-22,-10){\line(1,0){13}}
\put(11,-10){\line(1,0){13}}

\qbezier(15,0)(28,10)(15,20)
\qbezier(11,0)(24,10)(11,20)
\qbezier(7,0)(20,10)(7,20)

\put(17,18){\vector(-1,1){2}}
\put(13,18){\vector(-1,1){2}}
\put(9,18){\vector(-1,1){2}}

\put(-3,0){\line(1,0){10}}

}

\qbezier(-11,0)(-28,10)(-11,20)
\qbezier(-7,0)(-24,10)(-7,20)
\qbezier(-3,0)(-20,10)(-3,20)

\put(-14,2){\vector(1,-1){2}}
\put(-10,2){\vector(1,-1){2}}
\put(-6,2){\vector(1,-1){2}}

\qbezier(15,0)(28,10)(15,20)
\qbezier(11,0)(24,10)(11,20)
\qbezier(7,0)(20,10)(7,20)

\put(17,18){\vector(-1,1){2}}
\put(13,18){\vector(-1,1){2}}
\put(9,18){\vector(-1,1){2}}

\put(-3,0){\line(1,0){10}}

\put(-3,0){\vector(-1,-1){10}}
\put(-7,0){\vector(-1,-1){10}}
\put(-11,0){\vector(-1,-1){10}}

\put(17,-10){\vector(-1,1){10}}
\put(21,-10){\vector(-1,1){10}}
\put(25,-10){\vector(-1,1){10}}

\put(12,-9){\line(1.5,1){10}}
\put(-6,-7){\line(-1.5,1){10}}
}

\put(0,105){
\put(48,-67){\mbox{$=$}}

\put(5,0){
\put(77,-65){\line(1,0){12}}
\put(61,-65){\line(1,0){12}}

\put(70,-50){\vector(0,-1){30}}
\put(67,-50){\vector(0,-1){30}}
\put(64,-50){\vector(0,-1){30}}

\put(80,-80){\vector(0,1){30}}
\put(83,-80){\vector(0,1){30}}
\put(86,-80){\vector(0,1){30}}
}

\put(105,-67){\mbox{$+\ \ \ \Phi_n^{[3]}$}}

\put(15,0){
\put(155,-63){\line(0,1){12}}
\put(155,-79){\line(0,1){12}}

\put(135,-60){\vector(1,0){40}}
\put(135,-57){\vector(1,0){40}}
\put(135,-54){\vector(1,0){40}}

\put(175,-70){\vector(-1,0){40}}
\put(175,-73){\vector(-1,0){40}}
\put(175,-76){\vector(-1,0){40}}
}

}

\put(17,105){

\put(185,-67){\mbox{$+\ \ \ \Psi_n^{[3]}$}}

\put(5,0){
\put(225,-50){\vector(1,0){40}}
\put(265,-75){\vector(-1,0){40}}

\qbezier(225,-56)(254,-63)(225,-69)
\qbezier(265,-56)(236,-63)(265,-69)
\put(230,-68){\vector(-1,-0.25){6}}
\put(260,-57){\vector(1,0.25){6}}

\qbezier(225,-53)(260,-63)(225,-72)
\qbezier(265,-53)(230,-63)(265,-72)
\put(230,-70.5){\vector(-1,-0.3){6}}
\put(260,-54.5){\vector(1,0.3){6}}

\put(256,-61){\line(0,1){15}}
\put(256,-79){\line(0,1){15}}
\put(232,-61){\line(0,1){15}}
\put(232,-79){\line(0,1){15}}
}
}

\put(57,105){
\put(242,-67){\mbox{$+\ \ \ {\cal X}_n^{[3]}$}}

\put(65,0){
\put(225,-50){\vector(1,0){40}}
\put(225,-53){\vector(1,0){40}}
\put(265,-75){\vector(-1,0){40}}
\put(265,-72){\vector(-1,0){40}}

\qbezier(225,-56)(254,-63)(225,-69)
\qbezier(265,-56)(236,-63)(265,-69)
\put(230,-68){\vector(-1,-0.25){6}}
\put(260,-57){\vector(1,0.27){6}}

\put(256,-61){\line(0,1){15}}
\put(256,-79){\line(0,1){15}}
\put(232,-61){\line(0,1){15}}
\put(232,-79){\line(0,1){15}}
}
}
}

\end{picture}
\caption{\footnotesize
The horizontal AP lock tangle carrying representation $[3]$ and the {\bf vertical chain operator} obtained by vertical iteration of this lock. 
A chain tangle can be closed in two different ways giving rise to the unknot and to the 2-component 2-strand AP torus links
$APT[2,2n]$, see Fig.\,\ref{fig:chain-[3]-closures}.
%With the exception of Hopf at $n=1$, they are different from the more familiar parallel torus links,
%which are more similar to torus 2-strand {\it knots} with odd number of intersections.
%Two-strand torus {\it knots} are also bipartite (as being rational, see Theorem 2 in Section~\ref{sec:bipknots}) -- but depicted by more sophisticated diagrams.
Since the decomposition of a chain is just the same as for the single AP lock
(the only substitutions are $\phi_3 \rightarrow \Phi_n^{[3]}$, $\psi_3 \rightarrow \Psi_n^{[3]}$, $\chi_3 \rightarrow {\cal X}_n^{[3]}$),
it is especially simple to make further compositions.
}\label{fig:vertchain-[3]}
\end{figure}
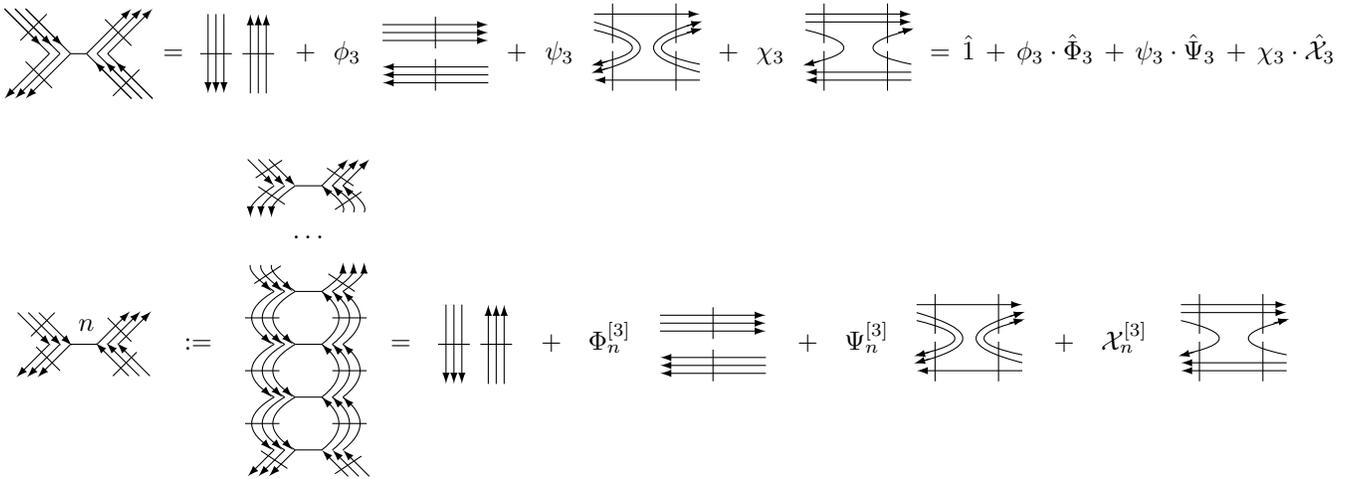

\noindent Again, the planar expansion of an antiparallel vertical chain from Fig.\,\ref{fig:vertchain-[3]} is 
\begin{equation}\label{chain-[3]}
    (\hat{1} + \phi_3 \cdot \hat{\Phi}_3 + \psi_3 \cdot \hat{\Psi}_3 + \chi_3 \cdot \hat{\cal X}_3)^n = \hat{1} + \Phi_n^{[3]} \cdot \hat{\Phi}_3 + \Psi_n^{[3]} \cdot \hat{\Psi}_3 + {\cal X}_n^{[3]} \cdot \hat{\cal X}_3\,.
\end{equation}
In order to find the coefficients $\Phi_n^{[3]}$, $\Psi_n^{[3]}$, ${\cal X}_n^{[3]}$, one needs to sum up the monomials from the expansion of the l.h.s. of~\eqref{chain-[3]} proportional to $\hat{\Phi}_3$, $\hat{\Psi}_3$, $\hat{\cal X}_3$ correspondingly. Thus, we provide several preliminary calculations in short. 

The first important observation is that all the operators $\hat{\Phi}_3$, $\hat{\Psi}_3$, $\hat{\cal X}_3$ are commutative. Second, similarly to Section~\ref{sec:chain}, it is easy to get sure that for $\sum_i k_i \neq 0$
\begin{equation}\label{mon-1-[3]}
\hat\Phi_3^{k_1}\hat\Psi_3^{l_1}\hat{\cal X}_3^{p_1} \ldots
\hat\Phi_3^{k_m} \hat\Psi_3^{l_{m}}\hat{\cal X}_3^{p_m} = \hat\Phi_3^{\sum_i k_i} \hat\Psi_3^{\sum_i l_i} \hat{\cal X}_3^{\sum_i p_i} = \left(\frac{[N+2]}{[3]}\right)^{\sum_i l_i} \left(\frac{[N+2]}{[3]}\cdot \frac{[N+1]}{[2]}\right)^{\sum_i p_i} D_{[3]}^{\sum_i k_i-1}\,\hat\Phi_3\,, %\sim \hat\Phi_3
\end{equation}
where we extensively utilised the reduction formulas from Fig.\,\ref{fig:red-cycle-[3]}. Other monomials have more involved expansions, and can be analogously calculated with the use of the tricks from Figs.\,\ref{fig:proj-prop-2},\,\ref{fig:red-cycle-[3]}. Here they are:
\begin{equation}\label{mon-2-[3]}
\begin{aligned}
    \hat\Psi_3^n &= \left(\frac{[N+4]}{[3]^2}\right)^{n-1}\left(\hat\Psi_3-\frac{[2]^2}{[N+2]} \hat{\cal X}_3 + \frac{[2]}{[N+1][N+2]} \hat\Phi_3\right)+\\
    &+\frac{[2]^2}{[N+2]}\left(\frac{[2][N+3]}{[3]^2}\right)^{n-1}\hat{\cal X}_3-\frac{1}{[N]}\left(\frac{[2]^2}{[N+2]}\left(\frac{[2][N+3]}{[3]^2}\right)^{n-1}-\frac{[2]}{[N+1]}\left(\frac{[N+2]}{[3]}\right)^{n-1}\right)\hat\Phi_3\,, \\
    \hat{\cal X}_3^n &=\left(\frac{[2][N+3]}{[3]^2}\cdot \frac{[N+2]}{[2]^2}\right)^{n-1}\left\{\hat{\cal X}_3+\frac{1}{[N]}\left(\left(\frac{[3][N+1]}{[N+3]}\right)^{n-1}-1\right)\hat\Phi_3\right\}, \\
    \hat\Psi_3^n \hat{\cal X}_3^k &= \left(\frac{[2][N+3]}{[3]^2}\right)^n \left(\hat{\cal X}_3^k - \frac{1}{[N]}\left(\frac{[N+2]}{[3]}\cdot \frac{[N+1]}{[2]}\right)^{k-1}\hat\Phi_3\right)+\frac{1}{[N]}\left(\frac{[N+2]}{[3]}\cdot \frac{[N+1]}{[2]}\right)^{k-1}\left(\frac{[N+2]}{[3]}\right)^n\hat\Phi_3
\end{aligned}
\end{equation}
where in the last expression $n\neq 0$ and $k\neq 0$. Accurately summing up contributions of all the monomials~\eqref{mon-1-[3]},~\eqref{mon-2-[3]}, we finally get
\begin{equation}\label{chain-coef}
\begin{aligned}
    \Phi_n^{[3]}&=(A^3q^{6})^{2n}\cdot D_{[3]}^{-1}\left\{\lambda_\varnothing^{2n} - \frac{[3][N+1]}{[N+3]}\cdot\lambda_{\rm adj}^{2n}+\frac{[3][N]}{[N+4]}\cdot \lambda_{[4,2^{N-2}]}^{2n}-\frac{[N][N+1]}{[N+3][N+4]}\cdot \lambda_{[6,3^{N-2}]}^{2n}\right\}, \\
    {\cal X}_n^{[3]}&=(A^3q^{6})^{2n}\cdot\frac{[2][3]^2}{[N+2][N+3]}\left\{\lambda_{\rm adj}^{2n}-\frac{[2][N+3]}{[N+4]}\cdot \lambda_{[4,2^{N-2}]}^{2n}+\frac{[N+2]}{[N+4]}\cdot \lambda_{[6,3^{N-2}]}^{2n}\right\}, \\
    \Psi_n^{[3]}&=(A^3q^{6})^{2n}\cdot\frac{[3]^2}{[N+4]}\left\{\lambda_{[4,2^{N-2}]}^{2n}-\lambda_{[6,3^{N-2}]}^{2n}\right\}.
\end{aligned}
\end{equation}
Here we denote 
\begin{equation}\label{eigen-[3]}
\begin{aligned}
    \lambda_{[6,3^{N-2}]}^2&=A^{-6} q^{-12}=\left(A^{-3} q^{-6}\right)^2\,, \\
    \lambda_{[4,2^{N-2}]}^2&=A^{-6} q^{-12}\left(1+\psi_3\cdot\frac{[N+4]}{[3]^2}\right)=\left(A^{-2} q^{-2}\right)^2\,, \\
    \lambda_{\rm adj}^2&=A^{-6} q^{-12}\left(1+\psi_3\cdot\frac{[2][N+3]}{[3]^2}+\chi_3\cdot \frac{[2][N+3]}{[3]^2}\cdot \frac{[N+2]}{[2]^2}\right)=A^{-2}\,, \\
    \lambda_{\varnothing}^2&=A^{-6} q^{-12}\left(1+\phi_3\cdot D_{[3]}+\psi_3\cdot\frac{[N+2]}{[3]}+\chi_3\cdot \frac{[N+1]}{[2]}\cdot \frac{[N+2]}{[3]}\right)=1\,,
\end{aligned}
\end{equation}
where final answers for $\lambda_\varnothing$, $\lambda_{\rm adj}$, $\lambda_{[4,2^{N-2}]}$ are the same as in~\eqref{eigen}, but their intermediate bipartite forms are different. These answers are in correspondence with the evolution method results. In this particular case of 2-strand braids, each strand carries the representation $[3]$. Due to the diagonality of the $\cal R$-matrix in a basis of irreducible representations, answers are expressed through the powers of the $\cal R$-matrix eigenvalues.

The coefficients~\eqref{chain-coef} obey the following relations:
\begin{equation}
\begin{aligned}
    \Phi_{n+m}^{[3]}&=\Phi_n^{[3]}+\Phi_m^{[3]}+\Phi_n^{[3]}\Phi_m^{[3]}D_{[3]}+\frac{[N+2]}{[3]}\left(\Phi_n^{[3]}\Psi_m^{[3]}+\Phi_m^{[3]}\Psi_n^{[3]}\right)+\\
    &+\frac{[N+1]}{[2]}\cdot \frac{[N+2]}{[3]}\left(\Phi_n^{[3]}{\cal X}_m^{[3]}+\Phi_m^{[3]}{\cal X}_n^{[3]}\right)+\frac{1}{[3]^2}\left(\Psi_n^{[3]}{\cal X}_m^{[3]}+\Psi_m^{[3]}{\cal X}_n^{[3]}\right)+\frac{[N+2]}{[3]^2}{\cal X}_n^{[3]}{\cal X}_m^{[3]}\,, \\
    {\cal X}_{n+m}^{[3]}&={\cal X}_{n}^{[3]}+{\cal X}_{m}^{[3]}+\frac{[2][N+3]}{[3]^2}\cdot\frac{[N+2]}{[2]^2}{\cal X}_{n}^{[3]}{\cal X}_{m}^{[3]}+\frac{[2]^2}{[3]^2}\Psi_n^{[3]}\Psi_m^{[3]}+\frac{[2][N+3]}{[3]^2}\left(\Psi_n^{[3]}{\cal X}_m^{[3]}+\Psi_m^{[3]}{\cal X}_n^{[3]}\right), \\
    \Psi_{n+m}^{[3]}&= \Psi_{n}^{[3]}+\Psi_{m}^{[3]}+\frac{[N+4]}{[3]^2}\Psi_{n}^{[3]}\Psi_{m}^{[3]}\,,
\end{aligned}
\end{equation}
and in particular,
{\small \begin{equation}\nn
\begin{aligned}
    0&=\phi_3 + \bar\phi_3 + \phi_3 \bar\phi_3 D_{[3]}+\frac{[N+2]}{[3]}\left(\phi_3 \bar\psi_3 + \bar\phi_3 \psi_3\right)+\frac{[N+1]}{[2]}\cdot \frac{[N+2]}{[3]}\left(\phi_3\bar\chi_3+ \bar\phi_3 \chi_3\right)+\frac{1}{[3]^2}\left(\psi_3 \bar\chi_3 + \bar\psi_3 \chi_3\right)+\frac{[N+2]}{[3]^2}\chi_3 \bar\chi_3\,, \\
    0&= \chi_3 + \bar\chi_3 + \frac{[2][N+3]}{[3]^2}\cdot\frac{[N+2]}{[2]^2} \chi_3 \bar\chi_3 + \frac{[2]^2}{[3]^2} \psi_3 \bar\psi_3 + \frac{[2][N+3]}{[3]^2} \left(\psi_3 \bar\chi_3 + \bar\psi_3 \chi_3\right)\,, \\
    0&= \psi_3 + \bar\psi_3 + \frac{[N+4]}{[3]^2} \psi_3 \bar\psi_3\,,
\end{aligned}
\end{equation}}
which are analogues of~\eqref{coef-rel-[1]},~\eqref{rel-(m+n)},~\eqref{rel-(m+n)-part}. 

Two different closures of the vertical chain from Fig.\,\ref{fig:vertchain-[3]} give rise to unknots and antiparallel torus links $APT[2,2n]$, see Fig.\,\ref{fig:chain-[3]-closures}. Their HOMFLY polynomials are obtained from those ones for the unknot and the Hopf link from Fig.\,\ref{fig:singlelockclosures3} under the change $\phi_3 \rightarrow \Phi_n^{[3]}$, $\psi_3 \rightarrow \Psi_n^{[3]}$, $\chi_3 \rightarrow {\cal X}_n^{[3]}$. Explicit answers are given in the next subsections. Further examples can be straightforwardly calculated by gluing chain operators together and using the planar expansion from Fig.\,\ref{fig:vertchain-[3]}. We do not provide such examples explicitly in what follows.

\begin{figure}[h!]
\begin{picture}(100,175)(40,-72)

\put(150,50){

\put(12,5){\mbox{$n$}}

\put(15,0){
\put(-20,17){\line(1,-1){17}}\put(-20,17){\vector(1,-1){14}}   \put(3,0){\vector(1,1){17}}
\put(-3,0){\vector(-1,-1){17}}   \put(20,-17){\line(-1,1){17}} \put(20,-17){\vector(-1,1){14}}
\put(-3,0){\line(1,0){6}}
\put(-23,17){\line(1,-1){17}}\put(-23,17){\vector(1,-1){14}}   \put(6,0){\vector(1,1){17}}
\put(-6,0){\vector(-1,-1){17}}   \put(23,-17){\line(-1,1){17}} \put(23,-17){\vector(-1,1){14}}
\put(-26,17){\line(1,-1){17}}\put(-26,17){\vector(1,-1){14}}   \put(9,0){\vector(1,1){17}}
\put(-9,0){\vector(-1,-1){17}}   \put(26,-17){\line(-1,1){17}} \put(26,-17){\vector(-1,1){14}}
\put(-20,4){\line(1,1){10}}  \put(-20,-4){\line(1,-1){10}}
%\put(19,6){\line(-1,1){10}}  \put(19,-6){\line(-1,-1){10}}
\qbezier(-20,17)(-40,42)(0,42) \qbezier(20,17)(40,42)(0,42)
\qbezier(-20,-17)(-40,-42)(0,-42) \qbezier(20,-17)(40,-42)(0,-42)
\qbezier(-23,17)(-43,45)(0,45) \qbezier(23,17)(43,45)(0,45)
\qbezier(-23,-17)(-43,-45)(0,-45) \qbezier(23,-17)(43,-45)(0,-45)
\qbezier(-26,17)(-46,48)(0,48) \qbezier(26,17)(46,48)(0,48)
\qbezier(-26,-17)(-46,-48)(0,-48) \qbezier(26,-17)(46,-48)(0,-48)
}

\put(55,-2){\mbox{$=$}}

\put(80,-2){\mbox{$D_{[3]} \ \ + \ \ \Phi_n^{[3]}\cdot D_{[3]}^2 \ \ + \ \ \Psi_n^{[3]}\cdot K^{(1)}_3 \ \ + \ \ {\cal X}_n^{[3]}\cdot K^{(2)}_3$}}
}

%-----------------------------------

\put(152,-45){

\put(-3,5){\mbox{$n$}}

\put(-20,17){\line(1,-1){17}}\put(-20,17){\vector(1,-1){14}}   \put(3,0){\vector(1,1){17}}
\put(-3,0){\vector(-1,-1){17}}   \put(20,-17){\line(-1,1){17}} \put(20,-17){\vector(-1,1){14}}
\put(-3,0){\line(1,0){6}}
\put(-23,17){\line(1,-1){17}}\put(-23,17){\vector(1,-1){14}}   \put(6,0){\vector(1,1){17}}
\put(-6,0){\vector(-1,-1){17}}   \put(23,-17){\line(-1,1){17}} \put(23,-17){\vector(-1,1){14}}
\put(-26,17){\line(1,-1){17}}\put(-26,17){\vector(1,-1){14}}   \put(9,0){\vector(1,1){17}}
\put(-9,0){\vector(-1,-1){17}}   \put(26,-17){\line(-1,1){17}} \put(26,-17){\vector(-1,1){14}}

\put(-20,4){\line(1,1){10}}  
\put(20,4){\line(-1,1){10}}  

\qbezier(-20,17)(-40,42)(-40,0) \qbezier(-20,-17)(-40,-42)(-40,0)
\qbezier(-23,17)(-37,36)(-37,0) \qbezier(-23,-17)(-37,-36)(-37,0)
\qbezier(-26,17)(-34,30)(-34,0) \qbezier(-26,-17)(-34,-30)(-34,0)
\qbezier(20,17)(40,42)(40,0) \qbezier(20,-17)(40,-42)(40,0)
\qbezier(23,17)(37,36)(37,0) \qbezier(23,-17)(37,-36)(37,0)
\qbezier(26,17)(34,30)(34,0) \qbezier(26,-17)(34,-30)(34,0)

\put(55,-2){\mbox{$=$}}

\put(80,-2){\mbox{$D_{[3]}^2 \ \ + \ \ \Phi_n^{[3]}\cdot D_{[3]} \ \ + \ \  \Psi_n^{[3]}\cdot K^{(2)}_3 \ \ + \ \ {\cal X}_n^{[3]} \cdot K^{(1)}_3$}}
}

\end{picture}
\caption{\footnotesize  The two closures of a multiple lock: unknots in the first line
and $APT[2,2n]$ link in the second line.
Recall that in representation $R=[3]$, transverse lines denote projectors $P_{[3]}$.
} \label{fig:chain-[3]-closures}
\end{figure}

\subsection{Unknots from 2-strand braids}

We have already calculated the HOMFLY polynomial for the unknot being the closure of a single lock tangle~\eqref{unknot-[3]}:
\begin{equation}
    H^{\rm unknot}_{[3]} = \frac{1}{q^{12}A^6} \left(D_{[3]} + \phi_3\cdot D_{[3]}^2 + \psi_3\cdot K^{(1)}_3 + \chi_3\cdot K^{(2)}_3 \right) =
D_{[3]}\,.
\end{equation}
Due to the previous subsection, it is enough to change $\phi_3 \rightarrow \Phi_n^{[3]}$, $\psi_3 \rightarrow \Psi_n^{[3]}$, $\chi_3 \rightarrow {\cal X}_n^{[3]}$ and the framing factor to obtain the expression for the unknot made from $n$ AP locks as presented in Fig.\,\ref{fig:chain-[3]-closures}\,:

\begin{equation}
\begin{aligned}
    H^{\rm unknot}_{[3]} &= (q^{6}A^3)^{-2n} \left(D_{[3]} + \Phi_n^{[3]}\cdot D_{[3]}^2 + \Psi_n^{[3]}\cdot K^{(1)}_3 + {\cal X}_n^{[3]}\cdot K^{(2)}_3 \right)=\\
    &=(q^{6}A^3)^{-2n} \cdot D_{[3]}\left(1+\phi_3\cdot D_{[3]}+\psi_3\cdot\frac{[N+2]}{[3]}+\chi_3\cdot \frac{[N+1]}{[2]}\cdot \frac{[N+2]}{[3]}\right)^n =
D_{[3]}\,.
\end{aligned}
\end{equation}

\subsection{Antiparallel torus links from 2-strand braids}

An antiparallel torus link is another closure of the chain operator as shown in Fig.\,\ref{fig:chain-[3]-closures}. Thus, the expression for the HOMFLY polynomial in the representation $[3]$ follows: 
\begin{equation}
\begin{aligned}
    H_{[3]}^{APT[2,2n]}&=\frac{1}{(A^3q^6)^{2n}}\cdot \Tr (1+ \phi_3\cdot \hat\Phi_3+\psi_3\cdot \hat \Psi_3 + \chi_3 \cdot \hat {\cal X}_3)^n=\frac{1}{(A^3q^6)^{2n}}\left(D_{[3]}^2+\Phi_n^{[3]}D_{[3]}+\Psi_n^{[3]}K^{(2)}_3+{\cal X}_n^{[3]}K_{3}^{(1)}\right)=\\
    &=1 + \frac{1}{A^{2n}}\frac{\{Aq\}\{A/q\}}{\{q\}^2} + \frac{1}{(q^2A^2)^{2n}}\frac{\{Aq^3\}\{A\}^2\{A/q\}}{\{q\}^2\{q^2\}^2}+\frac{1}{\left(A^3 q^6\right)^{2n}}\frac{\{A q^5\}\{A q\}^2\{A\}^2\{A/q\}}{\{q\}^2\{q^2\}^2\{q^3\}^2}=\\
    &=1+\lambda_{\rm adj}^{2n}D_{\rm adj}+\lambda_{[4,2^{N-2}]}^{2n}D_{[4,2^{N-2}]}+\lambda_{[6,3^{N-2}]}^{2n}D_{[6,3^{N-2}]}\,,
\end{aligned}
\end{equation}
and we have obtained the usual character expansion~\cite{mironov2013character,mironov2012character,itoyama2012character} result.

%\newpage???

\setcounter{equation}{0}
\section{Generic $R=[r]$}\label{sec:rep-[r]}

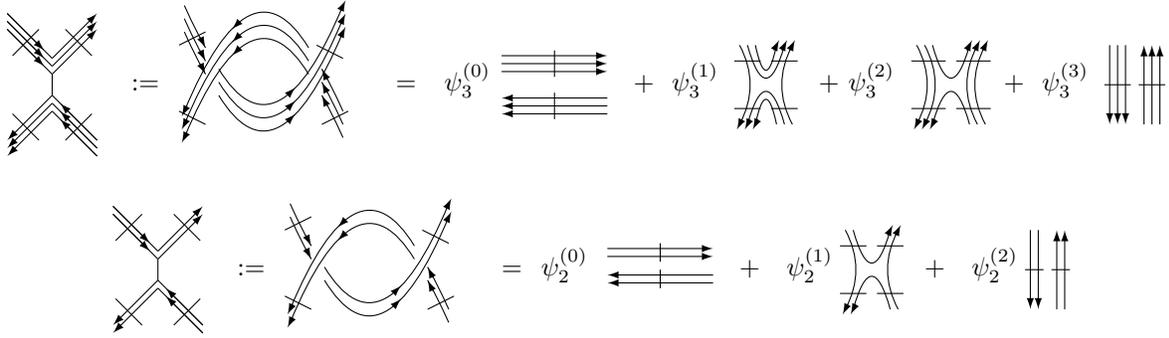
\begin{figure}[h!]
\begin{picture}(100,150)(20,-40)

\put(-40,70){
\put(100,0){
\put(-17,21){\line(1,-1){17}}\put(-17,24){\line(1,-1){17}} \put(-17,27){\line(1,-1){17}}
\put(-17,21){\vector(1,-1){14}} \put(-17,24){\vector(1,-1){14}}   \put(-17,27){\vector(1,-1){14}}
\put(0,4){\vector(1,1){17}} \put(0,7){\vector(1,1){17}} \put(0,10){\vector(1,1){17}}
\put(0,-4){\vector(-1,-1){17}}   \put(17,-21){\line(-1,1){17}} \put(17,-21){\vector(-1,1){14}}
\put(0,-7){\vector(-1,-1){17}}   \put(17,-24){\line(-1,1){17}} \put(17,-24){\vector(-1,1){14}}
\put(0,-10){\vector(-1,-1){17}}   \put(17,-27){\line(-1,1){17}} \put(17,-27){\vector(-1,1){14}}
\put(0,4){\line(0,-1){8}}

\put(-15,11){\line(1,1){10}}  \put(-15,-11){\line(1,-1){10}}
\put(15,11){\line(-1,1){10}}  \put(15,-11){\line(-1,-1){10}}

\put(30,-2){\mbox{$:=$}}

\qbezier(50,20)(55,9)(58,4) \qbezier(63,-4)(85,-40)(110,20)
\put(56,8){\vector(1,-2){2}} \put(90,-13){\vector(1,1){2}} \put(109,18){\vector(1,2){2}}
\qbezier(50,-20)(75,40)(97,4)  \qbezier(102,-4)(105,-9)(110,-20)
\put(104,-8){\vector(-1,2){2}} \put(70,13){\vector(-1,-1){2}} \put(51,-18){\vector(-1,-2){2}}

\qbezier(50,25)(55,14)(58,9) \qbezier(63,1)(85,-35)(110,25)
\put(56,13){\vector(1,-2){2}} \put(90,-8){\vector(1,1){2}} \put(109,23){\vector(1,2){2}}
\qbezier(50,-15)(75,45)(97,9)  \qbezier(102,1)(105,-4)(110,-15)
\put(104,-3){\vector(-1,2){2}} \put(70,18){\vector(-1,-1){2}} \put(51,-13){\vector(-1,-2){2}}

\qbezier(50,30)(55,19)(58,14) \qbezier(63,6)(85,-30)(110,30)
\put(56,18){\vector(1,-2){2}} \put(90,-3){\vector(1,1){2}} \put(109,28){\vector(1,2){2}}
\qbezier(50,-10)(75,50)(97,14)  \qbezier(102,6)(105,1)(110,-10)
\put(104,2){\vector(-1,2){2}} \put(70,23){\vector(-1,-1){2}} \put(51,-8){\vector(-1,-2){2}}

\put(48,15){\line(2,1){10}}  \put(48,-10){\line(2,-1){10}}
\put(110,10){\line(-2,1){10}}  \put(112,-10){\line(-2,-1){10}}

\put(130,-2){\mbox{$=$}}

\put(100,65){

\put(45,-67){\mbox{$\  \psi_3^{(0)}$}}
\put(70,-60){\vector(1,0){40}} \put(70,-57){\vector(1,0){40}}  \put(70,-54){\vector(1,0){40}}
\put(110,-70){\vector(-1,0){40}}  \put(110,-73){\vector(-1,0){40}} \put(110,-76){\vector(-1,0){40}}
\put(90,-52){\line(0,-1){10}}   \put(90,-68){\line(0,-1){10}}

\put(-70,0){
\put(190,-67){\mbox{$+\ \  \psi_3^{(1)}$}}
\qbezier(230,-50)(238,-65)(230,-80) \put(231,-78){\vector(-1,-2){2}}
\qbezier(250,-50)(242,-65)(250,-80) \put(249,-52){\vector(1,2){2}}
%\put(230,-50){\vector(0,-1){30}}
%\put(250,-80){\vector(0,1){30}}
\qbezier(233,-50)(240,-75)(247,-50)  \put(246,-52){\vector(1,2){2}}
\qbezier(236,-50)(240,-69)(244,-50)  \put(243,-52){\vector(1,2){2}}
\qbezier(236,-80)(240,-61)(244,-80)  \put(237,-78){\vector(-1,-2){2}}
\qbezier(233,-80)(240,-55)(247,-80)  \put(234,-78){\vector(-1,-2){2}}
\put(228,-56){\line(1,0){10}}   \put(242,-56){\line(1,0){10}}
\put(228,-74){\line(1,0){10}}   \put(242,-74){\line(1,0){10}}
}

\put(190,-67){\mbox{$+\  \psi_3^{(2)}$}}
\qbezier(230,-50)(238,-65)(230,-80) \put(231,-78){\vector(-1,-2){2}}
\qbezier(227,-50)(235,-65)(227,-80) \put(228,-78){\vector(-1,-2){2}}
\qbezier(250,-50)(242,-65)(250,-80) \put(249,-52){\vector(1,2){2}}
\qbezier(253,-50)(245,-65)(253,-80) \put(252,-52){\vector(1,2){2}}
%\put(230,-50){\vector(0,-1){30}}  \put(227,-50){\vector(0,-1){30}}
%\put(250,-80){\vector(0,1){30}}   \put(253,-80){\vector(0,1){30}}
\qbezier(233,-50)(240,-75)(247,-50)  \put(246,-52){\vector(1,2){2}}
\qbezier(233,-80)(240,-55)(247,-80)  \put(234,-78){\vector(-1,-2){2}}
\put(225,-56){\line(1,0){13}}   \put(242,-56){\line(1,0){13}}
\put(225,-74){\line(1,0){13}}   \put(242,-74){\line(1,0){13}}

\put(140,0){
\put(120,-67){\mbox{$+\ \ \psi_3^{(3)}$}}
\put(160,-50){\vector(0,-1){30}} \put(163,-50){\vector(0,-1){30}} \put(166,-50){\vector(0,-1){30}}
\put(173,-80){\vector(0,1){30}}  \put(176,-80){\vector(0,1){30}} \put(179,-80){\vector(0,1){30}}
\put(158,-65){\line(1,0){10}}   \put(171,-65){\line(1,0){10}}
}
}}
}

%----------------------------------------------------

\put(100,0){
%\put(-20,17){\line(1,-1){17}}\put(-20,17){\vector(1,-1){14}}   \put(3,0){\vector(1,1){17}}
%\put(-3,0){\vector(-1,-1){17}}   \put(20,-17){\line(-1,1){17}} \put(20,-17){\vector(-1,1){14}}
%\put(-3,0){\line(1,0){6}}
%\put(-23,17){\line(1,-1){17}}\put(-23,17){\vector(1,-1){14}}   \put(6,0){\vector(1,1){17}}
%\put(-6,0){\vector(-1,-1){17}}   \put(23,-17){\line(-1,1){17}} \put(23,-17){\vector(-1,1){14}}
%\put(-19,6){\line(1,1){10}}  \put(-19,-6){\line(1,-1){10}}
%\put(19,6){\line(-1,1){10}}  \put(19,-6){\line(-1,-1){10}}

\put(-17,21){\line(1,-1){17}}\put(-17,24){\line(1,-1){17}}   \put(0,4){\vector(1,1){17}}
\put(-17,21){\vector(1,-1){14}} \put(-17,24){\vector(1,-1){14}}
\put(0,7){\vector(1,1){17}}
\put(0,-4){\vector(-1,-1){17}}   \put(17,-21){\line(-1,1){17}} \put(17,-21){\vector(-1,1){14}}
\put(0,-7){\vector(-1,-1){17}}   \put(17,-24){\line(-1,1){17}} \put(17,-24){\vector(-1,1){14}}
\put(0,4){\line(0,-1){8}}

\put(-16,11){\line(1,1){10}}  \put(-16,-11){\line(1,-1){10}}
\put(16,11){\line(-1,1){10}}  \put(16,-11){\line(-1,-1){10}}

\put(30,-2){\mbox{$:=$}}

\qbezier(50,20)(55,9)(58,4) \qbezier(63,-4)(85,-40)(110,20)
\put(56,8){\vector(1,-2){2}} \put(90,-13){\vector(1,1){2}} \put(109,18){\vector(1,2){2}}
\qbezier(50,-20)(75,40)(97,4)  \qbezier(102,-4)(105,-9)(110,-20)
\put(104,-8){\vector(-1,2){2}} \put(70,13){\vector(-1,-1){2}} \put(51,-18){\vector(-1,-2){2}}

\qbezier(50,25)(55,14)(58,9) \qbezier(63,1)(85,-35)(110,25)
\put(56,13){\vector(1,-2){2}} \put(90,-8){\vector(1,1){2}} \put(109,23){\vector(1,2){2}}
\qbezier(50,-15)(75,45)(97,9)  \qbezier(102,1)(105,-4)(110,-15)
\put(104,-3){\vector(-1,2){2}} \put(70,18){\vector(-1,-1){2}} \put(51,-13){\vector(-1,-2){2}}

\put(48,15){\line(2,1){10}}  \put(48,-10){\line(2,-1){10}}
\put(110,10){\line(-2,1){10}}  \put(112,-10){\line(-2,-1){10}}

\put(130,-2){\mbox{$=$}}

\put(100,65){
\put(45,-67){\mbox{$ \psi_2^{(0)}$}}
\put(70,-60){\vector(1,0){40}} \put(70,-57){\vector(1,0){40}}
\put(110,-70){\vector(-1,0){40}}  \put(110,-67){\vector(-1,0){40}}
\put(90,-55){\line(0,-1){7}}   \put(90,-65){\line(0,-1){7}}

\put(-70,0){
\put(190,-67){\mbox{$+\ \ \ \psi_2^{(1)}$}}

\qbezier(230,-50)(238,-65)(230,-80) \put(231,-78){\vector(-1,-2){2}}
\qbezier(250,-50)(242,-65)(250,-80) \put(249,-52){\vector(1,2){2}}
%\put(230,-50){\vector(0,-1){30}}
%\put(250,-80){\vector(0,1){30}}
\qbezier(233,-50)(240,-75)(247,-50)  \put(246,-52){\vector(1,2){2}}
\qbezier(233,-80)(240,-55)(247,-80)  \put(234,-78){\vector(-1,-2){2}}
\put(228,-56){\line(1,0){10}}   \put(242,-56){\line(1,0){10}}
\put(228,-74){\line(1,0){10}}   \put(242,-74){\line(1,0){10}}
}

\put(70,0){
\put(120,-67){\mbox{$+\ \ \ \psi_2^{(2)}$}}
\put(160,-50){\vector(0,-1){30}} \put(163,-50){\vector(0,-1){30}}
\put(170,-80){\vector(0,1){30}}  \put(173,-80){\vector(0,1){30}}
\put(158,-65){\line(1,0){7}}   \put(168,-65){\line(1,0){7}}
}
}}

\end{picture}
\caption{\footnotesize
The universal notation for the coefficients $\psi_r^{(i)}$ in the planar decomposition.
The vertical lock tangle projected to the symmetric representation $[r]$
and its planar decomposition.
For a vertical lock there are
$(r-i)$ horizontal lines and $i$ vertical ones.
The very first coefficient for all $r$ lines horizontal is $\psi_r^{(0)} = 1$.
} \label{projerlock}
\end{figure}

\noindent Planar decomposition for an arbitrary symmetric representation $R=[r]$ contains $(r+1)$ terms.
The unified notation for the coefficients should be clear from  Fig.\,\ref{projerlock}.
They are equal to
\be\label{psi_r^(i)}
\psi_r^{(i)} =  q^{\frac{(4r-i-3)i}{2}} \frac{(A\{q\})^i}{[i]!} \left(\frac{[r]!}{[r-i]!}\right)^2,
\ee
and the two coefficients at the boarders of the sum are
\be
\psi_r^{(0)} = 1, \ \ \ \ \ \ \phi_r := q^{\frac{3r(r-1)}{2}}A^r [r]!\{q\}^r = \psi_r^{(r)}\,.
\ee
For the mirror orientation we get
\be
{\bar\psi_r^{(i)}} =  q^{-\frac{(4r-i-3)i}{2}} \frac{\left(-\{q\}/A\right)^i}{[i]!} \left(\frac{[r]!}{[r-i]!}\right)^2.
\ee
The unknot from one lock and the Hopf link possess the following planar decompositions:
\be
H^{\rm unknot}_{[r]} = D_{[r]} = \frac{1}{q^{2r(r-1)}A^{2r}}\left(D_{[r]} + \phi_r D_{[r]}^2+ \sum_{i=1}^{r-1} \psi_{r}^{(i)}K_r^{(i)}\right),
\label{unknotr}
\ee
\be
H^{\rm Hopf}_{[r]}= \frac{1}{q^{2r^2}A^{2r}}\sum_{i=0}^r q^{2(r-i)(r-i+1)} D_{[2r-i,i]}
=   1 +  \sum_{i=1}^r \frac{1}{q^{2i(i-1)}A^{2i}}\frac{\{Aq^{2i-1}\}\{A/q\}}{\{q^i\}^2}\prod_{j=0}^{i-2} \frac{\{Aq^j\}^2}{\{q^{i+1}\}^2}
= \nn \\
=  \frac{1}{q^{2r(r-1)}A^{2r}} \left(D_{[r]}^2 + \phi_r D_{[r]} +  \sum_{i=1}^{r-1} \psi_{r}^{(r-i)}K_r^{(i)}\right),
\label{Hopfr}
\ee
with the projector combinations
\be\label{K_r^(i)}
K_r^{(i)} = \frac{[N+r-1]!}{[N-1]![r]!} \frac{[N+r-1]![r-i]!}{[N+r-i]![r]!}
= \prod_{j=0}^{r-1}\frac{\{Aq^j\}}{\{q^{j+1}\}} \prod_{j=r-i}^{r-1}\frac{\{Aq^j\}}{\{q^{j+1}\}}\,.
\ee
They correspond to two $i$-fold circles, surrounded by the $(r-i)$-fold oval,
as clear from Figs.\,\ref{fig:singlelockclosures},\,\ref{fig:singlelockclosures3}.
Note that the sequences of $\psi_r^{(i)}$ are the opposite in (\ref{unknotr}) and (\ref{Hopfr}).

Calculation of the vertical chain in an arbitrary representation $[r]$ is not so difficult as it may seem from straightforward computations of Sections~\ref{sec:chain},~\ref{sec:chain-[3]}. We know explicitly the answers for the $\cal R$-matrix eigenvalues for a generic representation:
\begin{equation}\label{eigen-answ}
    \lambda_i = \left(A
q^{i-1}\right)^{-i}
\end{equation}
where index $i$ corresponds to irreducible representations $[r]\otimes
\overline{[r]}=\bigoplus\limits_{i=0}^{r} [r+i,r^{N-2},r-i]$. The framing factor assigned to each crossing to restore topological invariance is equal to $A^{-r} q^{-r(r-1)}=\lambda_r$. The simplifying assumption is that the planar decomposition for these eigenvalues takes a triangular form, as it can be seen from the simplest examples~\eqref{eigen-[1]},~\eqref{eigen},~\eqref{eigen-[3]}. So that in general we write 
\begin{equation}
    \lambda_{r-k}^2=\lambda_r^2 \left(1+\sum\limits_{j=1}^k \psi_r^{(j)}B_{r,k}^{(j)}\right),
\end{equation}
and substituting~\eqref{eigen-answ}, we find the coefficients $B_{r,k}^{(j)}$\,:
\begin{equation}
    B_{r,k}^{(j)}=\frac{[k]!}{[k-j]!}\frac{[N+2r-k-1]!}{[N+2r-k-1-j]!}\left(\frac{[r-j]!}{[r]!}\right)^2.
\end{equation}
Now, we can compute coefficients $\Psi_n^{(i)[r]}$ of the planar decomposition of the chain tangle. Again, we use their triangularity property (see formulas~\eqref{chain-[1]},~\eqref{Phi_n^[2]},~\eqref{Psi_n^[2]},~\eqref{chain-coef} for the lowest representations), the correspondence $\Psi_1^{(i)[r]}=\psi_r^{(i)}$ with $\psi_r^{(i)}$ given by~\eqref{psi_r^(i)} and the condition that the HOMFLY polynomial for unknots is equal to $D_{[r]}$. As the result, we obtain
\begin{equation}\label{Psi_n^(i)[r]}
    \Psi_n^{(i)[r]}=\frac{1}{[i]!}\left(\frac{[r]!}{[r-i]!}\right)^2\sum\limits_{j=0}^i (-1)^{i+j}\frac{[i]!}{[i-j]![j]!}\frac{[N+2r-(2j+1)]}{\prod\limits_{k=0}^i [N+2r-j-1-k]}\left(\frac{\lambda_{r-j}}{\lambda_{r}}\right)^{2n}.
\end{equation}
This result allows to efficiently calculate HOMFLY polynomials in (anti-)symmetric representations\footnote{We remind that for computations for $R=[1^r]$, one just changes $q\rightarrow q^{-1}$ in all quantities regarding the representation $[r]$, in particular in formulas~\eqref{K_r^(i)} and~\eqref{Psi_n^(i)[r]}.} for bipartite links composed from chains. It is a straightforward and easy calculation, especially after explicitly provided examples, and thus, we do not write down here these simple computations.

\setcounter{equation}{0}
\section{Conclusion}

In this paper we have explained that the planar decomposition of antiparallel lock tangles \cite{ALM},
which leads to efficient Kauffman-like calculus of the HOMFLY polynomials for bipartite knots and links,
can be further extended to arbitrary symmetric representations.
This calculus is less elementary than in the fundamental case, because projectors from $[1]^{\otimes r}$ to $[r]$
should be included, and their combinations can be different for different knots/links.
This diminishes universality and straightforwardness of the formalism.
Still, it is relatively simple, easily applicable beyond arborescent calculus of \cite{mironov2015colored}
and conceptually interesting. We have computed the planar decomposition of chain tangles being iterations of $n$ locks in an arbitrary representation $[r]$. This allows one to fastly calculate the HOMFLY polynomial in any (anti-)symmetric representation for a bipartite link glued from these 2-braids.
Only the simplest examples of unknots, 2-strand AP links, twist knots and double braid knots are explicitly provided for illustration,
but many more can be easily calculated and will be presented elsewhere.

Being direct and non-trivial generalization of Kauffman bracket machinery,
this formalism can be also used in applications like \cite{MMMMMel}, \cite{AndMor}
and especially in their version in \cite{Mel}. Moreover, the results of this paper can shed light on the intricate design of colored Khovanov(--Rozansky) polynomials~\cite{queffelec2016sln,robert2020closed,khovanov2000categorification,hedden2009khovanov,mackaay2007bar,beliakova2008categorification,caprau2014cohomology,ito2014colored,rozansky2014khovanov,willis2021khovanov,rose2016deformations,anokhina2024towards}. Namely, one can organize cycle decomposition in the case of an arbitrary representation $[r]$ in a hypercube of resolutions, and try to assign differentials between spaces corresponding to each combination of cycles. The limit to the colored HOMFLY polynomial is known due to our study.

\section*{Acknowledgements}

Our work is supported by the RSF grant No.23-41-00049.

\printbibliography

\end{document}